\documentclass[12pt]{article}
\usepackage{latexsym,amssymb}

\def\ide{{\mathbf{1}}}
\def\ii{{ \mathrm{i}}}
\def\ee{{\mathrm{e}}}
\def\tilde{\widetilde}
\def\hat{\widehat}
\def\n010{N^{0,1,0}}
\def\tn010{\tilde{N}^{0,1,0}}
\def\det{\mathrm{det}}
\def\lap{\boxtimes}
\def\vev#1{\langle #1 \rangle}

\def\sn{\mathcal{N}}

\renewcommand{\arraystretch}{1.3}
\newcommand{\ft}[2]{{\textstyle\frac{#1}{#2}}}
\def\tilde{\widetilde}

\def\1bar{1\hskip -.275cm -}
\def\2bar{2\hskip -.275cm -}
\def\3bar{3\hskip -.275cm -}
\newcommand{\size}{\tiny}
\newsavebox{\uuunit}
\sbox{\uuunit}
    {\setlength{\unitlength}{0.825em}
     \begin{picture}(0.6,0.7)
        \thinlines
        \put(0,0){\line(1,0){0.5}}
        \put(0.15,0){\line(0,1){0.7}}
        \put(0.35,0){\line(0,1){0.8}}
       \multiput(0.3,0.8)(-0.04,-0.02){10}{\rule{0.5pt}{0.5pt}}
     \end {picture}}

\makeatletter
\@addtoreset{equation}{section}
\makeatother

\begin{document}
\begin{titlepage}
\begin{flushright}
hep-th/0005220 \\
\end{flushright}
\vskip 2cm
\begin{center}
{{\Large \bf Shadow multiplets in AdS$_4$/CFT$_3$ and the super-Higgs
mechanism}\,\,$^\dagger$}\\
\vskip 2cm
{\large M. Bill\'o$^1$, D. Fabbri$^1$,  P. Fr\'e$^1$, P. Merlatti$^1$
and A. Zaffaroni$^2$} \\
\vfill
{ \sl $^1$ Dipartimento di Fisica Teorica, Universit\'a di Torino, \\ via P.
Giuria 1,
I-10125 Torino, \\
and \\
 Istituto Nazionale di Fisica Nucleare (INFN) - Sezione di Torino,
Italy \\
$^2$ Istituto Nazionale di Fisica Nucleare (INFN) - Sezione di Milano,
Italy }
\end{center}
\vfill
\begin{abstract}
{We discuss a general pairing that occurs in compactifications of M--theory on
$\mathrm{AdS}_4 \times X^7$ backgrounds  between massless ultra short
multiplets and their massive shadows, namely certain universal  long multiplets
with fixed protected dimensions. In particular we consider the shadow   of the
short graviton multiplet in $\mathcal{N}=3$ compactifications. It turns out to
be a massive spin $\ft 32$ multiplet with scale dimension $E_0=3$ and with the
quantum numbers of a superHiggs multiplet. Hence each $\mathcal{N}=3$
$\mathrm{AdS}_4 \times X^7$ vacuum is actually to be interpreted as a
spontaneously broken phase of an ${\cal N}=4$ theory. Comparison with standard
gauged $\mathcal{N}=4$ supergravity in $4$ dimensions reveals  the unexpected
bound $E_0 < 3$ on the dimension of the broken gravitino multiplet. This hints
to the existence of new  versions of extended supergravities, in particular
$\mathcal{N}=4$ where such upper bounds are evaded and where all possible vacua
have a reduced supersymmetry $\mathcal{N}_0 < \mathcal{N}$. We name them shadow
supergravities. In particular, using arguments based on the solvable Lie
algebra parametrization of the scalar manifold, we discuss the possible
structure of shadow $\mathcal{N}=4$ supergravity. Using our previous results on
the  SCFT dual of the $\mathrm{AdS}_4 \times N^{0,1,0}$ vacuum we discuss the
SCFT realization of the universal $\mathcal{N}=3$ shadow multiplet. RG flows
from an $\mathcal{N}=4$ to an $\mathcal{N}=3$ phase are  ruled out by the fact
that the $\mathcal{N}=4$ vacuum is at infinite distance in moduli space,
denoting the presence of a topology change.}
\end{abstract}
\vspace{2mm} \vfill \hrule width 3.cm
{\footnotesize
 $^ \dagger $ \hskip 0.1cm Supported by   EEC  under TMR contract
 ERBFMRX-CT96-0045}
\end{titlepage}
\section{Introduction}
\label{introduction}
The study of Anti-de-Sitter compactifications of M-theory or type II string
theories is a fascinating subject, resurrected to new life after the Maldacena
conjecture \cite{review}. The profound interplay between AdS compactifications
and conformal field theories has led to a new understanding of several results
derived in the eighties in the context of the Kaluza-Klein program. In this
paper, motivated by field theory considerations, we shall discuss some general
features of Freund-Rubin compactifications that were previously overlooked and
we are led to some unexpected and stimulating conclusions  about the existence
of so far unknown aspects of both supergravity and superconformal field
theories. Shadowing of long multiplets behind short ones in  Freund-Rubin
compactifications, and shadow supergravities that have not been constructed but
whose existence is suggested by many considerations, are the main subject of
this paper. In parallel, we shall discuss implications of our results for
CFT's.
\par
The natural arena for our discussion are M-theory or Type II string backgrounds
of the form AdS$_{p+2}\times X^{d-p-2}$ in d-dimensions (where $X_{d-p-2}$ is
an Einstein manifold), and the associated dual CFT's. Even if we mainly discuss
the case of AdS$_4$, some of our results apply to AdS$_5$ as well with minor
modifications. In previous papers \cite{adscftcheckers,noialtro}, we proposed
candidate dual CFT's for some supersymmetric AdS-compactification of M-theory
on coset spaces, discussing the $\mathcal{N}=2$ cases $Q^{1,1,1}$ and $M^{1,1,1}$
and the $\mathcal{N}=3$ solution $\n010$. The remaining supersymmetric case
$V^{5,2}$ has been discussed in \cite{poliv52}. In all these cases, as well as in
the case of the firstly discussed type IIB compactification AdS$_5\times T^{1,1}$
\cite{witkleb,gubser,gubserkleb,sergiotorino}, the comparison between the KK
spectrum and the spectrum of conformal composite operators of the CFT is
completely successful and the analysis of baryonic operators corresponding to
wrapped 5-branes gives results in agreement with quantum field theory
expectations. The analysis of the spectrum reveals that all these
compactifications share common features. One notably one, firstly noticed in
\cite{gubser}, is the existence of long multiplets with protected rational
dimensions. Following a suggestion in \cite{sergiotorino}, these multiplets can
often be written in the CFT as tensor product of short and massless
multiplets, so that they can be identified on the quantum field theory side.
Their protected dimension suggests the existence of some non-renormalization
theorem.
\par
The first purpose of this paper is  to illustrate an intriguing general feature
of Kaluza Klein spectra for compactifications of type AdS$_4\times X_7$ when
some $\mathcal{N}$--extended supersymmetry is present. As a consequence of the
mass formulae derived in the eighties \cite{univer}, there is a symmetry
structure that so far was not   appropriately noticed and explored. Indeed, the
supermultiplets are paired by curious relations that to each multiplet
associate another \emph{shadow} multiplet of different spin and of different
type, but with masses exactly predicted by those of the parent multiplet. The
shadows of short multiplets are generically long ones, but with rational
conformal dimensions and explain why the appearance of protected multiplets is
so common is AdS-compactifications on Einstein manifolds.
\par
Of special interest is the long shadow multiplet of the massless graviton. Such
shadow multiplet is \emph{universal} since it has the volume of the internal
manifold as one of its scalar component and, moreover, it has a
\emph{universal} structure that is independent from the detailed geometry of
$X_7$. In $\mathcal{N}=2$ compactifications, the massless graviton shadow is a
long, massive, \emph{vector} multiplet in the same $R$--symmetry representation
as the graviphoton. In $\mathcal{N}=3$ compactifications, instead, the shadow of
the massless graviton multiplet is a long, massive \emph{gravitino}. In all
Freund-Rubin compactification, the mass of the scalar associated with the
volume (which is paired to the massless graviton) is such that the
corresponding CFT operator has dimension 6. This is closely reminiscent of a
similar phenomena  in type II AdS$_5$ compactifications, where the volume of
the internal manifold corresponds to the CFT operator $F^4$, with dimension 8,
which is known to satisfy some non-renormalization theorem.
\par
We shall be mainly concerned with the $\mathcal{N}=3$ compactification on $\n010$,
where all the previous phenomena can be also discussed from  a different
perspective. All the components of the long gravitino multiplet are constructed
with \emph{constant} harmonics, suggesting the existence of a consistent
truncation to a theory with only a massless graviton and a massive gravitino.
This fact is extremely intriguing since it seems to indicate that the $\mathcal{N}=3$
compactification of M--theory should admit an interpretation as a spontaneously
broken phase of some suitable $\mathcal{N}=4$ theory. Exploring this possibility is the
main focus of the second part of this present paper, where we analyze the
theory from the perspective of four-dimensional gauged supergravity. We shall
look for an $\mathcal{N}=4$ gauged supergravity with an $\mathcal{N}=3$ critical point containing a
massless graviton and a massive broken gravitino with the quantum numbers
corresponding to those of the $\n010$ compactification. We would like to
stress that the analysis of the low-energy supergravity and its critical points
is at the basis of the study of the deformations of the dual CFT and the
induced RG flow \cite{RG}. In this context, using older results in
\cite{yasuda}, the phase space and the RG trajectories for the $\n010$
compactification have been partially studied in \cite{rey}, without looking for
$\mathcal{N}=4$ supersymmetry. We shall complete the reduction of the Lagrangian by
adding the scalars corresponding to the internal photon, not discussed in
\cite{yasuda,rey}, and we shall compared with the known version of $\mathcal{N}=4$ gauged
supergravities \cite{deRoo:1985np}. Here we shall discover some intriguing
facts.
\par
The existing versions of $\mathcal{N}=4$ gauged supergravity exhibit the spontaneous
symmetry breaking $\mathcal{N}=4\rightarrow \mathcal{N}=3$ that we are looking for
\cite{wageman}. This was indeed the original motivation for this paper.
Perhaps more surprising, there is a continuum of $\mathcal{N}=3$ critical points, with
arbitrary value for mass of the gravitino but strictly below the predicted
value for $\n010$. The  $\n010$ critical point can be realized only at
the boundary of the moduli space, at infinite distance from the $\mathcal{N}=4$ point.
The interpolation between the $\n010$ compactification and its parent
$\mathcal{N}=4$ theory necessarily involves some change of topology.
\par
{}From the point of view of four-dimensional supergravity, this result is
intriguing and suggests the existence of overlooked supergravities. The
unitarity bound for the $E_0$ (dimension of the operator dual to the lowest
component) quantum number for a gravitino multiplet is $E_0>1$. In the known
supergravities, there is a variety of $\mathcal{N}=3$ theories with all possible values
$E_0<3$. $E_0=3$ is, of course, the value for $\n010$ and for the
\emph{universal} volume multiplet in Freund-Rubin compactifications. There is
no theoretical reason for an upper bound on $E_0$. We suggest that
four-dimensional supergravities are missing where the values $E_0>3$ are
realized. We shall argue that these supergravities have been overlooked in the
past. The reason is that they should admit $\mathcal{N}=4$ supersymmetry but no $\mathcal{N}=4$
vacuum, leaving the freedom of relaxing the constraints imposed by global
symmetries. We shall give some arguments based on the coset structure of the
scalar manifold and analogies with similar situations with missing BPS states
in supergravity theories \cite{Ferrara:2000ki}.
\par
Finally, we should mention that we expect that there are certainly many other
results concerning CFT's and RG flows, following from our supergravity
analysis, than those presented in this paper. We leave them for future work.
\par
The paper is organized as follows. Section 2 fixes the conventions for
Freund-Rubin compactifications. Section 3 deals with the explicit construction
of the shadow supermultiplets, with particular emphasis for the case
$\n010$. Section 4 discusses the $\mathcal{N}=3$ critical points of four-dimensional
supergravity, the explicit reduction from 11 to 4 dimensions  and the upper
bound on $E_0$. Section 5, which is the conjectural part of this paper,
suggests and motivate the search for new shadow supergravities. Section 6
presents for completeness the explicit form of the long gravitino multiplet of
$\n010$, using results obtained in \cite{noialtro}. Section 7 contains some
conclusions and a short discussion of the CFT interpretation of our
supergravity results. Finally, the appendices contain conventions and useful
formulae.
\section{Freund--Rubin compactifications of 11D supergravity}
For $D=11$ supergravity \cite{CreJul,ricpie11} our basic conventions are those
of the geometric formulation \cite{ricpie11} and for Kaluza Klein
compactifications we follow the conventions of
\cite{noi321,spectfer,univer} (see appendix
 \ref{appendo}).
The bosonic action of 11D supergravity reads:
\begin{equation}
\label{boslag11}
S = {1\over \kappa_{11}^2} \int \mathcal{R}\, \det\, V  -
{1\over 16 \kappa_{11}^2} \int F \wedge {}^* F -
{1\over 96 \kappa_{11}^2} \int F\wedge F\wedge A~,
\end{equation}
where $\mathcal{R}$ is the scalar curvature, $V^{M}$ ($M=0,\ldots 10$)
are the vielbein, $A$ is a three-form and $F$ its four-form field-strength.
\par
The equations of motion that follow from eq. (\ref{boslag11}) are
\begin{eqnarray}
\label{boseom11}
\mathcal{R}^{M}_{~N} & = &
6 F^{M P_1P_2 P_3}
F_{N P_1 P_2 P_3} -{1\over 2} \delta^{M}_{~N} F^2  ~;\\
D_{M} F^{M P_1 P_2 P_3} & = &
{1\over 96} \varepsilon^{P_1 P_2 P_3 N_1\ldots P_8}
F_{N_1\ldots N_4 }\, F_{N_5\ldots N_8}
\label{maxw11d}
\end{eqnarray}
where $\mathcal{R}^{M}_{~N}$ is the Ricci tensor. Through all the paper,
we use ``flat''  indices,
namely we write all tensor components with respect to
the vielbein basis.
\par
The supersymmetry transformation of the gravitino $\psi_{M}$ is
\begin{equation}
\label{susygrav11}
\delta_\epsilon \psi_M = D_M \epsilon -
\left({\ii\over 3} F_{P_1 P_2 P_3 M}
\Gamma^{P_1 P_2 P_3} -
{\ii\over 8} F^{P_1 P_2 P_3 P_4}
\Gamma_{P_1 P_2 P_3 P_4 M}\right)
\epsilon~.
\end{equation}
\paragraph{Freund-Rubin compactifications}
Freund-Rubin \cite{freurub} (FR) compactifications are
solutions of the field equations of $11D$ supergravity (\ref{boseom11}) in
which the space-time $\mathcal{M}_{11}$ has the factorized form
\begin{equation}
\label{11=4+7}
\mathcal{M}_{11}=\mathrm{AdS}_4\times X^7~,
\end{equation}
the relevant one for investigating the AdS/CFT correspondence between
$3D$ superconformal gauge theories on the $M2$--brane world volume  and
supergravity in $4D$ anti de Sitter space.
The only non-vanishing components of $F$ are the $4$-dimensional ones:
\begin{equation}
F_{abcd}=e\, \varepsilon_{abcd}~.
\label{FreundRubin}
\end{equation}
The parameter $e$ sets the scale for both the 4-dimensional and the
$7$-dimensional cosmological constant (also $X^7$ must be an Einstein space):
\begin{equation}
\label{cosmol4e7}
\mathcal{R}_{ab} = -24\, e^2\, \eta_{ab}~,\hskip 0.8cm
\mathcal{R}_{\alpha\beta} = 12\, e^2\, \eta_{\alpha\beta}~.
\end{equation}
Greek letters $\alpha,\beta,\ldots$ will always be reserved to flat
$7$-dimensional indices, while Latin letters $a,b\ldots$ will stand
for $4$-dimensional flat indices.  We denote by $B^{\alpha\beta}$
the spin connection one-form, and by $B^\alpha$ the vielbein on $X^7$.
Furthermore  we decompose the 11-dimensional
gamma matrices ($\Gamma^M$) as tensor products of $4$-dimensional
($\gamma^a$) and
 $7$-dimensional ones ($\tau^\alpha$) as follows:
\begin{equation}
\label{gamma11=4+7}
\Gamma^a = \gamma^a\times \ide_7~,\hskip 0.8cm
\Gamma^\alpha = \gamma^5\times \tau^\alpha~.
\end{equation}
\par
In the Freund--Rubin  background the gravitino is set to zero; we can search
for preserved supersymmetries by requiring that its variation vanishes as well.
According to (\ref{11=4+7}) we write the $11$-dimensional spinor parameter
$\epsilon(x,y)$ as the tensor product of a $4$- and a $7$-dimensional spinor,
$\epsilon(x,y) = \epsilon(x)\times \eta(y)$. To every independent
$7$-dimensional (commuting) spinor parameter $\eta(y)$ that satisfies
\begin{equation}
\label{kilspineq7dim}
D\eta(y)\equiv(d - {1\over 4} B^{\alpha\beta}\tau_{\alpha\beta})\eta(y) =
e\, B^\alpha \tau_\alpha\eta(y)
\end{equation}
there is associated a residual supersymmetry in the $\mathrm{AdS}_4$ space.
\par
Indeed the vanishing of (\ref{susygrav11}) for the gravitino $\psi_a(x)$
in the background (\ref{FreundRubin}) reduces to the vanishing of its
supersymmetry transformation in $\mathrm{AdS}_4$:
\begin{equation}
\label{susygravads4}
D_a^{(\mathrm{AdS})}\epsilon(x)\equiv
(\partial_a - {1\over 4}\omega^{bc}_{~~a} \gamma_{bc} -
2\, e\, \gamma_5\gamma_a)\epsilon(x)=0~,
\end{equation}
whose integrability
is guaranteed by the expression of the AdS$_4$ curvature,
$R^{ab}_{~~cd}=-16\,e^2\,\delta^{ab}_{cd}$, that
corresponds to the Ricci tensor (\ref{cosmol4e7}).
\section{Shadow multiplets in Kaluza Klein theory}
\label{genshadow}
Consider a generic Freund--Rubin compactification of M--theory on
$\mathrm{AdS}_4\times X^7$. The fluctuations of the eleven--dimensional fields
around such a background can be expanded into harmonics on the compact
$7$--manifold and the linearized field equations can be suitably diagonalized
into eigenmodes of definite mass. The resulting general formulae were derived
in \cite{spectfer,bosmass} and organized into a systematic way in
\cite{univer}.
\par
In the present section we deal with an intriguing general
feature of such Kaluza Klein spectra when some $\mathcal{N}$--extended supersymmetry is
present. In that case all states  organize themselves into supermultiplets of
the relevant supersymmetry algebra namely $\mathrm{Osp}(\mathcal{N}\vert 4)$. However
this is not the end of the story. As a consequence of the results of
\cite{univer}, there is a further symmetry structure that so far was not
appropriately noticed and explored. Indeed, going beyond the implications of
pure superalgebra representation theory, the supermultiplets are further paired
by curious relations that to each multiplet associate another \emph{shadow}
multiplet of different spin and of different type, but with masses fixed
in terms of those of the parent multiplet.
\par
In general, the shadows of short multiplets are long ones. Yet, because of the
shadowing relation their conformal dimensions, derived from those of the parent
multiplets, are rational. Hence the appearance of rational long multiplets
\cite{adscftcheckers,sergiotorino,sanssergio} is partially explained by the
shadowing relation.
\par
Of special interest are the long shadow multiplets of massless ultrashort
multiplets. Such shadow multiplets have a \emph{universal} structure that is
independent from the detailed geometry of $X^7$ and simply follows from the
structure of the $D=11$ field equations plus the constraints of supersymmetry.
\par
In $\mathcal{N}=2$ compactifications, the massless graviton multiplet has always a
shadow which is a long, massive, \emph{vector} multiplet in the same
$R$--symmetry representation as the graviphoton. In $\mathcal{N}=3$ compactifications,
instead, the shadow of the massless graviton multiplet is a long, massive
\emph{gravitino}. This fact is of   utmost interest since it seems to indicate
that any $\mathcal{N}=3$ compactification of M--theory on $\mathrm{AdS}_4 \times X^7$
should admit an interpretation as a spontaneously broken phase of some suitable
$\mathcal{N}=4$ theory. Exploring this possibility is the main focus of the present
paper and will lead to some unexpected and stimulating conclusions  hinting to
the existence of so far unknown aspects of both supergravity and superconformal
field theories.
\par
Also massless vector multiplets have interesting shadows but, both in $\mathcal{N}=2$
and in $\mathcal{N}=3$ compactifications, these shadows are long, rational,
\emph{graviton} multiplets. This excludes an interpretation as  Higgs or
super-Higgs phenomena. The same is true of compactifications with $\mathcal{N}>4$, like
the maximally extended case of $\mathrm{AdS}_4 \times S^7$. Here the shadow
multiplets of the massless graviton multiplet are all massive short graviton
multiplets since the entire spectrum is composed of short multiplets
\cite{gunawar}. Obviously there is no super-Higgs mechanism in such cases.
\par
In section \ref{shagene} we illustrate the general mechanism  of  shadow
multiplet generation. Then in section \ref{miracle} we focus on the special
$\mathcal{N}=3$ case and we discuss the universal structure of the massive gravitino
multiplet generated by the shadow of the $\mathcal{N}=3$ massless graviton multiplet.
\subsection{Shadow multiplet generation}
\label{shagene}
 Since our argument on the existence of shadow multiplets relies heavily on the
Fermi--Bose harmonic relations constructed in \cite{univer}, we  have to recall
the notations and the main results of that paper.
\par
It is possible (see eq.s (3.21) of \cite{univer}) to parametrize the Kaluza
Klein expansion of the field fluctuations in terms of 4-dimensional fields that
are mass eigenstates. For the fluctuations $h_{MN}$ of the metric one sets
\begin{eqnarray}
h_{ab}\left(x,y\right)
&=& \Big( h_{ab}^I\left(x\right)
    - \frac{3}{M_{(0)^3}+32}D_{(a}D_{b)}
    \left[(2+\sqrt{M_{(0)^3}+36}\,) S^I\left(x\right)\right.
\nonumber\\
&+& \left. (2-\sqrt{M_{(0)^3}+36}\,)\Sigma^I\left(x\right)\right]
    + \ft54 \delta_{ab}
    \left[(6-\sqrt{M_{(0)^3}+36}\,)S^I\left(x\right)\right.
\nonumber\\
&+& \left. (6+\sqrt{M_{(0)^3}+36}\,)\Sigma^I\left(x\right)\right]
    \Big) \, Y^I \left(y\right)\,,
\label{kkexpansion}\\
h_{a\beta}\left(x,y\right)
&=& \big[(\sqrt{M_{(1)(0)^2}+16}-4)A_a^I\left(x\right)
\nonumber\\ +
&+& (\sqrt{M_{(1)(0)^2}+16}+4)W_a^I\left(x\right)\big] \, Y^I_\beta
    \left(y\right) \,,
\label{kkexpansion2}\\
h_{\alpha\beta}\left(x,y\right)
&=& \phi^I\left(x\right) Y^I_{(\alpha\beta)}\left(y\right) -
    \delta_{\alpha\beta}
    \left[(6-\sqrt{M_{(0)^3}+36}) S^I\left(x\right)\right.
\nonumber \\
&+& \left.  (6+\sqrt{M_{(0)^3}+36})\Sigma^I
    \left(x\right)\right] \, Y^I\left(y\right)~.
\label{kkexpansion3}
\end{eqnarray}
For the fluctuations $a_{MNR}$ of the three form field, one has
\begin{eqnarray}
\label{kkexpansionA}
a_{abc}\left(x,y\right)
&=& 2 \, \varepsilon_{abcd} \,D_d (S^I\left(x\right)+
    \Sigma^I\left(x\right)) Y^I\left(y\right) \,,
\nonumber \\
a_{ab\gamma}\left(x,y\right)
             &=& \ft23 \, \varepsilon_{abcd} \,
            (D_c A_d^I\left(x\right) + D_c W_d^I\left(x\right))\,
        Y_\gamma^I\left(y\right) \,,
\nonumber \\
a_{a\beta\gamma}\left(x,y\right)
             &=& Z_a^I\left(x\right) Y^I_{[\beta\gamma]}\left(y\right) \,,
\\
a_{\alpha\beta\gamma}\left(x,y\right)
            &=& \pi^I\left(x\right) Y^I_{[\alpha\beta\gamma]}\left(y\right)~.
\end{eqnarray}
Finally, for the fluctuations of the gravitino field,
\begin{eqnarray}
\psi_a\left(x,y\right)
            &=& \Big(
                \chi_a^I\left(x\right) +
        \frac{\ft47 M_{(1/2)^3}+8}{M_{(1/2)^3}+8}
                \big[D_a \lambda_L^I\left(x\right) \big]_{3/2}
\nonumber\\
&-&    (6+\ft37 M_{(1/2)^3})\gamma_5\gamma_a \lambda_L^I
\left(x\right)
           \Big) \, \Xi^I\left(y\right) \,,
\label{kkexpansionpsi}\\
\psi_\alpha(x,y) &=& \lambda_T^I\left(x\right) \Xi_\alpha^I\left(y\right) +
                \lambda_L^I\left(x\right)
                \big[ \nabla_\alpha \Xi ^I\left(y\right) \big]_{3/2}~.
\label{kkexpansionlam}
\end{eqnarray}
We denote by $x$ the coordinates of four dimensional space, while $y$ are
the coordinates on the compact $7$--manifold $X^7$.
\par
For the harmonics on $X^7$ that appear in the KK expansion
(\ref{kkexpansion}-\ref{kkexpansionpsi}) we have followed the conventions of
\cite{univer}. These conventions are summarized in Table \ref{tabinva}, and
consist in the following.
\renewcommand{\arraystretch}{1}
\begin{table}
\begin{center}
{\small
\begin{tabular}{cclll}
\hline\hline
Irrep & Dimension & Operator & Harmonic & Eigenvalue \\
\hline
\null & \null & \null & \null & \null \\
$\left [ 0 , 0 , 0 \right]$ & ${\bf 1}$ & $\lap_{\, (0)^3}$
& ${ Y}^I$  & $M_{\left(0\right)^3}$ \\
\null & \null & \null & \null & \null \\
$\left [ 1 , 0 , 0 \right ]$ & ${\bf 7}$ & $\lap_{\, (1)(0)^2}$
& ${Y}^I_{\alpha}$ & $M_{\left(1\right)\left(0\right)^2}$ \\
\null & \null & \null & \null & \null \\
$\left [ 1 , 1 , 0 \right]$ & ${\bf 21}$ & $\lap_{\, (1)^2(0)}$
& ${Y}^I_{\left[\alpha\beta\right]}$ & $M_{\left(1\right)^2\left(0\right)}$ \\
\null & \null & \null & \null & \null \\
$\left[ 1 , 1 , 1 \right]$ & ${\bf 35}$ & $\lap_{\, (1)^3} $
& ${Y}^I_{\left[\alpha\beta\gamma\right]}$ & $ M_{\left(1\right)^3}$ \\
\null & \null & \null & \null & \null \\
$\left [ 2 , 0 , 0 \right]$ & ${\bf 27}$
& $\lap_{\, (2)(0)^2}$ & ${Y}^I_{\left(\alpha\beta\right)}$
& $ M_{\left(2\right)\left(0\right)^2}$\\
\null & \null & \null & \null & \null \\
$\left [ \ft 1 2 , \ft 1 2 , \ft 1 2 \right]$ & ${\bf 8}$
& $\lap_{\, (\ft 12)^3}$
& $\Xi^I$ & $M_{\left(1\over 2\right)^3}$\\
\null & \null & \null & \null & \null \\
$\left [ \ft 3 2 , \ft 1 2 , \ft 1 2 \right]$ & ${\bf 48}$
& $\lap_{\, (\ft{3}{2})(\ft{1}{2})^2}$ & $\Xi^I_\alpha$
& $M_{\left(3\over 2\right)\left(1\over 2\right)^2}$
\\
\hline\hline
\end{tabular}
}
\end{center}
\caption{Conventions for the harmonics on $X^7$.}
\label{tabinva}
\end{table}
The harmonics on $X^7$ are  grouped into seven infinite towers
corresponding to as many irreducible representations of the tangent
group $\mathrm{SO}(7)$ that appear in the decomposition $4\oplus 7$ of eleven
dimensional tensors and spinors. Since $\mathrm{SO}(7)$ has rank 3, its {\sl
irreps} are labeled by three numbers $[\lambda _1 ,\lambda _2 ,\lambda
_3]$ that we take to be the Young labels\footnote{
That is, for bosonic tensors with symmetry
represented by a Young tableaux, $\lambda_i$ is the number of boxes in the
$i$-th row of the tableaux. For gamma--traceless irreducible spinor tensors
$\lambda^i$ is $1/2$ plus the number of boxes.}.
\par
The explicit form of the invariant operators appearing in table
(\ref{tabinva}) is given below:
\par
\noindent
- Laplacian on scalars:
\begin{equation}
\lap_{\, (0)^3} Y_{\alpha}\equiv   \,D^\mu \,
D_{ \mu } \, Y  \, = \,
    \, \square   \, Y_{\alpha} \, = \,
 M_{\, (0)^3} \,
Y_{\alpha}~;
\label{unozero}
\end{equation}
- Hodge--de Rham operator on one--forms:
\begin{equation}
\lap_{\, (1)(0)^2} Y_{\alpha}\equiv 2 \,D^\mu \,
D_{[\mu } \, Y_{\alpha ]} \, = \,
  \left ( \, \square +24 \, e^2 \,\right) \, Y_{\alpha} \, = \,
 M_{\, (1)(0)^2} \,
Y_{\alpha}~;
\label{unolinea}
\end{equation}
- Hodge--de Rham operator on two--forms:
\begin{equation}
\lap_{\, (1)^20} Y_{[\alpha\beta]}\equiv 3
\,D^{\mu} \, D_{[\mu } Y_{\alpha \beta ]} =
 \left[ \left ( \, \square +40 \, e^2 \,\right) \,
\delta_{\alpha\beta}^{\lambda\mu} - 2\,
C_{\alpha\beta}^{\phantom{\alpha\beta}\lambda\mu} \, \right] \,
Y_{[\lambda\mu]} \, = \, M_{\, (1)^20} \, Y_{[\alpha\beta]}~;
\label{unosquare}
\end{equation}
- $*d$   operator on three forms:
\begin{equation}
\lap_{\, (1)^3} \, Y_{[\alpha\beta\gamma]}\,
\equiv \, \ft{1}{24}
\epsilon_{\alpha\beta\gamma\lambda\mu\nu\rho}
D^\lambda Y^{\mu\nu\rho}\equiv \star \, d
Y_{[\alpha\beta\gamma]}=M_{\, (1)^3}\,
Y_{[\alpha\beta\gamma]}~;
\label{1cube}
\end{equation}
- Lichnerowitz operator on symmetric tensors:
\begin{equation}
\lap_{\, (2)(0)^2} Y_{(\alpha\beta)}   =
 \left[ \left ( \, \square +40 \, e^2 \, \right) \,
\delta_{(\alpha\beta)}^{(\lambda\mu)} - 4\,
C_{(\alpha\beta)}^{\phantom{(\alpha\beta)}
(\lambda\mu)} \, \right] \, Y_{(\lambda\mu)} \, = \,
 M_{\, (2)(0)^2} \,
Y_{(\alpha\beta)}~;
\label{duelinea}
\end{equation}
- ``de Sitter" Dirac operator:
\begin{equation}
\lap_{\, (\ft{1}{2})^3} \Xi \equiv \, \tau ^\mu \, \nabla_\mu \, \Xi \, = \,
\tau^\mu \, \left(
D_\mu  - \, e \, \tau _\mu \right) \, \Xi \, = \,
\left( \, {\slash \!\!\!\! D} - 7\, \, e \, \right)
\, \Xi \, = \, M_{\, (\ft{1}{2})^3} \, \Xi~;
\label{unmezcube}
\end{equation}
- ``de Sitter" Rarita Schwinger operator:
\begin{equation}
\lap_{\, (\ft{3}{2})(\ft{1}{2})^2} \, \Xi_\alpha \, \equiv \,
\tau_{\alpha\mu\nu} \nabla^\mu \,\Xi^\nu - \ft{5}{7}    \tau_\alpha
\,\tau^{\mu\nu}\,\nabla_\mu \,\Xi_\nu \, = \,
\left(\, {\slash \!\!\!\! D} -
5 \, e \, \right )\, \Xi_\alpha \, = \, M_{\, (\ft{3}{2})(\ft{1}{2})^2}
\,\Xi_\alpha~,
\label{raritsch}
\end{equation}
where the conventions for $7$ geometry are those
of \cite{univer} (see Appendix \ref{appendo}). In particular,
\begin{equation}
  \mathcal{C}^{\alpha \beta}_{~~\mu \nu } =
  \mathcal{R}^{\alpha \beta }_{~~\mu \nu
  } - 4 e^2 \, \delta ^{\alpha \beta}_{\mu \nu}
\label{weyltens}
\end{equation}
is the Weyl tensor on the Einstein manifold $X^7$, with Ricci tensor
as in eq. (\ref{cosmol4e7})
\par
Measuring everything in units of the Freund--Rubin scale, \emph{i.e.} setting
$e=1$, the masses of the mass eigenstates appearing in the Kaluza Klein
expansion (eq.s (\ref{kkexpansion}-\ref{kkexpansionpsi})) are expressed in terms
of the eigenvalues of the appropriate operators as follows:
\begin{eqnarray}
m_h^2 & = & M_{\left(0\right)^3}\,, \label{gmassh} \\ m_{\Sigma}^2&=&
M_{\left(0\right)^3} +176+24\sqrt{ M_{\left(0\right)^3}+36}\,,
\label{gmassigma}\\ m_S^2 &=& M_{\left(0\right)^3} +176-24\sqrt{
M_{\left(0\right)^3}+36}\,,\label{gmassesse}\\ m_{\phi}^2 &=&
M_{\left(2\right)\left(0\right)^2} \,,\label{gmassfi}\\ m_{\pi}^2 &=& 16\left(
M_{\left(1\right)^3}-2\right)\left( M_{\left(1\right)^3}-1\right)
\,,\label{gmasspi}\\
m_W^2 &=& M_{\left(1\right)\left(0\right)^2} + 48 + 12
\sqrt{M_{\left(1\right)\left(0\right)^2}+16}\,,\label{gmassW}\\ m_A^2 &=&
M_{\left(1\right)\left(0\right)^2} + 48 - 12
\sqrt{M_{\left(1\right)\left(0\right)^2}+16}\,,\label{gmassA}\\ m_Z^2 & = &
M_{\left(1\right)^2\left(0\right)} \,,\label{gmassZ}\\ m_{\lambda_L} & = &
-\left( M_{\left(1\over 2\right)^3} +16\right)\,,
\label{gmasslamL}\\
m_{\lambda_T} & = & M_{\left(3\over 2\right)\left(1\over
2\right)^2}+8\,,\label{gmasslamT}\\ m_{\chi} & = & M_{\left(1\over
2\right)^3}\,.\label{gmasschi}
\end{eqnarray}
\par
Each mass eigenmode corresponds to an irreducible unitary representation (UIR)
of the anti de Sitter group $\mathrm{SO}(2,3)\sim \mathrm{Sp}(4,\mathbb{R})$
which is labeled by two weights, the $\mathrm{SO}(3)$ spin $s$ (which is $s=0$
for scalars, $s=\ft 12$ for spinors  and so on up to $s=2$ for massive
gravitons) and the energy eigenvalue $E$ which satisfies the unitarity lower
bound
\cite{heidenreich,frenico,multanna}:
\begin{equation}
  E \, \ge \, s +1 \, .
\label{eneboundo}
\end{equation}
\par
In the alternative three-dimensional conformal interpretation of  the
$\mathrm{SO}(2,3)$ group each UIR is a primary conformal field with scale
dimension $\Delta=E$ and Lorentz $\mathrm{SO}(1,2)$ spin $s$ (compare with
\cite{susp} for further details).
\par
In \cite{univer} the relations between the masses (\ref{gmassh}
--\ref{gmasschi}) and the conformal dimensions are given in the following
form:
\begin{eqnarray}
m_{\left(0\right)}^2&=&16\left(E_{\left(0\right)}-2\right)
\left(E_{\left(0\right)}-1\right) \,,\nonumber\\
|m_{\left(1\over 2\right)}|&=&4E_{\left(1\over 2\right)}-6\,,\nonumber\\
m_{\left(1\right)}^2&=&16\left(E_{\left(1\right)}-2\right)
\left(E_{\left(1\right)}-1\right) \,,\nonumber\\
|m_{\left(3\over 2\right)}+4|&=&4E_{\left(3\over 2\right)}-6~,
\label{massenergy}
\end{eqnarray}
where $ m_{\left(s\right)}$ and $E_{\left(s\right)} $ denote the mass and
energy of a field of spin $s$
\footnote{The reader should be careful in comparisons with other
papers and take into account that the definition of mass utilized in
this paper is that of supergravity \cite{castdauriafre}. Specifically
the mass squared of scalars is defined as the deviation from a
conformal invariant equation, the mass of the gravitino is defined as
the deviation from a Rarita Schwinger equation with supersymmetry,
the mass squared of a spin one field is defined as the deviation from
a gauge invariant equation.}.
\paragraph{Supersymmetry and shadow multiplets}
Let us consider unitary irreducible representations of the
superalgebra
$\mathrm{Osp}(\mathcal{N}\vert 4)$, {\it i.e.} the  supersymmetric
extension
of $\mathrm{SO}(2,3)$ with  $\mathcal{N}$ supercharges. Each of them
is a  supermultiplet, represented by a suitable
constrained superfield, that contains fields whose spins $s$ and dimensions $E$ are
related to each other.  Via eq.s
(\ref{gmassh},\ref{gmasschi},\ref{massenergy}), such relations translate into
relations on the eigenvalues of the various Laplace Beltrami operators
(\ref{unozero},\ref{raritsch}).
\par
\begin{figure}
\centering
\begin{picture}(268,200)
\put (15,100){\circle{30}}
\put (12,97){\shortstack{\large{$2$}}}
\put (12,72){\shortstack{\large{$h$}}}
\put (60,100){\vector (-1,0){30}}
\put (75,100){\circle{30}}
\put (72,97){\shortstack{\large{$3\over 2$}}}
\put (72,72){\shortstack{\large{$\chi$}}}
\put (85.61,110.61){\vector (1,1){21.22}}
\put (85.61,89.39){\vector (1,-1){21.22}}
\put (117.43,142.43){\circle{30}}
\put (114.43,139.43){\shortstack{\large{$1^-$}}}
\put (114.43,114.43){\shortstack{\large{$Z$}}}
\put (117.43,57.57){\circle{30}}
\put (114.43,54.57){\shortstack{\large{$1^+$}}}
\put (104.43,29.57){\shortstack{\large{$A,W$}}}
\put (132.43,142.43){\vector (1,0){30}}
\put (132.43,57.57){\vector (1,0){30}}
\put (125.04,68.18){\vector (2,3){42}}
\put (177.43,142.43){\circle{30}}
\put (174.43,139.43){\shortstack{\large{$1\over 2$}}}
\put (174.43,114.43){\shortstack{\large{$\lambda_T$}}}
\put (177.43,57.57){\circle{30}}
\put (174.43,54.57){\shortstack{\large{$1\over 2$}}}
\put (174.43,29.57){\shortstack{\large{$\lambda_L$}}}
\put (192.43,142.43){\vector (1,0){30}}
\put (188.04,131.82){\vector (1,-1){32.5}}
\put (188.04,68.18){\vector (1,1){32.5}}
\put (192.43,57.57){\vector (1,0){30}}
\put (237.43,142.43){\circle{30}}
\put (234.43,139.43){\shortstack{\large{$0^+$}}}
\put (259.43,139.43){\shortstack{\large{$\phi$}}}
\put (237.43,100){\circle{30}}
\put (234.43,97){\shortstack{\large{$0^-$}}}
\put (259.43,97){\shortstack{\large{$\pi$}}}
\put (237.43,57.57){\circle{30}}
\put (234.43,54.57){\shortstack{\large{$0^+$}}}
\put (259.43,54.57){\shortstack{\large{$S,\Sigma$}}}
\end{picture}
\caption{\emph{``Internal''} supersymmetry relations between the harmonics of
Kaluza Klein fields: for each pair of fields linked by an arrow the
corresponding towers of harmonics can be related to each other by
multiplications with a Killing spinor $\eta^A$.}
\label{molecola}
\end{figure}
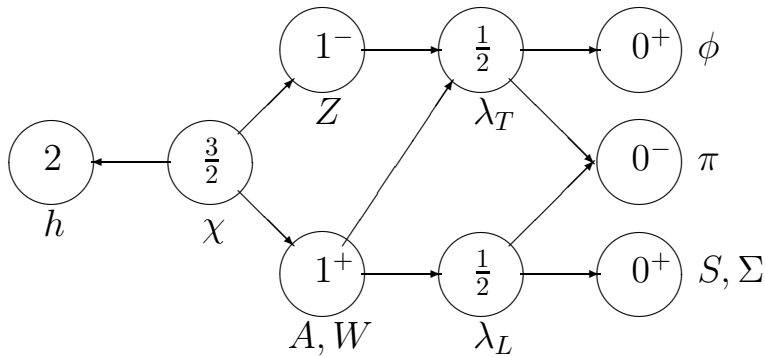

These relations hint to a sort of mirror image of the supersymmetry algebra
that is realized on the internal compact manifold $X^7$. This idea was
thoroughly analyzed in \cite{univer} and traced back to the existence of
(commuting) Killing spinors $\eta^A$ ($A=1,\ldots \mathcal{N}$) satisfying the Killing
spinor equation eq. (\ref{kilspineq7dim}), where $\mathcal{N}$ is the number of
preserved supersymmetries of the $\mathrm{AdS}_4 \times X^7$ compactification
one considers. By means of the Killing spinors $\eta^A$, to each harmonic $Y$
that is an eigenstate of a bosonic  Laplacian $\lap_{\, n_1,n_2,n_3}$ one can
associate another  fermionic harmonic $\Xi$ that is an eigenstate of a
fermionic laplacian $\lap_{\, n_1\pm \ft 12 ,n_2\pm \ft 12,n_3\pm \ft 12}$ with
suitably related eigenvalues $M_{\, n_1,n_2,n_3}$ and $M_{\, n_1\pm \ft 12
,n_2\pm \ft 12,n_3\pm \ft 12}$. These pairs of related harmonics were
explicitly constructed in \cite{univer} and follow the schematic pattern given
in fig. \ref{molecola}.
\par
As already stressed fifteen  years ago in \cite{univer}, these relations are
differential geometric identities on the compact $7$--manifold $X^7$ that are
required by consistency with the structure of UIR.s of the superconformal group
$\mathrm{Osp}(\mathcal{N}\vert 4)$. Because of this, it may at first sight appear that
the universal mass relations analyzed in \cite{univer} do not contain further
physical information besides the implications of supersymmetry. However, this
is not the case, because of another aspect of eq.s (\ref{gmassh}
--\ref{gmasschi},\ref{massenergy}), whose consequence is precisely the
existence of shadow multiplets.
\par
The point is that the relations (\ref{gmassh}-\ref{gmasschi}) between masses
(or conformal weights) and eigenvalues of the internal Laplacian is quadratic
rather than linear. Indeed the same harmonic always  plays a double role since
it appears in the expansion of two quite different Kaluza Klein fields.
Combined with the supersymmetry relations produced by Killing spinors, this has
the curious consequence that each supersymmetry multiplet of the Kaluza Klein
spectrum is associated with another one made, so to say, by the second roots of
the quadratic relations.
\paragraph{Examples}
We can illustrate the basic structure of this mechanism with two simple
examples that will be quite relevant in our subsequent discussion.
\par
Let us
observe that the same scalar harmonic $Y^I(y)$ is associated both to
the graviton field $h_{ab}(x)$, eq. (\ref{kkexpansion}) and to
a scalar field $\Sigma^I(x)$, eq. (\ref{kkexpansion2}). Using
the mass relations (\ref{gmassh} --\ref{gmasschi}), we see that the shadow
scalar of a  ``parent'' graviton has mass
\begin{equation}
  m_\Sigma^2 =m_h^2 +176 +24 \sqrt{m_h^2+36}
\label{shasca}
\end{equation}
In the case of the massless graviton, $m_h^2=0$, corresponding to the constant
harmonic $Y=1$, its shadow scalar $\Sigma$ has (squared) mass
\begin{equation}
  m^2_\Sigma = 320~.
\label{massigma}
\end{equation}
Using eq.s (\ref{massenergy}) this implies
\begin{equation}
  E_\Sigma = 6~.
\label{Esigma}
\end{equation}
We conclude that in every $\mathrm{AdS}_4 \times X^7$ compactification there is
always a \emph{universal} scalar mode of conformal dimension $E=6$ that is the
shadow of the graviton. Its geometric origin is apparent from the second of
equations (\ref{kkexpansion}): it is just the ``breathing'' mode corresponding
to an overall dilatation of the internal manifold $X^7$. The interesting
question is which supermultiplet this universal breathing mode belongs to in
the case of supersymmetric compactifications. We will tackle this question in
the next Section.
\par
As it follows by inspection of eq.s (\ref{kkexpansion2}), to the same vector
harmonic $Y^I_\alpha $ we associate two vector fields: one, $A_a$, with mass
given by eq. (\ref{gmassA}), the other, $W_a$, with mass given by eq.
(\ref{gmassW}). The very initial idea of Kaluza Klein theory is that the
isometries of the internal manifold give rise to massless gauge field in the
compactified $4$--dimensional theory. When we start from $D=11$ M--theory there
is a further aspect. Indeed to each Killing vector of the internal compact
manifold we associate two rather than one vector fields. In addition to the KK
massless gauge boson, we have its shadow  massive vector. It also belongs
to the adjoint representation of the isometry group, and it has fixed mass and
dimension:
\begin{equation}
  m^2_W = 192 \, \Rightarrow \, E_W =5~.
\label{emme2w}
\end{equation}
Once again the natural question is which multiplet do these shadow vectors
belong to in supersymmetric compactifications.
\par
Before addressing this question, let us consider one more example involving
fermionic fields. Comparing eq.s (\ref{kkexpansionpsi}) and
(\ref{kkexpansionlam}), we see that the same spinor harmonic $\Xi^I$ appears
both in the expansion of the gravitino $\psi_a(x,y)$ and as coefficient of
the expansion of a longitudinal spin--1/2 field $\lambda_L(x)^I$. Using
eq.s (\ref{gmasslamL},\ref{gmasschi}) the relation between the masses of the
spin--$3/2$  and spin--$1/2$ modes pertaining to the same harmonic is:
\begin{equation}
  m_\chi=-m_{\lambda_L} -16~.
\label{mchimlam}
\end{equation}
Applying eq. (\ref{mchimlam}) to the case $m_{\lambda_L}=0$, we see that
each massless spin $\ft 12$ particle of this type generates a
shadow massive gravitino with mass:
\begin{equation}
  m_\chi = -16 \, \Rightarrow \, E_\chi= \frac 9 2 ~.
\label{mchi92}
\end{equation}
Conversely, every massless gravitino produces a shadow
massive spin--1/2 field with mass:
\begin{equation}
  m_{\lambda _L} = -16 \, \Rightarrow \, E_{\lambda _L}= \frac {11}2~.
\label{mlam112}
\end{equation}
\subsection{The universal super-Higgs multiplet that shadows
the massless $\mathcal{N}=3$ graviton multiplet}
\label{miracle}
We come now to answer the questions posed in the previous section by focusing
on the very special case of $\mathcal{N}=3$ compactifications. We want to show that, in
such cases, the shadow of the massless graviton multiplet is a long massive
gravitino multiplet with a fixed energy, namely $E_0=3$.
\par
The  $\mathrm{Osp}(3\vert 4)$ supermultiplets relevant to Kaluza Klein
theory were analyzed in \cite{osp34},  and they  were classified using the
following nomenclature. By
\begin{equation}
  SD(s_\mathrm{max},E_0,J_0\vert 3)
\label{genenot3}
\end{equation}
we denote an $\mathrm{Osp}(3\vert 4)$ unitary irreducible representation (UIR)
with maximal spin $s_{max}$, Clifford vacuum energy $E_0$ and
$\mathrm{SU}(2)_R$ isospin $J_0$. The same notation can be extended to
all $\mathrm{Osp}(\mathcal{N}\vert 4)$ representations; for example, we denote
the $\mathcal{N}=2$ UIRs by
\begin{equation}
   SD(s_\mathrm{max},E_0,y_0\vert 2)~,
\label{genenot2}
\end{equation}
where $y_0$ is the Clifford vacuum hypercharge.
\par
Using the just introduced notation, Kaluza-Klein compactifications establish  a
very general shadowing relation between the following two representations that,
algebraically, are totally unrelated:
\begin{equation}
  SD(2,\ft 32,0\vert 3) \,\stackrel{\mbox{shadow}}{\longrightarrow} \,
  SD(\ft 32 , 3,0\vert 3)~.
\label{shadow23/2}
\end{equation}
\paragraph{The massless graviton multiplet}
The graviton multiplet $SD(2,\ft 32,0\vert 3)$ is {\sl ultrashort}
and corresponds to the gauge multiplet of $\mathcal{N}=3, D=4$
supergravity. Its algebraic structure, derived in \cite{osp34}, is
displayed in the first three columns of Table \ref{ushorgrav}.
\renewcommand{\arraystretch}{1.5}
\begin{table}
  \begin{center}
{\small
 \begin{tabular}{ccc|cc}
      \hline\hline
      \multicolumn{3}{c|}{$SD(2,3/2,0\vert 3)$}& \multicolumn{2}{c}
      {Kaluza Klein origin}\cr
      \hline
      Spin & Energy & Isospin  & Field & Harmonic\cr
      \hline\hline
      $2$ & $3$ & $0$ & $h_{ab}$ & $Y=1$\cr
      $\ft32$
         & $\ft52$ & $1$& $\chi^A_a$ &
           $\Xi^A =\eta^A$ (Killing spinor) \cr
      $1$  & $2$ &
      $1$ &$A_a^A$&
      $Y^A_\alpha =\ft 12 \, \epsilon^{ABC} \,
      {\bar \eta}^B \, \tau_\alpha  \eta^C \equiv k^A_\alpha $ \cr
      $\ft12$  & $\ft32$ & $0$ & $ \lambda_L$ & $\Xi=\ft 13
      \,\epsilon^{ABC}\, \tau_\alpha \, \eta^A \,{\bar \eta}^B \,
      \tau^\alpha  \eta^C  \equiv \eta^0 $\cr
      \hline\hline
    \end{tabular}
    }
    \caption{The massless $\mathcal{N}=3$ graviton multiplet and its Kaluza Klein origin.
     }
    \label{ushorgrav}
  \end{center}
\end{table}
The last two columns of the same table are copied from Table VI of
\cite{univer} and describe the Kaluza Klein origin of the multiplet.
\par
As  stressed  in \cite{spectfer} and \cite{univer}, we do not need to know the
details of $X^7$ geometry. It suffices to know that it is a compact Einstein
manifold admitting three Killing spinors  {\sl i.e.} three solutions $\eta^A$
($A=1,2,3$) of eq. (\ref{kilspineq7dim}): in terms of these we can construct
all the harmonics of the graviton gauge multiplet. In particular we construct
the Killing vectors $k^A_ \alpha \equiv \ft 12 \, \varepsilon^{ABC} \, {\bar
\eta}^B \, \tau_\alpha  \eta^C$ that generate the  $\mathrm{SU}(2)_R$ factor in
the isometry group  $G=\mathrm{SU}(2)_R \times G^\prime$ of $X^7$.
\paragraph{The long gravitino multiplet}
Let us now go back to $\mathrm{Osp}(3\vert 4)$ representation theory.
According to the analysis of \cite{osp34} for each energy label
\begin{equation}
  E_0 > 1
\label{eoge1}
\end{equation}
we have a long multiplet $SD(\ft 32 , E_0 ,0 |3)$ whose general
structure is displayed in table \ref{brokengravitino}.
\renewcommand{\arraystretch}{1.5}
\begin{table}
 \size
  \begin{center}
  {\small
    \begin{tabular}{cccc|ccc}
      \hline\hline
      \multicolumn{4}{c|}{$SD(3/2,E_0>1,0\vert 3)$ } &
      \multicolumn{3}{c}{Massive spin $\ft 32$ multiplet} \cr
      \multicolumn{4}{c|}{in $\mathrm{AdS}_4$ space} &
      \multicolumn{3}{c}{in Minkowski space} \cr
      \hline
      Spin & Energy & Isospin & $\#$ of fields &
      Spin & $\mathrm{SU}(3)$ repr. & $\#$ of fields  \cr
      \hline\hline
      $\ft32$ & $E_0+\ft32$ & $0$ & $1$ &
      $\ft32$ & {\bf 1} & $1$ \cr
      \hline
      1 & $\begin{array}{c}E_0 + 2 \\ E_0 + 1\end{array}$ &
      $\begin{array}{c} 1 \\ 1\end{array}$ & 6 &
      1 &  $\begin{array}{c} \mathbf{3} \\ \mathbf{3}\end{array} $& 6 \cr
      \hline
       \null & $E_0+\ft52$ & $1$ & \null &
       \null & $\mathbf{3}$ & \null\cr
      $\ft12$ & $E_0+\ft32$ & $\cases{2\cr 1}$ & $14$ &
      $\ft12$ & $\mathbf{8}$ & 14\cr
      \null & $E_0+\ft12$ & $1$& \null &
      \null & $\bar\mathbf{3}$ & \null\cr
      \hline
      0 & $\begin{array}{c} E_0 + 3 \\ E_0 + 2 \\ E_0 + 1 \\ E_0\end{array}$ &
      $\begin{array}{c} 0 \\ \cases{2\cr 0} \\ \cases{2\cr 0} \\ 0\end{array}$ &
      14 &
      0  & $\begin{array}{c} \mathbf{1} \\ \mathbf{6} \\
      \bar\mathbf{6} \\ \bar\mathbf{1} \end{array}$ & 14 \cr
      \hline\hline
    \end{tabular}
  }
   \caption{The $\mathrm{Osp}(3\vert 4)$ admits a long gravitino representation
    $SD(\ft 32, E_0 >1, 0 \vert 3)$ which has just the same structure as a
    massive spin $3/2$ multiplet of Poincar\'e $\mathcal{N}=3$ supersymmetry. Indeed this
    long representation has a smooth Poincar\'e limit, with arbitrary value of
    $E_0 > 1$ leading to an arbitrary mass of the gravitino.}
    \label{brokengravitino}
  \end{center}
\end{table}
\par
\par
As shown in table \ref{brokengravitino}, this multiplet has a smooth Poincar\'e
limit and becomes the standard multiplet of a massive $\mathcal{N}=3$
gravitino  containing a total of $14$ scalars, $14$ spin--1/2 particles and
$6$ vector bosons. The relation between the isospin assignments in the anti de
Sitter theory and the $\mathrm{SU}(3)$ assignments in the Poincar\'e theory is
easily retrieved by recalling that the anti de Sitter $R$--symmetry group is
just the maximal $\mathrm{SO}(3)$ subgroup of $\mathrm{SU}(3)$ defined as the
set of $3 \times 3$ unitary matrices that are also real. Under this embedding,
the $\mathrm{SU}(3)$ representations (which we label, as in table
\ref{brokengravitino}, by their dimensions) decompose as follows:
\begin{eqnarray}
\label{picchio}
\mathbf{6}\, ,~\bar\mathbf{6} & \to & J=2 \oplus J=1~;\nonumber\\
\mathbf{3}\, ,~\bar\mathbf{3} & \to & J=1 ~;\nonumber\\
\mathbf{1}\, ,~\bar\mathbf{1} & \to & J=0 ~.
\end{eqnarray}
This is sufficient to retrieve all the $\mathrm{SU}(2)_R$ representations
appearing in the corresponding $\mathrm{Osp}(3 \vert 4 )$ supermultiplet.
Obviously the \emph{mass eigenstates} $E=E_0+2$ and $E=E_0+1$ of
isospin $J=2$ can be linear combinations of the two  $J=2$
representations obtained from the branching of the $\mathrm{SU}(3)$
representations $\mathbf{6}$
and $\bar\mathbf{6}$
Similarly, the two mass eigenstates
$E=E_0+3$ and $E=E_0$ of isospin $J=0$ can be linear combinations of
all the four $J=0$ representations obtained from the branching of
$\mathrm{SU}(3)$ representations. The coefficients of such linear
combinations, just as the value of the parameter $E_0$ characterizing
the entire multiplet, are dynamical quantities that depend on the
detailed structure of the interaction Lagrangian in $D=4$.
\paragraph{The super-Higgs phenomenon}
Indeed, in the context
of four dimensional field theory, the only way to introduce a consistent
coupling of a massive spin $3/2$ particle is via supersymmetry
breaking. One starts from a locally supersymmetric Lagrangian
admitting a non--supersymmetric (or partially supersymmetric vacuum).
The spectrum of the theory calculated in the broken phase around such
a vacuum will contain a massive gravitino whose mass depends continuously on the
{\sl vev}.s of the scalar fields responsible for the super-Higgs
mechanism. In particular an $\mathcal{N}=3$ massive gravitino can
emerge only  from a partial supersymmetry breaking $\mathcal{N}=4 \,
\to \,\mathcal{N}=3$. If this occurs in Minkowski space--time  all
the states in the multiplet have the same mass and a manifest $\mathrm{SU}(3)$
symmetry is preserved (right hand side of table
\ref{brokengravitino}); if the breaking occurs in $\mathrm{AdS}_4$ space--time
then the same states are organized in the pattern displayed on the
left--hand side of table \ref{brokengravitino} ($SD(\ft 32
,E_0,0\vert 3)$) and only $\mathrm{SO}(3)\subset \mathrm{SU}(3)$ is preserved.
Yet a local $\mathrm{SU}(3)\times U(1)$--symmetry is a necessary feature of
the $\mathcal{N}=3$ Lagrangian. In any case, in the $D=4$ theory the value of $E_0$
depends on the {\sl vev}.s of the scalar fields and is a continuous
deformable parameter.
\paragraph{The ``shadow'' gravitino multiplet}
Our main point in this section is the explicit illustration of
eq. (\ref{shadow23/2}). Indeed
using the double role of the harmonics associated with the states of
the massless $\mathcal{N}=3$ graviton multiplet (table \ref{ushorgrav})
and the universal relations derived in \cite{univer} we can construct an
$\mathcal{N}=3$ massive spin $\ft 32$
multiplet with predetermined energy (or conformal weight)
$E_0=3$.
\par
The suggested conclusion will be that any $\mathrm{AdS}_4 \times X^7$ compactification
of M--theory with $\mathcal{N}=3$ supersymmetry looks like  a broken phase of an
$\mathcal{N}=4$ theory. We shall have a lot more to say about this conclusion in later
sections but for the time being let us discuss how the shadow spin $\ft 32$
multiplet emerges in Kaluza Klein theory.
\renewcommand{\arraystretch}{1.5}
\begin{table}
  \begin{center}
{\small
\begin{tabular}{c|cccccc}
  \hline
  \hline
  Spin & E & $\mathrm{SU}(2)_R$ & Field & Mass &
  $M_{[\lambda_1,\lambda_2,\lambda_3]}$ & Harmonic \\
  \hline\hline
  $\frac{3}{2}$ & $\frac{9}{2}$ & $J=0$& $\chi^-$ & $m_\chi=-16$ &
  $M_{\,\left(\ft{1}{2}\right)^2}=-16$ &
   $\Xi=\eta^0$  \\
   \hline
  $1$ & $5$ & $J=1$ & $W$ & $m^2_W=192$ & $M_{\, 1(0)^2}=48$ &
  $Y_\alpha=k^A_\alpha$ \\
 \null & $4$ & $J=1$ & $Z$ & $m^2_Z=96$ & $M_{\, \left (1 \right )^2 0}=96$ &
 $Y_{[\alpha\beta]}=
  T^A_{[\alpha\beta]}$ \\
  \hline
  \null & $\frac{11}{2}$ & $J=1$ & $\lambda_L$ & $m_{\lambda_L}=-16$ &
  $M_{\, \left (\ft{1}{2} \right )^3 }$
  & $\Xi=\eta^A$ \\
  $\frac{1}{2}$ & $\frac{9}{2}$ &
  $\begin{array}{c}J=2 \\ J=0\end{array}$ &
  $\lambda_T$ & $m_{\lambda_T}=12$ &
  $M_{\, \ft{3}{2} \left (\ft{1}{2} \right )^2}=4$ &
  $\begin{array}{c}\Xi_\alpha=\widehat{\Omega}^{(AB)}_\alpha \\
                   \Xi_\alpha=\widehat{\Omega}^{(A)}_\alpha \end{array}$ \\
  \null & $\frac{7}{2}$ & $J=1$ & $\lambda_T$ & $m_{\lambda_T}=-8$ &
  $M_{\, \ft{3}{2}\left (\ft{1}{2}\right )^2}=-16$ &
  $\Xi_\alpha =\Theta^A_\alpha$ \\
  \hline
  \null & $6$ & $J=0$ & $\Sigma $ & $ m^2_\Sigma =320 $ &
  $ M_{\,\left ( 0 \right )^3}=0$ & $Y =1$ \\
  0 & $5$ &
  $\begin{array}{c}J=2 \\ J=0\end{array}$ &
  $\pi$ & $m^2_\pi =192$ & $M_{\,\left ( 1 \right )^3}=-2$ &
  $\begin{array}{c} Y_{[\alpha\beta\gamma]}=
   \widehat{\Upsilon}^{(AB)}_{[\alpha\beta\gamma]}\\
   Y_{[\alpha\beta\gamma]}=
   \widehat{\Upsilon}_{[\alpha\beta\gamma]}\end{array}$ \\
  \null & $4$ &
  $\begin{array}{c}J=2 \\ J=0\end{array}$ &
  $\phi$ & $m^2_\phi =96$ & $M_{\,2\left ( 0 \right )^2}=96$ &
  $\begin{array}{c} Y_{(\alpha\beta)}= H^{m}_{(\alpha\beta)} \\
   Y_{(\alpha\beta)}= H^0_{(\alpha\beta)}\end{array}$\\
  \null & $3$ & $J=0$ & $\pi$ & $m^2_\pi=32$ & $M_{\,\left ( 1 \right )^3}=3$ &
  $Y_{[\alpha\beta\gamma]} =\mathcal{Q}_{\alpha\beta\gamma}$ \\
\hline\hline
\end{tabular}
}
\caption{The harmonics of the universal long spin--$3/2$ multiplet appearing in
Kaluza Klein compactifications with $\mathcal{N}=3$ supersymmetry.
This multiplet suggests a super-Higgs mechanism $\mathcal{N}=4 \to \mathcal{N}=3$ }
\label{tabellas}
\end{center}
\end{table}
\par
The effect is summarized in table \ref{tabellas} where we list
the harmonics associated with each of the states of $SD(\ft
32,3,0\vert 3)$.
\par
All of the harmonics of $SD(\ft 32,3,0\vert 3)$ can be expressed in a universal
way (namely, independently from the details of the $X^7$ geometry); indeed
they all can be expressed purely in terms of the Killing spinors.
Let us write here such expressions, leaving, for the sake of clarity,
the account of some derivations to Appendix \ref{appenc}.
\par
$\cdot$ \emph{The massive gravitino harmonic}
The top state of the multiplet is given by the massive gravitino of
scale dimension $E =\ft 92$. This corresponds to an internal
harmonic $\Xi$ being an eigenstate of the operator (\ref{unmezcube})
with eigenvalue $M_{(\ft 12)^3}=-16$. Such a harmonic is
\begin{equation}
 \Xi \equiv \eta^0 =\ft{1}{3} \epsilon_{ABC} \,\tau_\alpha
  \eta^A \overline{\eta}^B \,\tau^\alpha \, \eta^C~,
\label{etazero}
\end{equation}
namely, by comparison with table \ref{ushorgrav}, the harmonic of the
massless dilatino sitting in the $\mathcal{N}=3$ graviton multiplet.
\par
$\cdot$ \emph{The harmonic of the massive $W$ vector}
The $\mathcal{N}=3$ massless graviton multiplet  contains  a triplet
of massless graviphoton gauging the $\mathrm{SU}(2)$ R--symmetry group. Their
harmonics are the $\mathrm{SO}(3)$ Killing vectors $k^A_\alpha$ admitting the
following expression in terms of Killing spinors:
\begin{equation}
  k^A_\alpha=\ft{1}{2} \, \epsilon^{ABC} \,
  \overline{\eta}^B \,\tau_\alpha \, \eta^C~.
\label{kivectors}
\end{equation}
By means of the shadowing mechanism described in eq.s
(\ref{gmassW},\ref{emme2w}), the $W$--vectors associated with the
harmonics (\ref{kivectors}) are the $s=1,J=1,E=5$ states of the
universal multiplet we are discussing.
\par
$\cdot$ \emph{The harmonic of the massive $Z$ vector}
The universal multiplet of table (\ref{brokengravitino}) contains a
$J=1$ triplet of massive vector fields with scale dimension $E=4$.
These take origin from the $Z$--sector of the Kaluza Klein expansion,
namely they are associated with $2$--forms on the compact
$7$--manifold $X^7$. The corresponding harmonic, that is an eigenstate
of the Hodge de Rham operator (\ref{unosquare}) with eigenvalue
$M_{(1)^2 0}=96$, is constructed in terms of Killing spinors as
follows:
\begin{equation}
  T^A_{\alpha\beta}\equiv 7 \, \overline{\eta}^A
  \,\tau_{\alpha\beta} \, \eta^0 +
\overline{\eta}^A\, \tau_{[\alpha} D_{\beta]}
\eta^0~.
\label{Tharmo}
\end{equation}
Eq. (\ref{Tharmo}) is an immediate consequence of eq.s (4.72b),(4.73)
and (4.74) of \cite{univer}. Such equations explain how to construct
a transverse $2$--form with eigenvalue
\begin{equation}
  M_{(1)^20}=( M_{(\ft 12)^3} + 4) \, ( M_{(\ft 12)^3} + 8)
\label{relaz1}
\end{equation}
from any spinor $\Xi$ with eigenvalue $ M_{(\ft 12)^3}$. In the case
$M_{(\ft 12)^3}=-16$, eq.(\ref{relaz1})  yields $ M_{(1)^20}=96$. So
via the universal relations of \cite{univer} also this state is in
the shadow of the massless graviton multiplet and, as it is
explicitly evident from combining equations (\ref{Tharmo}) with eq.
(\ref{etazero}), its harmonic is expressed solely in terms of Killing
spinors.
\par
$\cdot$ \emph{The harmonic of the massive $E=11/2$ spinors}
Next the universal multiplet of table \ref{brokengravitino} contains
a $J=1$ triplet of massive spinors with scale dimension $E= \ft
{11}2$. These are just the shadows of the massless gravitinos,
according to the mass relation discussed in eq.(\ref{mlam112}).
Indeed their harmonic is just given by the triplet of Killing
spinors
\begin{equation}
  \Xi = \eta^A~.
\label{xi=chispi}
\end{equation}
Reading table \ref{brokengravitino} from top to bottom the next item
to discuss should be the spinors with $E=\ft 92$. To understand their
structure, it is however more convenient to make a leap and go first
to the pseudo--scalars with $E=5$.
\par
$\cdot$ \emph{The harmonic of the massive $\pi$ scalars with $E=5$}
We leave the discussion of this case to Appendix \ref{appenc}. The
harmonics turn out to be given simply by
\begin{equation}
  \Upsilon^{(AB)}_{\alpha\beta\gamma}=
  \overline{\eta}^A \,\tau_{\alpha\beta\gamma} \,
  \eta^B~.
\label{upsilon}
\end{equation}
Since the three index $\tau$--matrix is
symmetric, the above expression is symmetric in the $\mathrm{SO}(3)_R$ indices
$A,B$. Decomposing into the traceless and trace parts we obtain the
$J=2$ and $J=0$ states.
\par
$\cdot$ \emph{The harmonic of the massive $E=9/2$  spinors }
We leave the expression of these harmonics to Appendix \ref{appenc}.
\par
$\cdot$ \emph{The harmonic of the massive $E=7/2$  spinors }
Leaving the discussion to Appendix \ref{appenc}, we give here only the
resulting expression of the harmonics in terms of Killing spinors:
\begin{equation}
  \Theta^A_\alpha \equiv \epsilon ^{ABC} \, \left(
  \,\frac{3}{16}\,\tau _{\alpha \mu \nu \rho
  } \, \eta^B \, D_\mu \, T^C_{\nu\rho}+\frac{9}{2} \,\tau
  _\mu \, \eta^B \, T^C_{\alpha \mu } \right)~,
\label{truccolo}
\end{equation}
the quantities $T^A_{\alpha\beta}$ having been defined in eq. (\ref{Tharmo}).
\par
$\cdot$ \emph{The harmonic of the massive $\pi$ scalars with $E=3$}
This pseudo-scalar particle corresponds to the Clifford vacuum of the entire
super-Higgs multiplet and, for this reason it is probably the most important
state in the whole multiplet. In section \ref{n010sec} we shall see
its $7$--dimensional geometric
interpretation in the case of the specific compactification
$X^7=\n010$  and later we shall discuss its superconformal correspondent in terms of
world--volume gauge fields. Here we want to stress its universal
character as a {\sl shadow} of the massless graviton multiplet. Hence
we present the construction of its harmonic in terms of Killing
spinors.
\par
We refer to eq.s (4.40b-4.43) of \cite{univer} which, applied to our
case, show how to construct a $3$--form of eigenvalue $M_{(1)^3}=3$
starting from the harmonic $\Theta^A_\alpha $ of eigenvalue  $M_{\ft 32(\ft
12)^2}=-16$. Indeed the relevant mass relation (4.43) of
\cite{univer} is:
\begin{equation}
  M_{(1)^3}=-\frac 14 (M_{\ft 32(\ft 12)^2}+4)~.
\label{reluncutremez}
\end{equation}
Hence, following the prescriptions of \cite{univer}, we consider the
combination
\begin{equation}
  \widehat{\Pi}^{AB}_{\alpha\beta\gamma} ={\bar \eta}^A \, \tau
  _{[\alpha\beta} \,\Theta^B_{\gamma]} +\frac 17 \,{\bar \eta}^A \, \tau
  _{[\alpha} \,D_\beta \,\Theta_{\gamma]}^B
\label{clifvacspi}
\end{equation}
which, by construction, is an eigenstate of the operator (\ref{1cube})
with eigenvalue $M_{(1)^3}=3$. A priori this object might contain
isospins $J=2,J=1$ and $J=0$ parts. However, explicit computation in a
$\tau$--matrix basis
shows that
\begin{equation}
   \widehat{\Pi}^{AB}_{\alpha\beta\gamma}= \delta^{AB} \,
   \mathcal{Q}_{\alpha \beta \gamma }~.
\label{tascapane}
\end{equation}
Hence, in perfect agreement with $\mathrm{Osp}(3\vert 4)$ representation
theory we  just have a spin $J=0$ state and nothing else.
\par
$\cdot$ \emph{The harmonic of the Lichnerowitz scalars with $E=4$}
The last harmonics to be discussed are those associated with the
$E=0,J=2\oplus J=0$ scalars. From the Kaluza Klein viewpoint these
states originate from traceless fluctuations of the internal $7$--dimensional
metric. Their harmonics are  therefore eigenfunctions of the
Lichnerowitz operator (\ref{duelinea}) with eigenvalue
$M_{2(0)^2}=96$. To derive their expression in terms of Killing
spinors it suffices to recall the mass relations (4.18-4.25) of
\cite{univer} that teach us how to construct an eigenstate  of the
Lichnerowitz operator with eigenvalue
\begin{equation}
  M_{2(0)^2} = ( M_{\ft 32 (\ft 12)^2}+4) \,( M_{\ft 32 (\ft
  12)^2}+8 )
\label{m2fromm32}
\end{equation}
starting from an eigenstate of the Rarita-Schwinger operator
(\ref{raritsch}) pertaining to the eigenvalue $M_{\ft 32 (\ft 12)^2}$.
In our case the fermionic harmonic to start from is clearly
$\Theta^A_\alpha$ given in eq. (\ref{truccolo}), which has $M_{\ft 32 (\ft
12)^2}=-16$. Hence, following the prescriptions of \cite{univer}, we
consider the following combination:
\begin{equation}
K^{AB}_{(\alpha \beta)} ={\bar \eta}^A \,
\tau _{(\alpha } \, \Theta^B_{\beta )} + \frac{1}{3}
\,{\bar \eta}^A \,D_{(\alpha } \,
\Theta^B_{\beta)}~,
\label{KABalbe}
\end{equation}
which by construction is an eigenstate of eigenvalue $M_{2(0)^2}=96$. One can
explicitly check that (\ref{KABalbe})is symmetric in the isospin indices $AB$,
so that it decomposes into a $J=2$ plus a $J=0$ representation. In particular,
the combination that is an isospin singlet is
\begin{equation}
\mathcal{H}^{0}_{(\alpha \beta ) } =
\frac{1}{288} \left(K^{1,1}_{(\alpha \beta )} + K^{2,2}_{(\alpha \beta ) }
+ K^{3,3}_{(\alpha \beta ) }\right)~.
\label{hh0}
\end{equation}
\par
This concludes our proof that in Kaluza Klein supergravity when we compactify
M--theory on $\mathrm{AdS}_4 \times X^7$ with $\mathcal{N}=3$ Killing spinors there is always a
universal massive spin--$3/2$ multiplet of scale dimension
\begin{equation}
E_0=3
\label{e0is3}
\end{equation}
which is just the necessary shadow of the massless graviton multiplet.
\subsection{Freund-Rubin compactification on $\n010$}
\label{n010sec}
The space $\n010$, a particular instance of the series of 7-dimensional coset
spaces named $N^{p,q,r}$ in the classification of \cite{castromwar}, is the  only
7-dimensional coset that, when used as a compactification manifold for 11D
supergravity, can preserve $\mathcal{N}=3$ supersymmetry \cite{castn010}. The complete
KK spectrum of the $\n010$ compactification has been derived in
\cite{n010massspectrum}, and its $\mathrm{Osp}(3|4)$ multiplet structure
elucidated in \cite{osp34}.
\par
In the perspective of the AdS/CFT correspondence, the dual of the $\n010$
compactification should be a $\mathcal{N}=3$ conformal field theory in $2+1$
dimensions. This theory is the subject of a paper \cite{noialtro} which is
somehow companion to the present one.
\par
The space $\n010$ can be simply defined as the coset space
\begin{equation}
\label{n010}
{G\over H} =
{\mathrm{SU}(3)\over \mathrm{U}(1)}~,
\end{equation}
where, using the Gell-Mann matrices $\lambda^\alpha$ as
$\mathrm{su}(3)$ generators, the quotient is taken  w.r.t.
the $\mathrm{U}(1)$ subgroup generated by $\lambda^8$.
The isotropy group of $\n010$ is $\mathrm{SU}(3)\times \mathrm{SU}(2)$;
the $\mathrm{SU}(2)$ factor is the normalizer of the $U(1)$ action and,
explicitly, it is generated by $\lambda^{1,2,3}$.
\par
Let $\Omega^A=(\Omega^\alpha,\Omega^8)$ be the Maurer-Cartan
forms for $\mathrm{SU}(3)$, namely let
$\Omega = \Omega^A \lambda^A = g^{-1} d g$, with
$g\in\mathrm{SU}(3)$, so that $d\Omega + \Omega\wedge\Omega = 0$.
The vielbein corresponding to a generic
$\mathrm{SU}(3)\times \mathrm{SU}(2)$-invariant
metric are obtained from the coset vielbein
$\Omega^\alpha$ ($\alpha=1,\ldots 7$)
by rescaling independently the two
groups associated to
$\lambda^{\dot\alpha}$ ($\dot\alpha=1,2,3$) and
$\lambda^{\tilde\alpha}$ ($\tilde\alpha=4,5,6,7$). Indeed such a
decomposition is respected both by the $\mathrm{U}(1)$ quotient and by the
$\mathrm{SU}(2)$ action. Thus we have\footnote{Due to a different
choice of structure
constant, our rescaling $\alpha$ is minus twice the one used in
\cite{castellani99}.}:
\begin{equation}
\label{n010viel}
B^\alpha = (\alpha^{-1} \Omega^{\dot \alpha},\beta^{-1}
\Omega^{\tilde \alpha})~.
\end{equation}
The spin connection $B^{\alpha\beta}$ and the curvature associated to
these vielbein are straightforwardly computed (see \cite{castellani99});
in particular, the Ricci
tensor is diagonal, with only two independent entries:
\begin{eqnarray}
\label{riccin010}
\mathcal{R}_{\dot \alpha\dot \beta} & = & {1\over 4}
\left(\alpha^2 + {1\over 2}
{\beta^4\over\alpha^2}\right)\, \eta_{\dot \alpha\dot \beta}~;
\\
\mathcal{R}_{\tilde \alpha\tilde \beta} & = & {3\over 4}
\left(\beta^2 - {1\over 4}
{\beta^4\over\alpha^2}\right)\, \eta_{\tilde \alpha\tilde \beta}~.
\end{eqnarray}
Requiring the space to be Einstein,
gives two possible values of the ratio
$\beta^2/\alpha^2$, namely $\beta^2/\alpha^2=2$ or $\beta^2/\alpha^2=2/5$.
If we furthermore require that the Ricci tensor has exactly the value required
by the Freund Rubin ansatz, eq. (\ref{cosmol4e7}),
we find four different possibilities, of which two preserve some supersymmetry.
\par
The ``standard'' $\n010$ metric is obtained with the following rescalings:
\begin{equation}
\label{standardn010}
\alpha=-4\,\, e~,\hskip 0.5cm \beta=\pm 4\sqrt{2}\,\, e~.
\end{equation}
It preserves $\mathcal{N}=3$ supersymmetry.
It is known \cite{castn010} that, when $\n010$ is realized as the
coset (\ref{n010}), its Killing spinors must actually be \emph{constant}.
With the rescalings (\ref{standardn010}), there
are 3 independent constant spinors $\eta^A$ ($A=1,2,3$)
that satisfy eq. (\ref{kilspineq7dim}), namely
\begin{equation}
\label{n010kilspin}
-{1\over 4} B^{\alpha\beta}_{~~\gamma} \tau_{\alpha\beta}\,\eta^A
 = e\,\tau_\gamma\,\eta^A~.
 \end{equation}
They transform as a \emph{triplet} under the $\mathrm{SU}(2)$
part of the isometry, which therefore truly acquires the role of the R-symmetry
group $\mathrm{SU}(2)_R$ for the 4-dimensional gauged supergravity that arises
from the compactification.
\par
There is a possible solution that differs from (\ref{standardn010})
only by the sign of the rescaling $\alpha$.
While the sign of $\beta$ is irrelevant, because $\beta$ appears quadratically
also in the spin connection, reversing the sign of $\alpha$
amounts to reversing the sign of the spin connection
(or, equivalently, to changing the orientation
of the manifold). This solution with opposite orientation
preserves no supersymmetry.
\par
Of the two solutions with  $\beta^2/\alpha^2=2/5$, the one with
\emph{opposite} orientation to the standard $\n010$ (\ref{standardn010}),
that we name $\tn010$:
\begin{equation}
\label{tn010}
\alpha={20\over 3}\,\, e~,\hskip 0.5cm \beta=\pm {4\over 3}\sqrt{10}\,\, e~,
\end{equation}
preserves $\mathcal{N}=1$ supersymmetry.
Indeed, one finds that the Killing spinor equation
\begin{equation}
\label{tn010kilspin}
-{1\over 4} \tilde B^{\alpha\beta}_{~~\gamma} \tau_{\alpha\beta}\,\eta^0
= e\,\tau_\gamma\,\eta^0~
\end{equation}
admits a single solution $\eta^0$, which is a
\emph{singlet} of $\mathrm{SU}(2)_R$.
\par
The solution differing from the $\tn010$ solution
(\ref{tn010}) by the sign of $\alpha$ (so that it has the same
orientation as the standard $\n010$) preserves no supersymmetry.
\subsubsection{Harmonics of the shadow multiplet for $\n010$}
\label{n010harm}
Freund Rubin compactification on $\n010$, with rescalings (\ref{standardn010}),
gives an $\mathcal{N}=3$ theory on AdS$_4$. As discussed in Section \ref{miracle}, in this
compactification a ``shadow'' massive gravitino
multiplet is generated. This suggests that truncating the theory to include
the fields of this massive gravitino could be consistent, the resulting
4-dimensional theory corresponding to a \emph{broken} $\mathcal{N}=4$ supergravity theory
on AdS$_4$.
\par
In section \ref{miracle} all the harmonics corresponding to
the fields sitting in the shadow multiplet of a generic $\mathcal{N}=3$ KK
compactification have been expressed in terms of the $\mathrm{SU}(2)_R$ triplet of
killing spinors $\eta^A$, whose existence is necessary for having residual $\mathcal{N}=3$
supersymmetry. The peculiarity of the $\n010$ compactification is that these
spinors are actually \emph{constant}. Thus all the harmonics of the shadow
multiplet will in fact correspond to \emph{constant} deformations of the
internal space $\n010$. Tensor products of these harmonics will involve again
only constant harmonics, without involving harmonics of other massive
multiplets, indicating again the possibility of a consistent truncation.
\par
Let us discuss here in detail the explicit form of those harmonics that will be
relevant for the effective 4-dimensional theory we will construct, or whose
geometrical interpretation can be somehow illuminating. We refer to
table \ref{tabellas} for notations.
\par
First of all, the harmonic of the massive gravitino,
given in eq. (\ref{etazero}), turns out to be exactly the $\eta^0$ of eq.
(\ref{tn010kilspin}), i.e. the constant spinor that would generate the single
supersymmetry of the $\tn010$ compactification.
\paragraph{Killing vectors} The Killing vectors $k^A$ of the $\mathrm{SU}(2)_R$
isometry represent, according to the discussion in section \ref{miracle},
the harmonic associated to the massive vector $W$, of energy $E=5$. From
their universal expression in terms of the Killing spinors $\eta^A$, eq.
(\ref{kivectors}), one gets, in our conventions,  the following
explicit components:
\begin{equation}
\label{kivn010}
k^{1}_\alpha = \delta_{\alpha,2}~,\hskip 0.5cm
k^{2}_\alpha = -\delta_{\alpha,1}~,\hskip 0.5cm
k^{3}_\alpha = \delta_{\alpha,3}~.
\end{equation}
The vectors $k^A=k^A_\alpha\partial_\alpha$
close the $\mathrm{SU}(2)_R$ isometry algebra,
and the $\mathrm{SU}(2)_R$ transformation of the vielbein
$B^\alpha$ is obtained via the Lie derivative:
\begin{equation}
\label{su2actviel}
\delta_A\, B^\alpha = l_{k^A} B^\alpha =
k^A_\gamma (B^\alpha_{\beta|\gamma} - B^\alpha_{\gamma|\beta})\,B^\beta
\equiv (J^A)^\alpha_{~\beta}\, B^\beta~,
\end{equation}
where the Lie derivative is obviously simplified by the fact that the Killing
vectors are constant in the vielbein basis. The action of $\mathrm{SU}(2)_R$ on
the vielbein $B^\alpha$, encoded in the matrices $(J^A)^\alpha_{~\beta}$, is
reducible, and splits into a $J=1$ representation (spanned by the first three
vielbeins $B^{\dot\alpha}$) and a complex $J=1/2$ representation spanned by the
last four ones, $B^{\tilde\alpha}$.
\par
\paragraph{Deformations of the internal metric}
The constant harmonic $Y=1$ corresponds to
a dilatation of the internal metric,
namely to a common rescaling of all the vielbeins. This mode is
associated to a  four dimensional scalar field of energy $E=6$,
named $\Sigma(x)$  according to the
notation of eq.(\ref{kkexpansion}) and Table \ref{tabellas}.
\par
The harmonics $H^0_{(\alpha\beta)}$ and $H^{(AB)}_{(\alpha\beta)}$ that give rise
to scalar fields of energy $E=4$, parametrize traceless deformations of the
7-dimensional metric, see the 3rd of eq.s (\ref{kkexpansion}).
They are eigenfunctions  of the Lichnerowitz operator (\ref{duelinea})
with eigenvalue $M_{2(0)^2}=96$.
They admit a universal expression in terms of Killing spinors
that was derived in section \ref{miracle}
(see eq.s (\ref{KABalbe}) and (\ref{hh0})).
\par
It is interesting to retrieve the geometric interpretation of these
tensors on the manifold $\n010$. To this effect we can consider
the Lichnerowitz operator (\ref{duelinea}) applied to
the space of constant, $H$-invariant, traceless symmetric $2$-tensors.
Via explicit calculations this space is found to be $9$-dimensional.
\par
Six states correspond to the eigenvalue $M_{2(0)^2}=96$.
One of these states is a singlet ($J=0$) of $\mathrm{SU}(2)_R$, while the others
correspond to $J=2$. The explicit expression of the $J=0$ eigenstate is
\begin{equation}
\label{h0n010}
H^0_{\alpha\beta}=
({4\over 3}\, \delta_{\dot\alpha\dot\beta},\,
-\delta_{\tilde\alpha\tilde\beta})~.
\end{equation}
It corresponds to a rescaling of the 7-dimensional vielbein
that preserves the volume of $\n010$: it is a
``squashing'' deformation. We will name $\phi(x)$
the corresponding scalar fields of energy $E=4$.
The $J=2$ eigenstates are non-diagonal, and
correspond thus to deformations of the
$\n010$ metric which are non-diagonal in the vielbein basis.
\par
Three states correspond to an eigenvalue $M_{2(0)^2}=0$.
They are organized in a triplet ($J=1$) of $\mathrm{SU}(2)_R$.
The corresponding massless scalars belong to the Betti vector
multiplet, namely to the additional massless vector multiplet that
originates from the fact that $\n010$ admits
a harmonic $2$--form \cite{univer,adscftcheckers}.
\par
If we allow both dilatation and squashing deformations, we see that
we can in fact rescale independently the first three vielbeins,
$V^{\dot\alpha}$, and the last four ones, $V^{\tilde\alpha}$.
As we already noticed, see eq. (\ref{n010viel}), this is the most general type
of rescaling that still gives a $\mathrm{SU}(2)_R$-invariant metric.
This what we expected, since we considered only the deformations which are
$\mathrm{SU}(2)_R$ singlets.
\par
Summarizing, if we include in the effective theory the
$\mathrm{SU}(2)_R$-invariant scalar fields $\Sigma$ and $\phi$, of $E=6$ and
$E=4$, belonging to the ``shadow'' gravitino multiplet, we are considering,
according to eq. (\ref{kkexpansion}), the following fluctuation of the internal
metric:
\begin{eqnarray}
\label{metricfluc}
h_{\alpha\beta}(x,y) & = & -12 \Sigma(x)\, \delta_{\alpha\beta} + \phi(x)
\,H^0_{\alpha\beta}
\\ \nonumber
& = & \left((-12\, \Sigma(x) +{4\over 3}\, \phi)\,\delta_{\dot\alpha\dot\beta},
 (-12\, \Sigma(x) -\, \phi)\,\delta_{\tilde\alpha\tilde\beta}\right)~.
\end{eqnarray}
This corresponds\footnote{%
Indeed, from eq. (\ref{viel7resc}) it follows
\begin{equation}
\label{vieltofluc}
\delta_{\alpha\beta} V^\alpha(x,y) V^\alpha(x,y)
=\left((\delta_{\alpha\beta} + h_{\alpha\beta}(x,y) + O(h^2)\right)
\bar V^\alpha(y) \bar V^\beta(y)~.
\end{equation}}
to a rescaling of the 7-dimensional vielbeins:
\begin{eqnarray}
\label{viel7resc}
B^{\dot\alpha}(x,y) & = & \ee^{-6\Sigma(x)+{2\over 3}\phi(x)}\,
\bar B^{\dot\alpha}(y)
\\
B^{\tilde\alpha}(x,y) & = & \ee^{-6\Sigma(x)-{1\over 2}\phi(x)}\,
\bar B^{\tilde\alpha}(y)~,
\end{eqnarray}
where on the r.h.s. the vielbein $\bar B^\alpha(y)$ are those of the standard
$\mathcal{N}=3$ $\n010$ solution. It is convenient for the discussion of the
effective action to introduce the following combinations:
\begin{eqnarray}
\label{scalred}
w(x) & = & -18 \Sigma(x) - 2 \phi(x)~,\nonumber\\
z(x) & = & -18 \Sigma(x) + {1\over 3} \phi(x)~, \label{wzscalred}
\end{eqnarray}
in terms of which the rescalings of the vielbein read
\begin{eqnarray}
\label{viel7rescwz}
B^{\dot\alpha}(x,y) & = & \ee^{{1\over 3}w(x)}\,
\bar B^{\dot\alpha}(y)
\\
B^{\tilde\alpha}(x,y) & = & \ee^{-{1\over 6}w(x)+{1\over 2}z(x)}\,
\bar B^{\tilde\alpha}(y)~.
\end{eqnarray}
\paragraph{Turning on an internal 3-form}
The harmonics $\hat\Upsilon_{[\alpha\beta\gamma]}$ and
$\mathcal{Q}_{[\alpha\beta\gamma]}$ both corresponds to switching on,
a the three-form field $A_{\alpha\beta\gamma}$ on the internal
7-dimensional space.
They are eigentensors of the ${}^*d$ operator on three-forms,
eq. (\ref{1cube}), with eigenvalue
$M_{(1)^3}=-2$, resp. $M_{(1)^3}=3$, and are $\mathrm{SU}(2)_R$ singlets.
They give rise to scalar fields which we name $\pi_1(x)$, resp. $\pi_2(x)$,
of energy $E=5$, resp. $E=3$. In terms of Killing spinors, we have
 $\hat\Upsilon_{[\alpha\beta\gamma]}$
$=\bar\eta^A\tau_{\alpha\beta\gamma}\eta^A$, see eq. (\ref{upsilon}),
while the expression of
$\mathcal{Q}_{[\alpha\beta\gamma]}$ is given through
eq.s (\ref{tascapane},\ref{clifvacspi},\ref{truccolo}).
For $\n010$, these harmonics are constant tensors.
They are found to be given by linear combinations of the two
$\mathrm{SU}(2)_R$ invariants of order $3$ one can construct
with the vielbeins, using the Killing vectors (\ref{kivn010})
and the matrices $J^A$ describing the $\mathrm{SU}(2)_R$ action
(\ref{su2actviel}):
\begin{eqnarray}
  \mathcal{Q}_{\alpha\beta\gamma}&=&
  8 \,k^A_\alpha \,k^B_\beta \, k^C_\gamma \,
  \epsilon_{ABC} - 6 \, k^A_{[\alpha} \,
  J^A_{\beta\gamma]}~,
\label{qinterpre}\\
  \widehat{\Upsilon}_{\alpha\beta\gamma}&=&
  -3 \,k^A_\alpha \,k^B_\beta \, k^C_\gamma \,
  \epsilon_{ABC} - 6 \, k^A_{[\alpha} \,
  J^A_{\beta\gamma]}~.
\label{upinterpre}
\end{eqnarray}
Explicitly, they have the following non-zero components:
\begin{equation}
\label{ucomp}
\hat\Upsilon_{[123]} = 3~,~~
\hat\Upsilon_{[147]} =\hat\Upsilon_{[165]} =\hat\Upsilon_{[246]} =
\hat\Upsilon_{[257]} =\hat\Upsilon_{[345]} =\hat\Upsilon_{[376]} = 1
\end{equation}
and
\begin{equation}
\label{qcomp}
\mathcal{Q}_{[123]}= 2~,~~
\mathcal{Q}_{[147]}=\mathcal{Q}_{[165]}=\mathcal{Q}_{[246]}=
\mathcal{Q}_{[257]}=\mathcal{Q}_{[345]}=\mathcal{Q}_{[376]}= -1~.
\end{equation}
An alternative basis for the deformations parametrized by $\hat\Upsilon$ and
$\mathcal{Q}$, which will prove more convenient for later calculations,
is provided by the tensors
$\mathcal{S} = (\hat\Upsilon + \mathcal{Q})/5$ and
$\mathcal{T} = (2\hat\Upsilon - 3 \mathcal{Q})/5$,
whose only non-vanishing components are
\begin{equation}
\label{STcomp}
\mathcal{S}_{[123]}= 1~,~~
\mathcal{T}_{[147]}=\mathcal{T}_{[165]}=\mathcal{T}_{[246]}=
\mathcal{T}_{[257]}=\mathcal{T}_{[345]}=\mathcal{T}_{[376]}= 1~.
\end{equation}
Namely, the only components are of type
$\mathcal{S}_{\dot\alpha\dot\beta\tilde\gamma}$ and
$\mathcal{T}_{\dot\alpha\tilde\beta\tilde\gamma}$.
\par
Summarizing, to include in the effective theory
the $\mathrm{SU}(2)_R$-invariant scalar fields $E=5$ and $E=3$,
belonging to the ``shadow''  gravitino multiplet,
we are considering, according to eq. (\ref{kkexpansion}), a
fluctuation\footnote{%
Notice that the field $A_{\alpha\beta\gamma}$ coincides with its fluctuation
$a_{\alpha\beta\gamma}$ appearing in eq. (\ref{kkexpansion}) as the background
value of $A_{\alpha\beta\gamma}$ is zero.}
of the 3-form field in the 7-dimensional directions:
\begin{equation}
\label{3formfluct}
A_{\alpha\beta\gamma}(x,y) =
\pi_1(x)\, \hat\Upsilon_{\alpha\beta\gamma} +
\pi_2(x)\, \mathcal{Q}_{\alpha\beta\gamma}~.
\end{equation}
A parametrization that is more convenient for deriving the effective action
is in terms of two different scalar fields $f(x)$ and $g(x)$, associated to
the two subset of components of the three-form $A$
that rescale under (\ref{viel7resc}) in a fixed way:
\begin{eqnarray}
\label{3formfluct2}
A_{\dot\alpha\dot\beta\dot\gamma}(x,y) & = &
f(x)\, \ee^{18\Sigma(x) - 2 \phi(x)}\,
\mathcal{S}_{\dot\alpha\dot\beta\dot\gamma}
= f(x)\, \ee^{w(x)}\,\mathcal{S}_{\dot\alpha\dot\beta\dot\gamma}~,
\nonumber\\
A_{\dot\alpha\tilde\beta\tilde\gamma}(x,y) & = &
g(x)\, \ee^{18\Sigma(x) + {1\over 3}\phi(x)}\,
\mathcal{T}_{\dot\alpha\tilde\beta\tilde\gamma}=
g(x)\, \ee^{z(x)}\,\mathcal{T}_{\dot\alpha\tilde\beta\tilde\gamma}~,
\end{eqnarray}
where we have explicitly taken into account the effect of the
rescaling (see eq. (\ref{viel7resc})) so that the
tensors $\mathcal{S}_{\dot\alpha\dot\beta\dot\gamma}$ and
$\mathcal{T}_{\dot\alpha\tilde\beta\tilde\gamma}$ in the r.h.s. remain the
constant ones of eq. (\ref{STcomp}). Eq. (\ref{3formfluct2}) also displays
the geometrical meaning of the rescalings $w$ and $z$.
Of course, $f$ and $g$ will no longer be mass eigenstates of the
effective theory. Their relations to the mass eigenstates $\pi_1,\pi_2$ follows
from the relations among $\mathcal{S},\mathcal{T}$ and
$\hat\Upsilon,\mathcal{Q}$.
\subsubsection{The shadow of the massless vector multiplets}
As we observed in section \ref{shagene}, see eq. (\ref{emme2w}), every massless
vector that appears in Kaluza--Klein theory has a massive vector companion,
named $W$, of mass $m_W^2=192$, in the adjoint representation of the
same gauge group.
\par
We have seen what is the role of the massive shadows of the graviphotons in an
$\mathcal{N}=3$ compactifications. They are part of a universal massive
spin--3/2 multiplet that hints to a hidden super-Higgs mechanism. However, in
addition to the graviphotons gauging the $R$--symmetry, we also have the
massless vectors that belong to massless vector multiplets and correspond to
\emph{flavor} symmetries of the dual conformal field theory.
In which kind of multiplets do these ``shadows'' of the flavor
currents fit?
\par
The answer to this question cannot be given in universal terms
since  flavor symmetries depend on the specific choice of the $X^7$ manifold.
Yet it can be inferred that the \emph{type} of multiplet to which such shadows
belong is a universal feature of Kaluza Klein compactifications.
\par
In the case of the manifold $\n010$, we can read off the desired answer from the complete
spectrum of $\mathrm{Osp}(3 \vert 4) \times \mathrm{SU}(3)$ supermultiplets derived in
\cite{osp34,n010massspectrum}.
\renewcommand{\arraystretch}{1.3}
\begin{table}
  \centering
    {\small
    \begin{tabular}{ccc}
      \hline\hline
      \multicolumn{3}{c}{$SD(2,5/2,1\vert 3)$ (in the $\mathbf{8}$ of
      $\mathrm{SU}(3)_{\mathrm{flavor}})$} \cr
      \hline
      Spin & Energy & Isospin  \cr
      \hline\hline
      $2$ & $4$ & $1$ \cr
      \hline
        $\ft 32$ &
    $\begin{array}{c}\ft 92 \\ \ft 72 \end{array}$ &
    $\begin{array}{c}1 \oplus 0 \\ 2\oplus 1 \oplus 0\end{array}$ \cr
      \hline
      1 &
      $\begin{array}{c} 5\\ 6 \\ 3 \end{array}$ &
      $\begin{array}{c} 0 \\ 2\oplus 1 \oplus 1 \oplus 0 \\
        2\oplus 1 \oplus 0\end{array}$  \cr
      \hline
      $\ft12$ &
      $\begin{array}{c} \ft 92 \\ \ft 72 \\ \ft 52 \end{array}$ &
      $\begin{array}{c} 1 \\ 2\oplus 1\oplus 1\oplus 0 \\ 1\end{array}$ \cr
      \hline
      $0$ &
      $\begin{array}{c}4\\ 3\end{array}$ &
      $\begin{array}{c}1 \\ 1\oplus 0\end{array}$ \cr
      \hline\hline
    \end{tabular}
    }
   \caption{In $\n010$ compactification, the
    shadow of the massless vector multiplet in the adjoint of
    $\mathrm{SU}(3)_{\mathrm{flavor}}$ is the exceptional
    short ${\cal N}=3$ graviton multiplet $SD(2,\ft 5 2 , 1 \vert 3)$
    with $E_0 =\ft52$ and  $J_0=1$.}
   \label{shadowsu3}
\end{table}
In particular, looking at eq. (8) of that
paper and at table \ref{shadowsu3}, we can consider the case
$k=1,j=0$. Correspondingly, we find a long
graviton multiplet $SD(2,E_0=\ft 52 , J_0= 1 \vert 3)$ in the adjoint
representation $\mathbf{8}$ of the flavor $\mathrm{SU}(3)$ group.
Indeed, recalling eq. (7) of \cite{osp34}, for $M_1=M_2=1$
and $J_0=1$ we have $H_0=64$,  hence $E_0=5/2$.
This long rational multiplet is displayed in table \ref{shadowsu3}
and includes both the $W$-shadows of the massless $\mathrm{SU}(3)$ gauge bosons
and the massive gravitinos that are shadows of the massless $\mathrm{SU}(3)$
gauginos. Since it goes up to spin $s_{\rm max}=2$ this long multiplet cannot
be used to describe any Higgs or super-Higgs mechanism. This is in agreement
with another feature that also points in the same direction.  Indeed,
differently from the harmonics of the universal super-Higgs multiplet, the
harmonics of this long multiplet are not \emph{constant}. So we can expect that
tensor products thereof  can contains the harmonics of other massive
multiplets. This excludes that the truncation of Kaluza Klein theory to
massless multiplets plus this particular long multiplet can be a consistent
one.
\subsection{Comparison with $\mathcal{N}=2$ compactifications}
The same shadowing mechanism may be directly tested also in the case of
${\cal N}=2$ compactifications using the results on Kaluza Klein
spectra obtained in \cite{m111,adscftcheckers,poliv52}.
For instance, looking at the spectrum of $M^{1,1,1}$ given in
\cite{m111}, one finds the existence of a long vector multiplet with
the structure displayed in table \ref{longvector}.
\begin{table}
\centering
{\small
\begin{tabular}{ccccc}
\hline\hline
Spin &
Energy &
Hypercharge&
Mass ($^2$)&
K.K. origin
\\
\hline
$1$      & $5$      & $0$       & $192$    &  $W$   \\
$\ft12$  & $ \ft{11}2$  & $-1$  & $-16$ &
                    $\lambda_L$   \\
$\ft12$  & $ \ft{11}2$  & $ 1$ &
                       $-16$ & $\lambda_L$    \\
$\ft12$  & $\ft{9}2$  & $-1$ & $12$ &   $\lambda_T$\\
$\ft12$  & $\ft{9}2$  & $ 1$ & $12$ &   $\lambda_T$ \\
$0$      & $6$      & $0$   & $320$  &  $\Sigma$   \\
$0$      & $5$      & $-2$ & $192$  & $\pi$ \\
$0$      & $5$      & $2$ & $192$  & $\pi$ \\
$0$      & $5$      & $0$   & $192$  & $\pi$  \\
$0$      & $4$      & $0$   & $96$  &  $\phi$  \\
\hline\hline
\end{tabular}
}
\caption{The universal long vector multiplet that in $\mathcal{N}=2$
compactifications is the shadow of the massless graviton multiplet}
\label{longvector}
\end{table}
Indeed, it suffices to look at eq. (3.20) of \cite{m111}, and set $M_1=M_2=J=0$
to realize that the corresponding long vector multiplet shown in table 3 of the
same paper has $E_0=4$, $y_0=0$ and exactly reproduces table \ref{longvector}
of the present paper. On the other hand, comparison of table \ref{longvector}
with table \ref{brokengravitino} reveals that the field components of this long
vector multiplet are just a subset of the universal gravitino shadow multiplet
of ${\cal N}=3$ compactifications. In particular, we observe the presence of
the $E=6$ scalar mode $\Sigma$ and of the massive $W$ vector with $E=5$ that is
the shadow partner of the $\mathrm{SO}(2)$ $R$-symmetry gauge field. Therefore
we can easily conclude that table \ref{longvector} displays the universal
structure of the shadow multiplet of the massless supergravity multiplet in
${\cal N}=2$ compactifications. The difference is that in this case this
universal multiplet reaches only spin $1$ and not spin $3/2$. Hence there is no
room for superHiggs interpretation in this class of compactifications.
\section{Partial breaking from $\bf \mathcal{N}=4$ to  $\bf \mathcal{N}=3$ and
the coset structure of supergravity}
\label{whye0}
As we argued in the previous sections, in an $\mathcal{N}=3$ compactification
on $\mathrm{AdS}_4 \times X^7$, the truncation of Kaluza Klein theory to  the
massless modes plus the massive gravitino multiplet should be consistent. The
corresponding $4$--dimensional Lagrangian should then describe a spontaneously
broken phase of $\mathcal{N}=4$ supergravity. Hence we turn our attention to
the partial supersymmetry breaking mechanism from $\mathcal{N}=4$ to
$\mathcal{N}=3$ in the context of $4$--dimensional field theory. Here we find
an intriguing surprise. Using the standard formulation of the $\mathcal{N}=4$
theory, which so far has been considered unique, it turns out that in an
$\mathrm{AdS}_4$-vacuum with $\mathcal{N}=3$ residual unbroken supersymmetry
there is an upper bound on the scaling dimension (or mass) of the broken
gravitino multiplet:
\begin{equation}
 E_0 \left(\vev \mu \right) <  3~,
\label{boundo}
\end{equation}
where $\vev\mu$ stands for the vacuum expectation values
(v.e.vs) of the scalar fields $\mu$, that we name ``moduli'',
in the theory.
\par
Actually, as we are going to explain, if we require $\mathcal{N}=4 \to \mathcal{N}=3$ partial
breaking, the true moduli space $\mathcal{M}^{\mathrm{p.b.}}$  of such vacua is
given by a small subset of scalar fields that are singlet under the residual
$\mathrm{SO}(3)_R$ R--symmetry group and under all unbroken gauge symmetries.
These moduli fill two copies of the upper complex plane equipped with the
Poincar\'e metric:
\begin{equation}
  \mathcal{M}^{\mathrm{p.b.}}\, = \,
  \frac{\mathrm{SU}(1,1)}{\mathrm{U}(1)} \otimes
  \frac{\mathrm{SO}(2,1)}{\mathrm{SO}(2)}~.
\label{modspace}
\end{equation}
\par
The coordinates of $\mathcal{M}^{\mathrm{p.b.}}$ are suitable combinations of
the four isospin singlets ($J=0$) with energies $E_0+3$, $E_0+2$, $E_0+1$ and
$E_0$ appearing in table \ref{brokengravitino}. From the point of view of
Kaluza-Klein theory, these coordinates parametrize the corresponding metric and
internal photon deformations we have already discussed. In the context of
$4$--dimensional $\mathcal{N}=4$ supergravity, the two factors of
$\mathcal{M}^{\mathrm{p.b.}}$ have different origins.
The manifold $ {\mathrm{SU}(1,1)}/{\mathrm{U}(1)}$ contains the scalar degrees
of freedom, $\mu_{\rm grav}$, of the $\mathcal{N}=4$ graviton multiplet, while
${\mathrm{SO}(2,1)}/{\mathrm{SO}(2)}$ is parametrized\footnote{%
Clearly, ${\mathrm{SO}(2,1)}/{\mathrm{SO}(2)}$  is locally equivalent to
$\mathrm{SU}(1,1)/\mathrm{U}(1)$ (or to
$\mathrm{SL}(2,\mathbb{R})/\mathrm{O}(2)$; however we will see later  that, at
the level of the embedding into the full moduli space, the correct description
is the first one.}
by a subset, $\mu_{\rm vec}$, of the $6 \times n$ scalars belonging to $n$
vector multiplets coupled to supergravity.
\par
The main issue of the present section is to show that, in the context of
standard $\mathcal{N}=4$ supergravity, the scaling dimension $E_0$ is actually a
function only of the vector multiplet moduli:
\begin{equation}
  E_0= E_0\left(\vev{\mu_{\rm vec}}\right)~,
\label{restric}
\end{equation}
and that it satisfies the bound in eq.(\ref{boundo}). Moreover, the
bound is saturated only at the boundary of the ``matter'' moduli space
$\mathrm{SO}(2,1)/\mathrm{O}(2)$. This means that, in the appropriate
Poincar\'e metric, the vacuum realized by Kaluza Klein theory, which is
perfectly regular, is at \emph{infinite distance} from all other vacua of
standard $\mathcal{N}=4$ supergravity. This suggests the existence of a more general
formulation of $\mathcal{N}=4$ supergravity able to accommodate also the vacuum that is
explicitly realized by Kaluza Klein theory. Since unitarity imposes only the
bound:
\begin{equation}
E_0 > 1~,
\label{unitarbu}
\end{equation}
it is strongly suggested that the bound (\ref{boundo}) simply divides the full
supergravity moduli space of in two or more open charts. The first chart, which
can be described by the standard Lagrangian, contains all the vacua for which
the massive gravitino corresponds to a unitary representation with $E_0 < 3$.
The other open charts, that should be covered by one or more ``shadow''
Lagrangians, should contain all the unitary representations with  $E_0 > E_{\rm
min}$, for some  $E_{\rm min} < 3$.
\par
In section \ref{hinti} we argue about the possible existence of
similar shadow extensions for all $\mathcal{N} \ge 4$ supergravities.
\subsection{General aspects of partial supersymmetry breaking}
\label{genaspect}
To discuss the partial breaking $\mathcal{N}=4 \to \mathcal{N}=3$ we begin by recalling some
very general aspects of the super-Higgs mechanism in extended supergravity that
were codified in the literature of the early  and middle eighties
\cite{susyb1,deRoo:1985np,susyb2,susyb3} (for a review see chapter II.8 of
\cite{castdauriafre}) and were further analyzed and extended in the middle
nineties \cite{Zinovev:1992mw,Ferrara:1996gu,Fre:1997js,Girardello:1997hf}.
\par
A  vacuum of supergravity is simply identified by a configuration of constant
v.e.v.s of the scalar  fields, $\vev\phi^i=\phi^i_0$,  that is an extremum of
the scalar potential:
\begin{equation}
  \left.\frac{\partial V}{\partial \phi ^i}\right|_{\phi=\phi_0} =  0~,
\label{extrepot}
\end{equation}
and a metric\footnote{%
Here we temporarily use the curved 4D space-time indices $\mu,\nu,...$ in place
of the ``flat'' ones $a,b,\ldots$  to  avoid possible conflicts of notations.}
$ds^2 = g_{\mu\nu} dx^\mu \otimes dx^\nu$ that is either that of Minkowski
space, if $V(\phi_0)=0$, or that of anti de Sitter space AdS$_4$, if $V(\phi_0)
< 0$, or that of de Sitter space dS$_4$, if $V(\phi_0) > 0$. Therefore, in
relation with the super-Higgs mechanism, there are just three relevant items of
the entire supergravity construction that we have to consider.
\begin{enumerate}
\item The \emph{gravitino mass matrix} $S_{AB}(\phi)$ , namely the
non-derivative scalar field dependent term that appears in the gravitino
supersymmetry transformation rule:
\begin{equation}
  \delta \psi _{A\vert \mu} = \mbox{derivative terms} \, + \,
  S_{AB} \left( \phi \right) \, \gamma_\mu \, \epsilon^B~,
\label{gravsusy}
\end{equation}
and reappears as a mass term in the Lagrangian:
\begin{equation}
  \mathcal{L}^{\rm SUGRA} \, = \, \dots \, + \, \mbox{const} \,\left( \,
  S_{AB} (\phi ) \, \psi^A_\mu \, \gamma^{\mu\nu} \, \psi^B_\nu \, +
  \,  S^{AB} (\phi ) \, \psi_{A\vert\mu} \, \gamma^{\mu\nu} \,
  \psi_{B\vert\nu} \,\right)
\label{gravmasmat}
\end{equation}
\item The \emph{fermion shifts}, namely the non-derivative scalar field
  dependent terms in the supersymmetry transformation rule of the spin $\ft 12$
  fields :
\begin{eqnarray}
  \delta \, \lambda^i_R & =& \mbox{derivative terms} \, + \,
  \Sigma_{A}^{\phantom{A} i} \left( \phi \right) \,  \epsilon^A~,
  \nonumber\\
  \delta \, \lambda^i_L & =& \mbox{derivative terms} \, + \,
  \Sigma^{A\vert i} \left( \phi \right) \,  \epsilon_A~.
\label{fermioshif}
\end{eqnarray}
\item The scalar potential itself, $V(\phi)$.
\end{enumerate}
These three items are related by a general supersymmetry Ward identity, firstly
discovered in the context of gauged $\mathcal{N}=8$ supergravity \cite{dewit1} and
later extended to all supergravities \cite{susyb1,susyb2,susyb3}, that, in the
conventions of \cite{deRoo:1985np,wageman,deroowag,wagthesis} reads as follows:
\begin{equation}
  3\,S_{AC} \, S^{CB} - \frac{1}{2} K_{i,j} \Sigma_{A}^{\phantom{A} i}
  \Sigma^{B \vert j} \, = \, -\delta_{A}^{B} \, V~,
\label{wardide}
\end{equation}
where $K_{i,j}$ is the kinetic matrix of the spin--1/2 fermions. The numerical
coefficients appearing in (\ref{wardide}) depend on the normalization of the
kinetic terms of the fermions, while $A,B,\dots = 1,\dots,\mathcal{N}$ are
$\mathrm{SU}(\mathcal{N})$ indices that enumerate the supersymmetry charges. We also
follow the standard convention that the upper or lower position of such indices
denotes definite chiral  projections of Majorana spinors, right or left
depending on the species of fermions considered\footnote{For instance, we have
$\gamma_5 \, \epsilon_A= \epsilon_A$ and $\gamma_5 \, \epsilon^A = -
\epsilon^A$.}.
The position denotes also the way of transforming of the fermion with respect
to $\mathrm{SU}(\mathcal{N})$, with lower indices in the fundamental and upper indices
in the fundamental bar. In this way we have $S^{AB} = \left(S_{AB} \right)
^\star$ and $\Sigma_A^{\phantom{A} i} =\left(\Sigma^{B \vert i} \right)
^\star$. Finally, the index $i$ is a collective index that enumerates all
spin--1/2 fermions present in the theory.
\par
In the case of $\mathcal{N}=4$ supergravity the spin--1/2 fermions  are the dilatino and
the gauginos. The dilatino $\chi^A$ belongs to the graviton multiplet, and has
a left-chiral projection in  the ${\bar {\bf 4}}$ of $\mathrm{SU}(4)$,
implying  a fermion shift:
\begin{equation}
  \delta \, \chi^A \, = \,\mbox{derivative terms} \, + \,
  \Sigma_{AB} \left( \phi \right) \,  \epsilon^B~.
\label{dilashif}
\end{equation}
The gauginos $\lambda_{A}^I$ belong to the vector multiplets and have a
left-chiral projection in the ${\bf 4}$ of $\mathrm{SU}(4)$; the index $I$ belongs to
the adjoint of the gauge group. Their fermion shifts are
\begin{equation}
  \delta \, \lambda_A^I \, = \,\mbox{derivative terms} \, + \,
  \Sigma_{A}^{\phantom{A} B\vert I} \left( \phi \right) \,  \epsilon_B~.
\label{gaushif}
\end{equation}
\par
A vacuum configuration $\phi_0$ that preserves $\mathcal{N}_0$ supersymmetries is
characterized by the existence of $\mathcal{N}_0$ vectors $\rho^A_{(\ell)}$
($\ell=1,\ldots,\mathcal{N}_0$) of $\mathrm{SU}(\mathcal{N})$, such that
\begin{eqnarray}
\label{breakpat}
S_{AB} \left( \phi_0 \right) \, \rho^A_{(\ell)} & = &
\ee^{\rm i\theta}\,
\sqrt{\ft{-V(\phi_0)}{3}} \, \rho_{A(\ell)}
~,\nonumber\\
\Sigma_A^{\phantom{A}i} \left( \phi_0 \right) \, \rho^A_{(\ell)} & = & 0~,
\end{eqnarray}
where $\theta$ is an irrelevant phase. Indeed, consider the spinor
\begin{equation}
  \epsilon ^A(x) = \sum_{\ell =1}^{\mathcal{N}_0} \, \rho^A_{(\ell)}
  \epsilon^{(\ell)}(x)~,
\label{Eadix}
\end{equation}
where $\epsilon^{(\ell)}(x)$ are $\mathcal{N}_0$ independent solutions of the equation
(\ref{susygravads4}) for covariantly constant spinors in $\mathrm{AdS}_4$ (or
Minkowski space) with  $4\, e= \sqrt{-V(\phi_0)/{3}}$.
Then it follows that under supersymmetry transformations of parameter
(\ref{Eadix}) the chosen vacuum configuration $\phi=\phi_0$ is
invariant\footnote{As already stressed, the v.e.v.s of all the fermions are
zero and equation (\ref{breakpat}) guarantees that they remain zero under
supersymmetry transformations of parameters (\ref{Eadix}).}. That such a
configuration is a true vacuum follows from another property proved, for
instance, in  \cite{susyb3}: all vacua that admit at least one vector
$\rho^A$ satisfying eq. (\ref{breakpat}) are automatically extrema  of the
potential, namely they satisfy eq. (\ref{extrepot}).
\par
An important general feature  of extended supergravities, $\mathcal{N}\ge 2$, is that
the fermion shifts and the gravitino mass--matrix are uniquely determined by
the \emph{gauging} of the theory and are proportional to the gauge group
coupling constants $g_i$.
Indeed there are very general formulae for these objects expressing them in
terms of geometrical data of the scalar manifold  and of the structure
constants of the gauge group (or of representation  matrices if, in addition to
vector multiplets, also hyper-multiplets are  present).  Hyper-multiplets are
present only for the case $\mathcal{N}=2$, whose most general  form and gauging is
discussed in \cite{bertolo} and whose partial breaking is discussed
\cite{Fre:1997js,Girardello:1997hf,Ferrara:1996gu,Zinovev:1992mw}.
\par
For $\mathcal{N}\ge 5$  hyper and vector multiplets are absent and the scalar
manifold is believed to be a uniquely fixed non--compact coset space. For
$\mathcal{N}=3,4 $ there are no hyper-multiplets and, in addition to the
graviton multiplet, there are at most vector multiplets. Also in this case,
the geometry of the scalar manifold is fixed to be  that of a  non--compact
coset space
\begin{equation}
\label{porcaloca}
\mathcal{M}_{(n)} = {G_{(n)}\over H_{(n)}}~,
\end{equation}
depending on the number $n$ of vector multiplets.  It is usually required that
the subgroup $H_{(n)}$, the maximal compact subgroup of $G_{(n)}$ is of the
form
\begin{equation}
\label{formH}
H_{(n)} = \mathrm{SU}(\mathcal{N})\times H^\prime_{(n)}~.
\end{equation}
This is due to the assumption, always made in supergravity constructions
\cite{Cremmer:1978tt,Cremmer:1977zt}, that an $\mathcal{N}$--extended locally
supersymmetric Lagrangian should have a \emph{local} $\mathrm{U}(1) \times
\mathrm{SU}(\mathcal{N})$  invariance, $\mathrm{U}(\mathcal{N})\sim
\mathrm{U}(1) \times \mathrm{SU}(\mathcal{N})$  being the automorphism group of
the supersymmetry algebra with $\mathcal{N}$  supercharges\footnote{A subtle
distinction occurs at the level of the $\mathrm{U}(1)$--factor that is missing
only in the $\sn=8$ case, where the graviton multiplet is $CPT$ self--adjoint.}.
This assumption, that leads to \emph{unique} choices of the  scalar manifolds
$\mathcal{M}_{(n)}$ manifolds and, correspondingly, to unique supergravity
Lagrangians, is motivated  by the requirement that all supercharges should be
treated on the same footing and that at least one vacuum with the full
$\mathcal{N}$--extended supersymmetry should be present in the theory one wants
to construct.
\par
The point we want to make in the present paper, which is strongly suggested by
the shadow mechanism we have discovered in Kaluza--Klein theory, is the
following.  The enforcement of $\mathrm{SU}(\mathcal{N})$ local symmetry may be
an unnecessary and too strong constraint for $\mathcal{N}$--extended local
supersymmetry. Breaking $\mathrm{SU}(\mathcal{N})$ treats the $\mathcal{N}$
supercharges on an unequal footing and prevents the existence of vacua
preserving all of them. Yet if we are looking for Lagrangians where one or more
supersymmetries are always broken in any choice of the vacuum, dismissing
$\mathrm{SU}(\mathcal{N})$ may be  the appropriate decision. Said in different
words, what we are possibly looking for is not the construction of
$\mathcal{N}$--extended supergravity,  rather it is the coupling of
$\mathcal{N}- \mathcal{N}_0$ \emph{massive} gravitino multiplets to
$\mathcal{N}_0$--extended supergravity. Since the only way to give mass to a
gravitino is via a super-Higgs mechanism, the two programmes amount to the same
thing.  Yet, in the second perspective, the need for $\mathrm{SU}(\mathcal{N})$
local symmetry disappears.  We should just be satisfied with
$\mathrm{SU}(\mathcal{N}_0)$ symmetry. We shall argue that this is the
necessary step in order to avoid the bound (\ref{boundo}) implied by the
standard $\mathrm{SU}(4)$--symmetric $\mathcal{N}=4$ theory.
\subsection{Standard $\mathrm{SU}(4)$ symmetric $\mathcal{N}=4$ supergravity}
Let us for the moment step back to the standard formulation of $\mathcal{N}=4$
supergravity and analyze its implications.
\par
The $\mathcal{N}=4$ graviton multiplet contains a scalar  and a pseudo-scalar
that can be combined into a complex field $S$ parametrizing the coset manifold
$\mathrm{SU}(1,1)/U(1)$. On the other hand, each $\mathcal{N}=4$ vector
multiplet contains $6$ scalars. In the standard coupling of $n$ vector
multiplets these $6 \times n$ scalars are identified with the coordinates of
the coset manifold $\mathrm{SO}(6,n)/\left( \mathrm{SO}(6)\times \mathrm{SO}(n)\right) $.
Hence in standard $\mathcal{N}=4$ supergravity the choice of the coset manifold
\footnote{The $\mathrm{SO}(6,n)$ symmetry  has been derived in direct
supergravity constructions using the conformal tensor calculus \cite{deroowag}
but it has also a string theory origin. Compactifying the heterotic string on
$T^6$ gives at the massless level $\mathcal{N}=4$ supergravity coupled to
$n=22$ vector multiplets.  The symmetry $\mathrm{SO}(6,22)$ arises
\cite{Narain:1986jj,Narain:1987am} as the symmetry of the lattice of momenta
and winding numbers.} (\ref{porcaloca}) is as follows
\cite{ericn4,deroowag,perret2}:
\begin{equation}
  \mathcal{M}^{(\mathcal{N}=4)}_{(n)}
  =\frac{\mathrm{SU}(1,1)}{\mathrm{U}(1)} \, \times
  \, \frac{\mathrm{SO}(6,n)}{\mathrm{SO}(6) \times \mathrm{SO}(n)}~.
\label{so6n/so6son}
\end{equation}
The $\mathrm{U}(4)=\mathrm{U}(1) \times \mathrm{SU}(4)$ symmetry arises because
of the $\mathrm{U}(1)$ denominator in the coset describing the graviton
multiplet scalar and because of the local isomorphism $\mathrm{SO}(6) \simeq
\mathrm{SU}(4)$ in the isotropy subgroup of the coset describing the matter
multiplet scalars.
\par
Let us notice, as it will be important later, that at the level of global
supersymmetry there is no difference between $\mathcal{N}=4$ and
$\mathcal{N}=3$ theories. The field content of the vector multiplets is the
same, and we always have $6 \times n$ scalars. The difference arises only at
the level of coupling to supergravity. As it was shown in
\cite{Castellani:1986ka}, if we couple the same gauge theory to the
$\mathcal{N}=3$, rather than to the $\mathcal{N}=4$, graviton multiplet the
scalars do not parametrize the coset manifold
$\mathrm{SO}(6,n)/(\mathrm{SO}(6)\times \mathrm{SO}(n))$ but have to be
interpreted as the coordinates of the manifold
\begin{equation}
 \mathcal{M}^{(\mathcal{N}=3)}_{(n)}
 =\frac{\mathrm{SU}(3,n)}{\mathrm{SU}(3) \times \mathrm{U(}1)
 \times \mathrm{SU}(n)}~,
\label{su3n/su3u1sun}
\end{equation}
which replaces (\ref{so6n/so6son}) and displays only a local $\mathrm{SU}(3)
\times U(1)$ symmetry. $\sn=3$ supergravity coupled to $n$ vector multiplets
both in un-gauged and gauged versions has been fully constructed in
\cite{Castellani:1986ka} using (\ref{su3n/su3u1sun}) as a starting point.
\par
We might follow the perspective advocated some lines above and try to further
couple to such a theory also a massive $\mathcal{N}=3$ spin--1/2 multiplet,
whose field content and $\mathrm{SU}(3)$ representation assignments have been
displayed in the right hand part of table \ref{brokengravitino}. Should such a
coupling necessarily involve an extension of the local symmetry from
$\mathrm{SU}(3)$ to $\mathrm{SU}(4)$  and of the duality symmetries from
$\mathrm{SU}(3,n)$ to $\mathrm{SO}(6,n)$?  This is what happens if we think of
such a coupling as a broken phase of  ``standard'' $\mathcal{N}=4$
supergravity. In this section we want to argue that what  enforces the bound
(\ref{boundo}) is precisely the $\mathrm{SO}(6,n)$ symmetry. This suggests the
existence of a shadow supergravity where such a symmetry is dismissed.
\paragraph{Symplectic embeddings and the de Roo and Wagemans formulation}
Let us now introduce  the formulae for the gravitino mass matrix
(\ref{gravsusy}) and for the fermion shifts (\ref{dilashif},\ref{gaushif}) as
they arise in standard $\mathcal{N}=4$ supergravity. For this we utilize the
formulation by de Roo and Wagemans \cite{deRoo:1985np,deroowag,wagthesis}, since
it is the most general one so far available. Indeed these authors have been the
only ones to realize the presence of some additional phase parameters that had
escaped notice in the other constructions (\cite{ericn4,perret2} or chapter
II.8 of \cite{castdauriafre}) and that turn out to be essential for the
partial  $\mathcal{N}=4 \to \mathcal{N}=3 $ breaking mechanism \cite{wageman}.
\par
Consider a family of  $\mathrm{SU}(1,1)$ matrices $\mathcal{S}(u_1,u_2)$
that parametrize the coset $\mathrm{SU}(1,1)/\mathrm{U}(1)$. Namely, set
\begin{equation}
  \mathcal{S}(u_i)\equiv
  \left( \matrix{ \varphi_1(u_i) & \varphi_2^*(u_i) \cr
  \varphi_2(u_i) & \varphi_1^*(u_i)\cr}\right)~,
\label{su11cos}
\end{equation}
with the constraint $|\varphi_1|^2 -|\varphi_2|^2 =1$. The choice of an
explicit coset parametrizations is conceptually irrelevant for the construction
of supergravity but there are special choices (or gauges) that can simplify
calculations in a substantial way and that we shall discuss in a moment.
\par
Similarly, let $L^\Lambda_{\phantom{\Lambda}\Gamma}(\phi_i)$,  function of $6n$
parameters $\phi_i$, be a family of $\mathrm{SO}(6,n)$  matrices parametrizing
the coset  $\mathrm{SO}(6,n)/(\mathrm{SO}(6)\times \mathrm{SO}(n))$. Thus,
\begin{equation}
\label{grouppro}
L^\Lambda_{\phantom{\Lambda}\Gamma} \,
L^\Sigma_{\phantom{\Sigma}\Delta}\, \eta_{\Lambda\Sigma} =
\eta_{\Gamma\Delta}~,
\end{equation}
where we use the Killing metric
\begin{equation}
\label{so6signa}
\eta_{\Lambda\Sigma}= \mbox{diag}(\underbrace{-,\dots,-}_6 ,
\underbrace{+,\dots,+}_n)~.
\end{equation}
Next we reassemble the coset elements $L^\Lambda_a$ ($a=1,\ldots 6$) into
\begin{equation}
  \Phi^\Lambda_{AB} = \ft 12 \, \sum_{x=1}^3 \left( L_x^{\phantom{x}\Lambda} \,
  J_{AB}^{(-)x}\, + \,{\rm i} \, L_{x+3}^{\phantom{x+3}\Lambda}\,
  J_{AB}^{(+)x}\,\right)
\label{Abdual}
\end{equation}
satisfying the reality condition
\begin{equation}
  \Phi^{\Lambda\vert{AB}} \equiv \left( \Phi^\Lambda_{AB}\right) ^\star = -
  \epsilon^{ABCD} \, \Phi^\Lambda_{CD}~.
\label{realselfcon}
\end{equation}
In eq. (\ref{Abdual}) we made use of the 't Hooft matrices $J^{(\pm)x}_{AB}$
($x=1,2,3$) which provide a complete basis for $4\times 4$ antisymmetric
matrices. They are (anti)self-dual: $J_{AB}^{(\pm)x} = \pm\epsilon_{ABCD}
J_{AB}^{(\pm)x}$;   the $J^{(+)}$ commute with the $J^{(-)}$ and both realize
the quaternion algebra.
\par
We consider the gauging of a  $6+n$--dimensional gauge group, in general
composed of several simple factors. The gravitino mass-matrix and the fermion
shifts determined by de Roo and Wagemans are then
\begin{eqnarray}
S_{AB} & = & -\frac{2}{3} \sum_{i=1}^{k} \, g_i \,
\left( e^{{\rm i} \theta_i} {\varphi}^{\star}_1 + e^{-{\rm i} \theta_i}
{\varphi}^{\star}_2 \right) \, \Phi_{AC}^{\Lambda_i} \, \Phi^{CD \vert
\Delta_i} \, \Phi_{DB}^{\Sigma_i}  \,
f_{\Lambda_i\Sigma_i}^{\phantom{\Lambda\Sigma}\Omega_i} \, \eta_{\Delta_i
\Omega_i}
\nonumber\\
\Sigma_{AB}&=& -\frac{4}{3} \sum_{i=1}^{k} \, g_i \, \left( e^{-{\rm i}
\theta_i} {\varphi}_1 - e^{{\rm i} \theta_i} {\varphi}_2 \right) \,
\Phi_{AC}^{\Lambda_i} \, \Phi^{CD \vert \Delta_i} \, \Phi_{DB}^{\Sigma_i}  \,
f_{\Lambda_i\Sigma_i}^{\phantom{\Lambda\Sigma}\Omega_i} \, \eta_{\Delta_i
\Omega_i}
\nonumber\\
\Sigma_{A}^{B\vert \Lambda}&=& -2  \sum_{i=1}^{k} \, g_i \, \left( e^{-{\rm i}
\theta_i} {\varphi}_1 - e^{{\rm i} \theta_i} {\varphi}_2 \right)\, \left( \,
f_{\Lambda_i\Sigma_i}^{\phantom{\Lambda\Sigma} \Lambda} \,
\Phi_{AC}^{\Lambda_i}\,  \Phi^{CB \vert \Sigma_i} \, +
\right.
\nonumber\\
& & \left. \qquad \quad \qquad \eta_{\Delta_i\Omega_i}\,
f_{\Lambda_i\Sigma_i}^{\phantom{\Lambda\Sigma}\Delta_i}\,\Phi_{CD}^{\Lambda}\,
\Phi^{CD \vert \Omega_i}\, \Phi_{AE}^{\Lambda_i}\, \Phi^{EB \vert \Sigma_i}\,
\right)~,
\label{wagegaushi}
\end{eqnarray}
where $g_i$ are the coupling constants associated to each simple factor in the
full gauge group and the constant phases $\theta_i$ are additional parameters,
also in one--to--one correspondence with such simple factors.
\subsubsection{Solvable Lie algebra parametrization of the scalar manifold}
\label{modsolv}
The choice of a convenient parametrization of the scalar manifold
might simplify computations and help to
characterize the geometrical meaning of the various fields. The so-called
solvable Lie algebra parametrization has proven useful in many questions
concerning extended supergravities
\cite{Andrianopoli:1997bq,Andrianopoli:1997zg,Andrianopoli:1998wi}, and we
shall apply it also to the scalar manifold of standard
$\mathcal{N}=4$ supergravity (\ref{so6n/so6son}), as this will prove useful
in the following.
\par
In the solvable Lie algebra approach, the non-compact scalar manifold $G/H$ is
obtained exponentiating    an appropriate solvable subalgebra
$\mathrm{Solv}(G/H)$ of the isometry algebra $\mathbb{G}$ (for a review see
\cite{Trigiante:1998vu}):
\begin{equation}
  \frac{G}{H}\simeq \exp\left(\mathrm{Solv}\left(\frac{G}{H}\right)\right)~.
\label{expsolv}
\end{equation}
The solvable Lie algebra $\mathrm{Solv}(G/H)$ is spanned by the non compact Cartan
generators $\mathcal{H}_i$ of $\mathbb{G}$ plus all the step operators
$E^\alpha$ associated with roots $\alpha$ that are not orthogonal to all the
non compact Cartan generators. This approach  has the distinctive advantage of
giving a unique group theoretical characterization to all the scalar fields
$\phi$ of supergravity and provides simple polynomial parametrizations of the
coset representatives $L(\phi) \in G$.
\par
Let us now discuss the solvable parametrization of the manifold
(\ref{so6n/so6son}).
\paragraph{The solvable Lie algebra of the coset ${\bf SU(1,1)/U(1)}$}
\label{su11cospar}
We write the three generators of the $\mathrm{sl}(2,\mathbb{R})$ Lie
algebra in the form:
\begin{equation}
l_0 = \left( \matrix{ {\frac{1}{2}} & 0 \cr 0 &
  -{\frac{1}{2}} \cr  }  \right)~,\hskip 0.2cm
l_+  =  \left( \matrix{ 0 & 1 \cr 0 & 0 \cr  }  \right)~,\hskip 0.2cm
l_{-} = \left( \matrix{ 0 & 0 \cr 1 & 0 \cr  } \right)~,
\label{l01m1}
\end{equation}
leading to the standard form of the commutation relations:
\begin{equation}
  \left [ l_0 , l_\pm \right ] = \pm \, l_\pm \quad , \quad \left [
  l_+ \, l_- \right ] = 2 \, l_0~.
\label{commusl2}
\end{equation}
The solvable Lie algebra generating the coset
$\mathrm{SL}(2,\mathbb{R})/\mathrm{O}(2)$ can be taken to be spanned by the two
non compact generators $l_0, l_-$ and we can write the coset element in the
form:
\begin{equation}
  \Lambda(\alpha,\beta)=\exp(\alpha \, l_0) \, \exp( \beta\, l_-) \, =
  \, \left(
  \matrix{ {\ee^{{\frac{\alpha}{2}}}} & 0 \cr
    \beta \,  \ee^{- \frac{\alpha}{2}}  &
  {\ee^{{\frac{-\alpha}{2}}}} \cr  }  \right)~.
\label{Lamcoset}
\end{equation}
We can convert an $\mathrm{SL}(2,\mathbb{R})$ matrix $\Lambda$ into an
$\mathrm{SU}(1,1)$ matrix $\mathcal{S}=\mathcal{C}\Lambda\mathcal{C}^{-1}$ by
means of a Cayley transformation:
\begin{equation}
  \mathcal{C}= \frac{1}{\sqrt{2}} \,\left(  \matrix{
    1 & -{\rm i} \cr
   1 & {\rm i} \cr
  } \right)~,
\label{cayley}
\end{equation}
and we obtain the solvable parametrization of the coset representative
(\ref{su11cos})
\begin{equation}
  \mathcal{S}
  \equiv \left( \matrix{ \varphi_1 & \varphi_2^* \cr
  \varphi_2 & \varphi_1^*\cr}\right)
  = \left(  \matrix{
  \cosh\frac{\alpha}{2} - \frac{\ii}{2} \, \beta \, \ee^{-\alpha/2} &
  \sinh\frac{\alpha}{2} - \frac{\ii}{2}\,\beta \,\ee^{-\alpha/2} \cr
  \sinh\frac{\alpha}{2} + \frac{\ii}{2}\,\beta\,\ee^{-\alpha/2} &
  \cosh\frac{\alpha}{2} + \frac{\ii}{2}\,\beta\,\ee^{-\alpha/2}}\right)
\label{su11cos2}
\end{equation}
that enters the potential and fermion shift formulae of standard $\mathcal{N}=4$
supergravity (\ref{wagegaushi}).
\paragraph{The solvable Lie algebra of the coset ${\bf SO(6,3)/SO(6)
\times SO(3)}$}
\label{so6ncos}
We should now consider the second factor in the scalar manifold
(\ref{so6n/so6son}), namely $\mathrm{SO}(6,n)$ modded by the action of the subgroup
$ \mathrm{SO}(6) \times \mathrm{SO}(n)$.
The rank (i.e., the number of non--compact generators)
of this non--compact coset is $3$ and its
dimension is $6n$. This means that the corresponding solvable Lie
algebra is spanned by $3$ Cartan generators and $6n -3$ step operators
$E^\alpha$. Anticipating a result that will be discussed in the next section,
since our main interest is in the partial breaking mechanism from
$\mathcal{N}=4$ to $\mathcal{N}=3$ we focus on the case $n=3$
which  suited to describe the minimal gauging that triggers such a breaking.
Most of the following formulae can in any case immediately be generalized
to any $n$.
\par
A generic element $\mathcal{A}$ of the Lie algebra $\mathrm{so}(6,3)$, using the
Killing metric (\ref{so6signa}),
has the following $3 \times 3$ block structure:
\begin{equation}
  \mathcal{A} \, = \, \left(
 \begin{array}{c|c||c}
 A_3 &  -B_3^T & B_2^T \\
 \hline
 B_3 &   A_2 & B_1^T \\
 \hline\hline
 B_2 & -B_1 & A_1
 \end{array} \right)
\label{so36genele}
\end{equation}
where $A_i = -A_i^T$ ($i=1,2,3$) span an $\mathrm{so}(3)_1 \times
\mathrm{so}(3)_2 \times \mathrm{so}(3)_3$ subalgebra
of $\mathrm{so}(6,3)$,
while the $B_i$ are generic. The orthogonal  splitting
$\mathbb{G}=\mathbb{H}\oplus\mathbb{K}$ of this algebra with respect to the
subalgebra  $\mathbb{H} \equiv \mathrm{SO}(3) \times \mathrm{SO}(6)$ is as
follows:
\begin{equation}
\mathbb{H} \ni
\left(
 \begin{array}{c|c||c}
 A_3 &  -B_3^T & 0 \\
 \hline
 B_3 &   A_2 & 0 \\
 \hline\hline
 0 & 0 & A_1
 \end{array} \right)~,\hskip 0.4cm
 \mathbb{K}  \ni
 \left(
 \begin{array}{c|c||c}
 0 &  0 & B_2^T \\
 \hline
 0 &   0 & B_1^T \\
 \hline\hline
 B_2 & -B_1 & 0
 \end{array} \right)~.
\label{ortodeco}
\end{equation}
The solvable presentation of the coset is
obtained by choosing coset representatives $L=\exp(\mathrm{Solv}(G/H))$,
instead of the choice $L_{\rm orth} = \exp(\mathbb{K})$ leading to the
standard off-diagonal parametrization.
The structure of the solvable Lie algebra is the following.
\par
The
three-dimensional non-compact Cartan subalgebra contains the matrices of the
form
\begin{equation}
\left(
 \begin{array}{c|c||c}
 0 &  0 & 0 \\
 \hline
 0 &   0 & H \\
 \hline\hline
 0 & H & 0
 \end{array} \right)~,
\label{cartano}
\end{equation}
where $H=\mathrm{diag}(h_1,h_2,h_3)$.
\par
The remaining $15$ dimensional space,
spanned by the step operators associated with positive roots that are not
orthogonal to the non--compact Cartan subalgebra,
is  given by the matrices either of the form
\begin{equation}
\label{tuttaalg}
\left(
 \begin{array}{c|c||c}
 0 &  -E^T & E^T \\
 \hline
 E &  0 & 0 \\
 \hline\hline
 E & 0 & 0
 \end{array} \right)~,
\end{equation}
with $E$ a generic $3\times 3$ matrix (this gives 9 parameters), or of the form
\begin{equation}
\label{tuttaalg2}
\left(
 \begin{array}{c|c||c}
 0 & 0 & 0 \\
 \hline
 0 &  U - U^T & U^T + D^T \\
 \hline\hline
 0 & U + D & D^T - D
 \end{array} \right)~,
\end{equation}
where $U$ is upper triangular and $D$ lower-triangular, accounting for 6
parameters.
\par
The entries of the matrices $H,E,U,D$ can be used as the coordinates of the
18-dimensional space containing the scalar fields of the 3 vector multiplets
that, as we shall see, participate to the $\mathcal{N}=4\to\mathcal{N}=3$ supersymmetry
breaking.
\subsection{Partial breaking to $\mathcal{N}=3$}
\label{partialbreaking}
If we consider the  $\mathcal{N}=3$ massive spin--3/2 multiplet of eq.
(\ref{brokengravitino}), it is easy to be convinced that the minimal number  of
vector multiplets one has to couple to $\mathcal{N}=4$ supergravity in order to produce
a spontaneous breaking from $\mathcal{N}=4$ to $\mathcal{N}=3$ is three.
\par
Indeed for each massive gravitino we have $6$ massive vectors and $14$ massive
scalars. In the super-Higgs mechanism, the $6$ vectors become massive by eating
up as many scalars: this means that, including the $3$ graviphotons that remain
massless in $\mathcal{N}=3$ supergravity, we need $20$ scalars and $9$ vectors to begin
with. This counting is in agreement with the coupling of $3$ vector multiplets
of $\mathcal{N}=4$ supersymmetry: the $9$ vectors are the $6$ graviphotons of   $\mathcal{N}=4$
supergravity plus the $3$ matter photons. The $20$ scalars are the $6 \times 3$
matter scalars plus the $2$ scalars contained in the $\mathcal{N}=4$ graviton
multiplet. The counting agrees also at the fermion level. Before breaking we
have $4$ dilatinos plus $4 \times 3=12$ gauginos, a total of $16$ spin--1/2
fermions. After breaking we remain with $1$ dilatino sitting in the $\mathcal{N}=3$
graviton multiplet and $14$ spinors sitting in the massive gravitino multiplet,
a total of $15$. The missing fermion is the \emph{goldstino} which has been
eaten up by the gravitino to become massive.
\par
A more detailed analysis that includes also the isospin representation
assignments displayed in table \ref{brokengravitino} gives additional
information. In order to realize the partial breaking $\mathcal{N}=4 \to \mathcal{N}=3$ in
AdS$_4$, the three $\mathcal{N}=4$ vector multiplets plus the $6$ graviphotons must be
used to gauge a minimal group:
\begin{equation}
  \mathcal{G}_{\rm min} =
  \mathrm{SO}(4) \times \mathrm{SO}(3) = \mathrm{SO}(3)_1 \times
  \mathrm{SO}(3)_{2} \times \mathrm{SO}(3)_{3}~.
\label{minimgaug}
\end{equation}
In AdS$_4$ the relevant supersymmetry algebra is
$\mathrm{Osp}(\mathcal{N}\vert 4)$ and the gravitinos are assigned to the vector
$\mathcal{N}$--dimensional representation of the R--symmetry
subalgebra $\mathrm{SO}(\mathcal{N})\subset\mathrm{Osp}(\mathcal{N}\vert 4)$.
With respect to the  $\mathrm{SU}(\mathcal{N})$ local
symmetry group of the supergravity Lagrangian,  the embedding of
$\mathrm{SO}(\mathcal{N})$ is the maximal one described by selecting those unitary,
uni-modular $\mathcal{N} \times \mathcal{N}$ matrices that are also real and hence orthogonal.
The partial breaking  $\mathcal{N}=4 \to \mathcal{N}=3$ is algebraically described by the
embedding of $\mathrm{Osp}(3\vert 4)$ into $\mathrm{Osp}(4 \vert 4)$:  in
particular the $R$--symmetry subgroup of $\mathrm{SO}(3)\subset
\mathrm{Osp}(3\vert 4)$ is the subgroup of  $\mathrm{SO}(4) \subset
\mathrm{Osp}(4\vert 4)$ such that the vector representation decomposes as
${\bf 4}\to {\bf 3} \oplus {\bf 1}$, the singlet being the massive gravitino,
the triplet containing the massless ones.  When we look at $\mathrm{SO}(4)$ as
$\mathrm{SO}(3)_1 \times \mathrm{SO}(3)_2$,  this embedding
corresponds to choosing
$ \mathrm{SO}(3)_R = \mbox{diag}(\mathrm{SO}(3)_1 \times \mathrm{SO}(3)_2)$.
\par
On the other hand, a look at table \ref{brokengravitino} shows that also the
massive vectors sitting in the massive gravitino multiplet are in the triplet
of the $R$--symmetry group $\mathrm{SO}(3)_R$. This implies that the three
matter multiplets coupled to $\mathcal{N}=4$ supergravity must gauge a third
$\mathrm{SO}(3)$ group and that the residual unbroken $R$--symmetry group
must be the diagonal subgroup of the three:
\begin{equation}
  \mathrm{SO}(3)_R = \mbox{diag} \,
  \left( \mathrm{SO}(3)_1 \times \mathrm{SO}(3)_2  \times \mathrm{SO}(3)_3
  \right)~.
\label{O3diag2}
\end{equation}
\par
So this argument uniquely selects the model that in standard $\mathcal{N}=4$
supergravity can realize the spontaneous breaking in AdS$_4$ to an $\mathcal{N}=3$
theory with a residual $\mathcal{G}^\prime$ unbroken gauge symmetry  and one
massive gravitino multiplet. It is the gauging of the  group
\begin{equation}
  \mathcal{G} =  \mathrm{SO}(3)_1 \times \mathrm{SO}(3)_2  \times \mathrm{SO}(3)_3
   \times \mathcal{G}^\prime~.
\label{fullgauge}
\end{equation}
\subsubsection{Moduli space of ${\bf \mathcal{N}=3}$ vacua}
Among the vector multiplet scalars,
only those of the 3 multiplets gauging the
$\mathrm{SO}(3)_3$ group in (\ref{fullgauge}) can actually be relevant for the
partial breaking mechanism. These 18 fields span the coset
$\mathrm{SO}(6,3)/(\mathrm{SO}(6)\times\mathrm{SO}(3))$, whose parametrization
we already discussed in sec. \ref{so6ncos}.  The residual unbroken gauge group
$\mathcal{G}^\prime$ is a mere spectator in this game, and we can forget about
the associated scalar fields.
\par
On the other hand, recalling table \ref{brokengravitino}, we know that, after
partial breaking, the massive gravitino multiplet contains  a total of $4$
scalar (or pseudo-scalar) fields that are singlet under the $\mathrm{SO}(3)_R$
R--symmetry group of $\mathcal{N}=3$ supergravity. These are the only fields
that can develop non vanishing v.e.vs in an $\mathcal{N}=3$ vacuum and
constitute the moduli space $\mathcal{M}^{\rm p.b.}$ we are looking for, and
that was anticipated in eq. (\ref{modspace}).
\par
Two of these fields are, of course,  the scalars of the $\mathcal{N}=4$
graviton multiplet, spanning the $\mathrm{SU}(1,1)/\mathrm{U}(1)$ coset space
that was already discussed. The other two moduli are the
$\mathrm{so}(3)_R$-singlets among the 18 vector-multiplet scalars. They belong
to the sub-manifold of $\mathrm{SO}(6,3)/(\mathrm{SO}(6) \times
\mathrm{SO}(3))$ spanned by the solvable Lie algebra generators that commute
with the diagonal subgroup (\ref{O3diag2}). Referring to eq.
(\ref{so36genele}), we see that the matrices of the $\mathrm{so}(3)_R$
subalgebra are obtained setting  $A_1=A_2=A_3=A$ and $B_1=B_2=B_3=0$. The
normalizer of this subalgebra in $\mathrm{so}(3,6)$ is an
$\mathrm{sl}(2,\mathbb{R})$ Lie algebra spanned by the following three
matrices:
\begin{equation}
  l_0=  \left(
 \begin{array}{c|c||c}
 \mathbf{0} &   \mathbf{0} & \mathbf{0} \\
 \hline
 \mathbf{0} &   \mathbf{0} & \mathbf{1} \\
 \hline\hline
 \mathbf{0} &   \mathbf{1} & \mathbf{0}
 \end{array} \right)~,~~
 l_- = \left(
 \begin{array}{c|c||c}
 \mathbf{0} &  \mathbf{1} & \mathbf{1} \\
 \hline
 \mathbf{-1} &  \mathbf{0} & -\mathbf{0} \\
 \hline\hline
 \mathbf{1} &  \mathbf{0} & \mathbf{0} \
 \end{array} \right)~,~~
l_+ = \left(
 \begin{array}{c|c||c}
 \mathbf{0} &  \mathbf{-1} & \mathbf{1} \\
 \hline
 \mathbf{1} &  \mathbf{0} & \mathbf{0} \\
 \hline\hline
 \mathbf{1} &  -\mathbf{0} & \mathbf{0}
 \end{array} \right)~,
\label{solv2}
\end{equation}
that satisfy the  $\mathrm{sl}(2,\mathbb{R})$ commutation relations in the
standard form (\ref{commusl2}). If we exponentiate the solvable Lie algebra
spanned by $l_0$ and $l_-$ we generate a coset, locally isomorphic to
$\mathrm{SL}(2,\mathbb{R})/\mathrm{O}(2)$, embedded into
$\mathrm{SO}(6,3)/(\mathrm{SO}(6) \times \mathrm{SO}(3))$, whose coordinates
are $R$--symmetry singlets and can develop non vanishing v.e.vs in the
$\mathcal{N}=3$ vacua. This coset is the second factor,
$\mathrm{SO}(2,1)/\mathrm{SO}(2)$, in the  moduli space $\mathcal{M}^{\rm
p.b.}$ of $\mathcal{N}=3$ vacua, eq. (\ref{modspace}).
\par
According to this analysis, we can introduce an $\mathrm{SO}(6,3)$ coset
representative $\hat L$ defined, in a similar way to eq.(\ref{Lamcoset}),
by
\begin{equation}
\label{cosdef1}
L =\exp(a \, l_0) \, \exp(b \, l_-)~.
\end{equation}
To better characterize the coset, it is however convenient
to perform a change of basis $\mathcal{B}$ such as to bring the Killing form
$\eta$ of eq. (\ref{so6signa}) into
\begin{equation}
\label{neweta}
\eta^\prime = \mathcal{B}\eta\mathcal{B}^{-1}=
\mathrm{diag}(+,-,-,+,-,-,+,-,-)~.
\end{equation}
The group $\mathrm{SO}(3,6)$ admits a $\mathrm{SO}(1,2)^3$ subgroup whose
matrices are block-diagonal in the rotated basis. The
$\mathrm{sl}(2,\mathbb{R})\sim\mathrm{so}(2,1)$ algebra of eq. (\ref{solv2})
is the Lie algebra of the diagonal subgroup
\begin{equation}
\mbox{diag} \, \left( \,\mathrm{SO}(1,2)_1 \times \mathrm{SO}(1,2)_{2} \times
  \mathrm{SO}(1,2)_{3}\right)~.
\label{diagso21}
\end{equation}
The coset representative (\ref{cosdef1}) in the new basis becomes thus
block-diagonal:
\begin{equation}
  L^\prime = \mathcal{B} \, L \, \mathcal{B}^{-1}=
  \mathrm{diag}(\mathcal{L},\mathcal{L},\mathcal{L})~.
\end{equation}
It is made of three copies of the same $\mathrm{SO}(1,2)$
matrix
\begin{equation}
  \mathcal{L}=\left( \matrix{ {\frac{1 + {b ^2} + {\ee^{2\, a}}}
    {2\,{\ee^a}}} &
  {\frac{-1 + {b ^2} + {\ee^{2\, a}}}
    {2\,{\ee^a}}} & {\frac{b}{{\ee^a}}} \cr
  {\frac{-1 - {b ^2} + {\ee^{2\, a}}}
    {2\,{\ee^a}}} &
  {\frac{1 - {b ^2} + {\ee^{2\, a}}}{2\,{\ee^a}}}
   & -{\frac{b}{{\ee^a}}} \cr b & b & 1
   \cr  }\right)~.
\label{so21coset}
\end{equation}
The above $\mathrm{SO}(1,2)$  matrix is quadratic in the entries $\ee^{\pm a/2}$ and
$b\ee^{-a/2}$ that would appear in a $\mathrm{SL}(2,\mathbb{R})$ matrix
constructed analogously to (\ref{Lamcoset}). This is quite natural since the $2
\times 2$ $\mathrm{SL}(2,\mathbb{R})$ matrices  provide the spinor
representation of $\mathrm{SO}(1,2)$, while the  $3 \times 3$ matrix
(\ref{so21coset}) is in the vector representation.  This shows that in the
matter sector the  coset generated by eq. (\ref{cosdef1}) is better interpreted
as the coset $\mathrm{SO}(1,2)/\mathrm{SO}(2)$.  \par This also suggests the
use of a  different parametrization that somehow simplifies the expression of
the coset representative. Indeed the $\mathrm{SO}(1,2)/\mathrm{SO}(2)$ coset is
an instance of an off--diagonal non--compact coset for which we can use a
standard projective parametrization. A coset representative in this
parametrization reads
\begin{equation}
\label{projpar}
\mathcal{L}^{\rm (proj)}=
   \frac{1}{\sqrt{1-v^2}}\,
   \left(
   \begin{array}{c|c}
     1 & \mathbf{v}^T   \\
    \hline
    \mathbf{ v} &  \sqrt{1-v^2} \, {\bf
    1}\, + \,  \frac{\mathbf{v} \mathbf{v}^T}{v^2}\,
    \left(1 -\sqrt{1-v^2} \,\right) \,
\end{array}\right)~,
\end{equation}
where the squared norm $v^2\equiv \mathbf{v}^T\mathbf{v}$ of the vector
$\mathbf{v}=(v_1,v_2)$ is bounded: $v^2<1$. The relation between the projective
coordinates $v_1,v_2$ and the coordinates $a,b$ of the solvable parametrization
is found to be the following:
\begin{equation}
  \ee^a = \pm \frac{1+v_1}{\sqrt{1-v_1^2 -v_2^2} }~,~~
  b=\pm \frac{v_2}{\sqrt{1-v_1^2 -v_2^2} }~,
\label{soluti}
\end{equation}
where either both plus or both minus signs are taken. The sign ambiguity
corresponds just to the usual double covering of the spinor representation.
\subsubsection{Discussion of $\mathcal{N}=3$ vacua and moduli dependence
of $E_0$}
Our final goal is to discuss the manifold of $\mathcal{N}=3$ vacua of standard
$\mathcal{N}=4$ supergravity and to derive the moduli dependence of the scale
dimension $E_0$, see eq. (\ref{restric}). What we have to do is just
to derive the dependence of the fermion shifts and mass matrices on
the $4$ scalar fields that are R--symmetry singlets and impose the
conditions for $\mathcal{N}=3$ unbroken supersymmetry. This can in principle
single out a sub-manifold of the moduli space $\mathcal{M}^{\rm
p.b.}$ or impose restrictions on the additional parameters $g_i$
and $\theta_i$ appearing in eq. (\ref{wagegaushi}). We will then
investigate the dependence of $E_0$ on the fields and
parameters that remain free.
\par
Hence, according to the discussion above,  we introduce a
new coset representative for $\mathrm{SO}(3,6)$ which is generated as follows:
\begin{equation}
  L^{\rm (proj)}=\mathcal{B}^{-1} \,
  \mathrm{diag}\left(\mathcal{L}^{\rm (proj)},\mathcal{L}^{\rm (proj)},
  \mathcal{L}^{\rm (proj)}\right)\,\mathcal{B}~.
\label{nuovcoss}
\end{equation}
The next point in the explicit search of $\sn=3$ vacua involves the
calculation of three complex quantities, introduced by Wagemans and de Roo
\cite{wageman,deroowag} and
named $Z_1,Z_2,Z_3$.
They
have the following definition in terms of the
matrix elements of the coset representative:
\begin{eqnarray}
  Z_1 &=& \mathcal{L}^{\rm (proj)}_{4,1}+{\rm i} \,
   \mathcal{L}^{\rm (proj)}_{7,1}~,\nonumber\\
  Z_2 &=&  \mathcal{L}^{\rm (proj)}_{4,4}+{\rm i} \,
   \mathcal{L}^{\rm (proj)}_{7,4}~\nonumber\\
 Z_3 &=&  \mathcal{L}^{\rm (proj)}_{4,7}+{\rm i} \,
   \mathcal{L}^{\rm (proj)}_{7,7}~.
\label{z1z2z3}
\end{eqnarray}
A simple explicit expression of these quantities is obtained introducing   a
polar parametrization of the vector $\mathbf{v}$  of the projective
presentation (\ref{projpar}). As noticed after eq. (\ref{projpar}), the vector
$\mathbf{v}$ ranges in the Poincar\`e disk  $\Delta\approx
\mathrm{SO}(1,2)/\mathrm{SO}(2)$. Hence we set
\begin{equation}
  v_1 + {\rm i} v_2 = \rho \, \ee^{{\rm i} \theta}
\label{polarpar}
\end{equation}
and we obtain:
\begin{eqnarray}
Z_1 & = & \frac{\rho}{\sqrt{1-\rho^2}} \, \ee^{{\rm i} \theta}~, \nonumber\\
Z_2 & = & \frac{1}{2}\left[ \left( 1+\frac{1}{\sqrt{1-\rho^2}}\right) -
 \left( 1-\frac{1}{\sqrt{1-\rho^2}}\right)\, \ee^{2{\rm i} \theta}
 \right ]~,\nonumber\\
Z_3 & = & \frac{\ii}{2}\left[ \left( 1+\frac{1}{\sqrt{1-\rho^2}}\right)
+
 \left( 1-\frac{1}{\sqrt{1-\rho^2}}\right)\, \ee^{{\rm i} \theta}
 \right ]~.
\label{polarze}
\end{eqnarray}
As we see, in the origin of the coset manifold, at $w=0$ we have
$Z_1=0, Z_2=1, Z_3={\rm i}$.
\par
As previously discussed, see eq. (\ref{breakpat}), to preserve $\mathcal{N}=3$
unbroken supersymmetries it is necessary to fix three out of the four
eigenvalues of the fermion shift matrices.
As shown by de Roo and Wagemans \cite{wageman}, these conditions are
satisfied when the three following quantities are equal:
\begin{equation}
g_1Z_1\left( \ee^{{\rm i} \theta_1} {\varphi}^\star_1+ \ee^{-{\rm i} \theta_1}
{\varphi}^\star_2 \right)=
g_2Z_2\left( \ee^{{\rm i} \theta_2} {\varphi}^\star_1+ \ee^{-{\rm i} \theta_2}
{\varphi}^\star_2 \right)=
g_3Z_3\left( \ee^{{\rm i} \theta_3} {\varphi}^\star_1+ \ee^{-{\rm i} \theta_3}
{\varphi}^\star_2 \right)\equiv\Theta~.
\label{N=3condition}
\end{equation}
In this case, the mass of the three gravitinos gauging the unbroken
supersymmetries is
\begin{equation}
  M_{(3)}=\vert\Theta\vert~,
\label{M3}
\end{equation}
while the mass of the ``broken gravitino'' is:
\begin{equation}
  M_{(1)}=\left|\Theta\right|
  \sqrt{\frac{8+\rho^4-\rho^2 (8+\rho^2)\cos 4 \theta}
{8+\rho^4-8\rho^2-\rho^4\cos 4 \theta}}~.
\label{M1}
\end{equation}
The ratio of the two masses does not depend on the vev.s of the
scalars in the graviton multiplet nor on the three free parameters $\theta_i$,
but only on the variables $\rho$ and $\theta$ that parametrize the matter coset.
It is straightforward to show that the ratio is bounded by three and
that this particular value is never reached in the bulk of the moduli space.
On the contrary, three is the limit of the mass ratio at the boundary of the
disk.
\begin{figure}
  \centering
  \input{epsf}
  \epsfxsize=9cm
  \epsffile{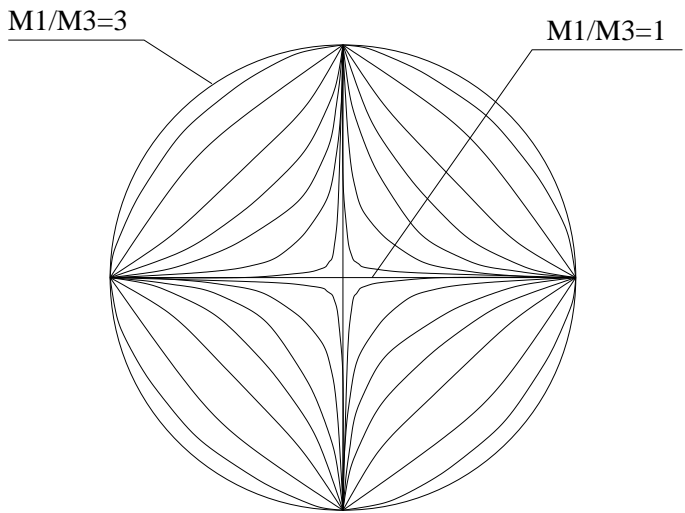}
  \caption{Projection of the moduli space
   $\mathcal{M}^{\rm p.b.}$ onto the disk $\Delta\approx
  \mathrm{SO}(1,2)/\mathrm{SO}(2)$.
  The lines of constant ratio $M_{(1)}/M_{(3)}$ are shown.
   }\label{thedisk}
\end{figure}
Figure \ref{thedisk} represents the projection of the moduli space on the
sub-factor $\Delta$.
Each point of the disk, namely each possible couple of vev.s for the matter
scalars, determines through eq. (\ref{N=3condition}) a specific set of
values for the remaining vev.s and parameters, that preserve ${\cal N}=3$
unbroken supersymmetries.
The lines of fixed ratio $M_{(1)}/M_{(3)}$ are shown.
\begin{figure}
  \centering
  \input{epsf}
  \epsfxsize=7cm
  \epsffile{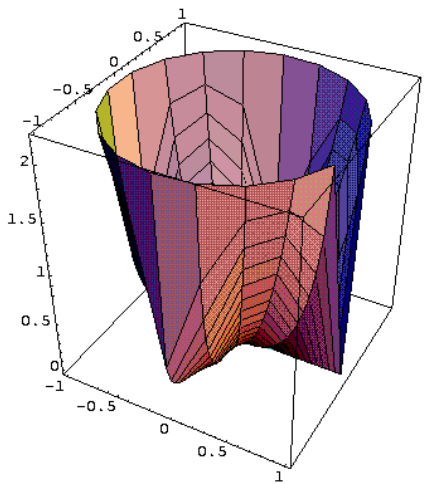}
  \caption{
  A three dimensional plot of the ratio $M_{(1)}/M_{(3)}$ over the disk $\Delta\approx
  \mathrm{SO}(1,2)/\mathrm{SO}(2)$
   }\label{thecup}
\end{figure}
Figure \ref{thecup} gives a three dimensional view of the same plot.
It is now interesting to remark the meaning of the ratio $M_{(1)}/M_{(3)}$.
Indeed for the normalization conventions of de Roo and Wagemans adopted
so far, the mass matrix eigenvalues $M$ of the gravitinos linearly
depend on their Casimir energy $E_{(3/2)}$ (see eq. \ref{massenergy}):
\begin{equation}
  M=\left|m_{(3/2)}+4\right|e=(4E_{(3/2)}-6)\, e\,.
\end{equation}
Now, the energy of the ``massless'' (i.e. $m_{(3/2)}=0$) gravitinos is
fixed (in units of the Freund Rubin scale, $e$):
$E^{\rm massless}_{(3/2)}=5/2\ \Rightarrow\ M_{(3)}=4$.
Hence the ratio
\begin{equation}
  \frac{M_{(1)}}{M_{(3)}}=\frac{4E_{(3/2)}-6}{4}=E_{(3/2)}-\frac{3}{2}=E_0
\label{M1toM3ratio}
\end{equation}
exactly gives the Clifford vacuum energy of the massive ``unbroken gravitino''
multiplet. The most interesting thing to note about this point, is that the
boundary value $E_0=3$ is exactly the energy of the shadow multiplet of the
$\n010$ compactification. This means that the dimensional reduction of $D=11$
supergravity on this background does not belong to the family of $\mathcal{N}=4$
four-dimensional supergravities known in the literature.
\subsection{Comparison with Kaluza Klein theory}
\label{compkk}
To be more precise about this last statement, we want now to derive the
effective four dimensional Lagrangian from the eleven dimensional supergravity
theory compactified on the $\n010$ coset space, taking into account (some of)
the modes belonging to the ``shadow'' massive gravitino multiplet. As already
said, we expect such a truncation to be a consistent one. Moreover, since the
only consistent way to give mass to a gravitino is via a super-Higgs mechanism,
we expect that the effective 4D theory actually be a $\mathcal{N}=4$
supergravity, of which the $\n010$ solution represent a $\mathcal{N}=3$
extremum.
\par
In order to make a complete comparison between Kaluza Klein theory  and the
standard broken  $\mathcal{N}=4$ supergravity, we also need the form in this
latter of the kinetic terms of the scalar fields parametrizing the moduli space
$\mathcal{M}^{\rm p.b.}$ of eq. (\ref{modspace}). The two factors in  eq.
(\ref{modspace}) both correspond to a copy of the upper half plane. The
$\sigma$-model metric for the scalars in each factor will therefore be the
Poincar\'e metric.
\par
In the solvable parametrization we adopt, this is seen as follows. Let us focus
on one of the factors, e.g. $\mathrm{SU}(1,1)/\mathrm{U}(1)$, containing the
scalars of the graviton multiplet; the formulae for the other factor will be
analogue. In terms of the coset representative (\ref{Lamcoset})
we calculate the left-invariant 1--form
\begin{equation}
 \Omega= \Lambda^{-1}d\Lambda=\left( \matrix{ {\frac{ {1}}{2}} d\alpha & 0
   \cr -  \beta\, {d\alpha}
       +  {d\beta}  &
  -{\frac{ {1}}{2}}d\alpha \cr  } \right)
\label{leftinv}
\end{equation}
and then we calculate the vielbein projecting onto the normalized non--compact
generators that span the subspace $\mathbf{k}$ orthogonal  to the compact
subalgebra $\mathbf{h}$. We obtain the zweibein 1--form
corresponding to the metric
\begin{equation}
  ds^2 =
{\frac{{{ {d\alpha}}^2} + {{\left( -\left( \beta\,
              {d\alpha} \right)  +
           {d\beta} \right) }^2}}{2}}~.
\label{2metric}
\end{equation}
The metric (\ref{2metric}) is exactly in the form
of the standard Poincar\'e metric
\begin{equation}
  ds^2 = \frac{dS \, d{\bar S}}{(\mbox{Im} S)^2}
\label{verpoin}
\end{equation}
on the upper half plane, upon setting
\begin{equation}
  S=
 \left( -{\rm i} \ee^{-\alpha} + \beta \, \ee^{-\alpha} \right) ^{-1}~.
\label{poincar}
\end{equation}
Looking also at the kinetic terms of
the vector fields in Wagemans and de Roo action
\cite{deRoo:1985np,wagthesis}, one can see that $S$ is exactly the
\emph{dilaton} field, namely it appears as a generalized field
dependent coupling constant in the standard way. By this we mean that
with the identification (\ref{poincar})
the kinetic matrix $\mathcal{N}_{\Lambda\Sigma}$ of the vector fields
has the standard form
\begin{equation}
  \mathcal{N}=\mbox{Re}S \, \eta + {\rm i} \, \mbox{Im}S \, \eta \, L
  \, L^T \, \eta
\label{senform}
\end{equation}
obtained through the Gaillard Zumino formula and through the
standard embedding of $SL(2,R) \times SO(6,n)$ into $Sp(2(n+6),R)$
as described in  lectures \cite{mylecture} (see eq. (105) of  that
paper).
\par
The metric for the R--symmetry neutral vector multiplet scalars that fill the
$\mathrm{SO}(2,1)/\mathrm{O}(2)$ factor in the moduli space $\mathcal{M}^{\rm
p.b.}$ is another copy of the Poincar\'e metric (\ref{2metric})
with $(a,b)$ replacing $(\alpha,\beta)$.
%
\subsubsection{Effective 4-dimensional theory for $\n010$ compactification}
\label{effn010}
To derive the effective theory for the $\n010$ compactification, we consider a
minimal set-up consisting of the inclusion only of the shadow massive scalars
which are singlets of the  $\mathcal{N}=3$ R-symmetry group $\mathrm{SU}(2)_R$;
let us denote them collectively by $\mu$. We shall in particular derive the
$\sigma$-model kinetic terms of the scalars as well as their potential
$V(\mu)$.
\par
We discussed in section \ref{n010harm} the singlet shadow scalars
and their geometric interpretation as moduli of the $\n010$
compactification. They are the ``breathing'' and ``squashing'' modes
$\Sigma(x)$ and $\phi(x)$, of energies $6$ and $4$, and the modes
$\pi_1(x)$ and $\pi_2(x)$ related to an internal 3-form, of energies
$5$ and $3$. These fields should correspond to mass eigenstates of the
mass matrix in the effective theory expanded around the $\n010$ vacuum.
\par
Including the fields $\Sigma$ and $\phi$ corresponds to rescaling as in eq.
(\ref{viel7resc}) the 7-dimensional vielbein.
The effective theory that includes rescalings of the internal vielbein
has been already considered in \cite{yasuda}, with the aim of checking the
stability of the compactification. The contributions from the internal 3-form
modes, eq. (\ref{3formfluct2}), were not considered in that paper.
\par
The basic techniques for the ``warped'' dimensional reduction,
corresponding to eq.s (\ref{viel7resc}) and (\ref{3formfluct2}),
of the 11D supergravity Lagrangian (\ref{boslag11}) go back to \cite{ss}.
To obtain a 4D supergravity in the Einstein frame, it is necessary to perform a
Weyl rescaling of the 4D metric. Thus we set
\begin{equation}
\label{weyl4d}
V^a(x) = \ee^{21\,\Sigma(x)}\,E^a(x)~,
\end{equation}
the vierbein $E^a$ corresponding to the Einstein frame metric.
\par
The kinetic part of the 4-dimensional action arise both from the
Einstein-Hilbert term in the 11D action (\ref{boslag11}) under the rescalings
(\ref{viel7resc}) and from the ``Yang-Mills'' part $F_{MNPQ}^2$ because of the
mixed terms $F_{a\beta\gamma\delta}$ due to the internal 3-form
(\ref{3formfluct2}). To exhibit the $\sigma$-model kinetic terms
corresponding to the expected structure, eq. (\ref{modspace}),
of the moduli space we leave the energy eigenstates
$\Sigma(x),\phi(x),$ $\pi_1(x),\pi_2(x)$ in favor of the combinations
$w(x)$ and $z(x)$, given in eq. (\ref{wzscalred}), of the rescalings,
and of the fields $f(x)$ and $g(x)$ of eq. (\ref{3formfluct2}).
\par
Then the kinetic action is
\begin{eqnarray}
\label{kin4d}
S_{\mathrm{kin}} & = & {1\over \kappa_4^2}\int
\Bigl[\mathcal{R} +
{1\over 4}(\partial_a w)^2 + {9\over 4} \ee^{-2\, w} (\partial_a f)^2
\nonumber\\
& + &
{3\over 2}(\partial_a z)^2 + {27\over 2} \ee^{-2\, z} (\partial_a g)^2
\Bigl]\,\mathrm{det}\, E~,
\end{eqnarray}
where the 4-dimensional gravitational coupling constant is defined
in terms of the 11-dimensional one and of the volume $V_7$ of the ``standard''
$\n010$ space by
\begin{equation}
\label{k4}
\kappa_{11}^2 = \kappa_{4}^2\, V_7~.
\end{equation}
\par
The 4D potential for the scalars gets contributions from
all the terms in the action (\ref{boslag11}).
The contributions from the Einstein-Hilbert term due to the ``internal''
curvature $\mathcal{R}^{(7)}$, describing the $x$-dependence of the volume
of the internal space, were already computed in \cite{yasuda}.
The kinetic terms for the 3-form, $(F_{MNRP})^2$, contribute both because of the
internal three-form
(\ref{3formfluct2}), yielding a term  $(F_{\alpha\beta\gamma\delta})^2$, and
of the  Freund-Rubin term $(F_{abcd})^2$. In fact
we insist on a Freund-Rubin  ansatz
\begin{equation}
\label{genFR}
F_{abcd}(x)=Q(x)\, \varepsilon_{abcd}~,
\end{equation}
where $F_{abcd}$ are the components of $F$ along the vielbeins $E^a$.
The 11D ``Maxwell'' equation of motion, eq.  (\ref{maxw11d}), requires that
$Q$ is no longer a constant, but rather that
\begin{equation}
\label{FRmod}
Q(x) = \ee^{-2\,z(x) -{1\over 3}w(x)}
\biggl(1 + {9\over 4} \bigl(f(x)\,g(x) +
{g^2(x)\over 2}\bigr)\biggr)\,\, e~.
\end{equation}
The above reduces to the usual FR ansatz, eq. (\ref{FreundRubin}),
when  $w=z=f=g=0$.
The ``Chern-Simons'' term in the action (\ref{boslag11})
contributes to the 4D scalar potential because of those terms that
can be written as being proportional to
\begin{equation}
\label{CScontr}
\varepsilon^{abcd}F_{abcd}\,\,
\varepsilon^{\alpha_1\ldots\alpha_3\beta_1\ldots\beta_4}
A_{\alpha_1\ldots\alpha_3}F_{\beta_1\ldots\beta_4}~.
\end{equation}
\par
Altogether, the potential appearing in the 4-dimensional action,
\begin{equation}
\label{defpot4d}
S_{\rm pot} = -{1\over \kappa_4^2}\int\, e^2\, V(\mu)\, \mathrm{det}E~,
\end{equation}
is the following:
\begin{eqnarray}
\label{pot4d}
V(w,z,f,g) & = &  -12\, \ee^{-w - 2 z} \left( 1 + 8 \ee^{w - z} - 2
\ee^{2(w- z)}\right) + 36 \, \ee^{-w-6z}
\biggl( 1 + {9\over 4}(fg + {g^2\over 2})\biggr)^2
\nonumber\\
& + & 216\, \ee^{-w - 4 z}\left( f^2 + 2 f g +
\left(1 + 6 \ee^{2(w - z)}\right) g^2\right) \nonumber\\
& + & 729\, \ee^{-w-6z} (g^2 + 2 f g)
\biggl( 1 + {9\over 4} (fg + {g^2\over 2})\biggr)~.
\end{eqnarray}
Extracting the quadratic part of the potential (\ref{pot4d}) around the
$\n010$ critical point $w=z=f=g=0$ we find the mass terms
\begin{equation}
\label{massterms}
24\, e^2\,(w^2 + 4\,  w z + 16\, z^2)
+4\, e^2\,(27 f^2 + 8 \, fg + {8\over 27} g^2)~.
\end{equation}
The eigenvalues of the mass matrix for the fields $w/\sqrt{2}$ and
$\sqrt{3}z$, those that have canonical kinetic terms in the $\n010$
vacuum, see eq. (\ref{kin4d}), correspond to AdS masses\footnote{The
definition of AdS$_4$ mass $m$ that we use is such that the free eq.
of motion of a scalar $\varphi(x)$ is
$(\square + \mathcal{R}^{(4)}/3 + m^2) \varphi =0$.
In the $\n010$ background this leads to the AdS squared mass $m^2$
being shifted of $32\, e^2$ with respect to the ``usual'' squared mass
$\bar m^2$, appearing as ${1\over 2}\bar m^2\varphi^2$ in the Lagrangian.}
\begin{equation}
\label{masssu}
m_\Sigma^2 = 320 \, e^2~,\hskip 0.8cm
m_\phi^2 = 96 \, e^2~,
\end{equation}
namely to energies $E=6$ and $E=4$,
as we expected from table \ref{brokengravitino}. The mass eigenstates are
indeed proportional respectively to the fields $\Sigma(x)$ and $\phi(x)$.
\par
The eigenvalues of the mass matrix for the canonically normalized fields
$3f/\sqrt{2}$ and $g/(3\sqrt{3})$ correspond to AdS
masses
\begin{equation}
\label{massfg}
m_{\pi_1}^2 = 192 \, e^2~,\hskip 0.8cm
m_{\pi_2}^2 = 32 \, e^2~,
\end{equation}
again in agreement with our expectations: the eigenstates are proportional
to $\pi_1(x)$ and $\pi_2(x)$, and correspond to energies $E=5$ and $E=3$.
\subsubsection{Comparison}
From the eleven dimensional compactification geometric point of view,
we have switched on 4D fields corresponding to rescaling in a particular
way (see eq. (\ref{viel7rescwz})) the internal vielbein and to turn on a
three form potential in the internal dimensions. This 11D point of view
has also suggested the combinations of scalar fields to use.
Indeed it is natural to switch on independently those components of the three
form which scale in a different way under the rescalings (\ref{viel7rescwz});
as already noticed, this is what gave rise to the definitions of the fields
$f$ and $g$.
Moreover if we want to pair each of them with the field that scales the relative
component of the three form (\ref{3formfluct2}), we see that the pairing of $f$
with $w$ and of $g$ with $z$ has a natural geometric interpretation.
\par
From the form of the effective action (\ref{kin4d})  this pairing is also
evident. In particular the kinetic terms suggest us that these two copies  of
fields parametrize two manifold that are both locally isometric  to
$\mathrm{SL}(2,\mathbb{R})/\mathrm{O}(2)$. Indeed we can interpret the kinetic
terms as the metric of this coset space in the solvable parametrization, see
eq. (\ref{2metric}). So it is now natural to ask if this 4D Lagrangian
suggested by M theory in 11D can be interpreted as a ``standard'' Lagrangian
describing the supersymmetry breaking  from $\sn=4$ to $\sn=3$.
\par
Having the same form of the kinetic terms in both the theories, what we have to
do is to look at the scalar potentials\footnote{The  explicit form of the
scalar potential in the ``standard'' de Roo and Wagemans' theory is rather
cumbersome and we do not report it explicitly here,  see
\cite{deroowag,deRoo:1985np,wageman,wagthesis}}. If we do this we can find
certain analogies between the two, but there is no way to match one with the
other. So we notice that the matching with the known supersymmetric
$\mathcal{N}=4$ theory spontaneously broken to $\mathcal{N}=3$ is impossible,
and these latter are the only known theories where a massive gravitino is
coupled to $\mathcal{N}=3$ massless gravitons.
\par
Now there are two possibilities. One is that the KK truncation
to the massive gravitino is an inconsistent one, but as previously discussed
there are strong arguments that this is not the case. The other is that
new ``shadow'' supergravities do exist.
\par
Indeed, in the KK compactification of M theory on the coset manifold $\n010$ we
cannot find the $\mathcal{N}=4$ theory realized at the massless level, since
this coset does not have the right isometries. Instead, we can have an
enhancement of symmetry if we take into consideration also some massive mode.
Then we can argue that in order  to find $\mathcal{N}=4$  supersymmetry
realized at the massless level in this context  we have to look for some change
of topology of the internal manifold.
\par
From the point of view of the effective four dimensional theory, this involves
something that happens at the boundary of the moduli space, where some
rescaling vanishes or goes to infinity. This is specular to what happens if we
try to match the standard $\mathcal{N}=4$ 4D Lagrangian with the result of KK
compactification. Also in that case, we have shown that the requested value of
the Casimir energy of the gravitino is never reached but it is the limiting
value on the boundary of the moduli space (of the matter fields).
\par
We conclude that a new shadow supergravity has to be built and that, if this
new theory has something to do with the standard one, then it can only  be at
the boundary of the two respective moduli spaces. At the boundary one can
guess the relation between the scalar fields. Indeed it is easy to see that if
we perform the change of parametrization (\ref{soluti}) on the fields $z$ and
$g$  the effective potential (\ref{pot4d}) is the same for both the choice of
the sign at the limiting value of $\left| v \right|=1$. This is not true for
the fields $w$ and $f$. So we can guess that in the $\mathcal{N}=4$ theory
parent to the $\n010$ compactification the fields $z$ and $g$ will be those
parametrizing the matter coset $\mathrm{SO}(1,2)/\mathrm{SO}(2)$, while $w$ and
$f$ will belong to the new massless graviton multiplet.
\par
Moreover, we can argue that the change of topology requested for the
interpolation between the $\n010$ compactification and this theory involves a
singularity of the vielbein $B^{\tilde{\alpha}}$, ($\tilde{\alpha}=4,5,6,7$),
as the match with standard $\mathcal{N}=4$ supergravity is possible only at the
boundary of the moduli space of the matter fields (related to $z$ and $g$) and
the action of the rescaling on the vielbein is as in (\ref{viel7rescwz}).
\section{On the suggested shadow extensions of $\bf \mathcal{N}\ge 4$ supergravities}
\label{hinti}
To understand the true origin of the bound (\ref{boundo}) on the conformal
dimension of the broken gravitino field, we need to adopt a viewpoint we have
already considered in our general discussion of partial supersymmetry breaking
(see section \ref{genaspect}). Namely we have to turn things around and
consider the coupling of a massive $\mathcal{N}=3$ gravitino multiplet to
$\mathcal{N}=3$ supergravity. Then we  can uniquely determine the scalar
manifold $\mathcal{M}^{\mathcal{N}=3}_{3/2}$ containing the $14$ scalars of
such a multiplet.
\par
Indeed, from table \ref{brokengravitino} we know that, of these $14$ scalars, 2
are $\mathrm{SU}(3)$--singlets, while 12 sit in the $\mathbf{6}$ and
$\bar\mathbf{6}$ representations.  The $\mathrm{SU}(3)$--singlets parametrize
the coset  $\mathrm{SU}(1,1)/\mathrm{U}(1)$ that, within the $\mathcal{N}=4$
context, is absorbed by the graviton multiplet. The remaining $12$ fields must
parametrize a non--compact 12-dimensional coset $G/H$ satisfying the following
conditions:
\begin{enumerate}
  \item The isotropy subgroup is $H=\mathrm{SU}(3) \times \mathrm{U}(1)$, since
  we look for a theory with manifest $\mathcal{N}=3$ supersymmetry and
  admitting $\mathcal{N}=3$ vacua. Since $\mbox{dim} H=9$, this implies that
  $\mbox{dim}G=21$. \item The coset generators $\mathbb{K}$ in the orthogonal
  splitting $\mathbb{G}=\mathbb{H}\oplus\mathbb{K}$ must transform in the
  $\mathbf{6}$ (or $\bar \mathbf{6}$) of $\mathrm{SU}(3)$, namely in the
  (conjugate) symmetric tensor representation.
\end{enumerate}
The unique solution of these constraints is $G=\mathrm{Usp}(3,3)$. Thus
altogether we have
\begin{equation}
   \mathcal{M}^{\mathcal{N}=3}_{3/2}=
   \frac{\mathrm{SU}(1,1)}{\mathrm{U}(1)}\times
   \frac{\mathrm{USp}(3,3)}{\mathrm{SU}(3) \times \mathrm{U}(1)}~.
\label{masgravmanif}
\end{equation}
\par
In sec. (\ref{partialbreaking}) we have discussed how the coupling of a massive
$\mathcal{N}=3$ gravitino to $\mathcal{N}=3$ supergravity can be obtained from
$\mathcal{N}=4$ supergravity, gauging the minimal group $\mathcal{G}_{\rm min}$
of eq. (\ref{minimgaug}),  via a super-Higgs phenomenon. In this case, before
breaking, the scalars fill the 20-dimensional\footnote{Beside the two fields
belonging to the $\mathcal{N}=4$ graviton multiplet, 6 scalar are eaten up by
as many vectors that become massive in the super-Higgs phenomenon.} space
$\mathcal{M}^{(\mathcal{N}=4)}_{(3)}$, see eq. (\ref{so6n/so6son}).
\par
If the super-Higgs mechanism of sec. (\ref{partialbreaking}) were really the
most general way of coupling a massive gravitino multiplet to $\mathcal{N}=3$
supergravity, we should be able to single out, within the scalar manifold of
$\mathcal{N}=4$ vector multiplets, namely the $\mathrm{SO}(3,6)/(\mathrm{SO}(3) \times
\mathrm{SO}(6))$ factor in $\mathcal{M}^{(\mathcal{N}=4)}_{(3)}$, a consistent
truncation to the scalar manifold of the gravitino multiplet
(\ref{masgravmanif}). Indeed it should be possible to rewrite the theory in
$\mathcal{N}=3$ language with no additional restrictions and span the whole
$14$--dimensional moduli space of such a coupling.
\par
But this is a crucial point: such a consistent truncation does not exist
in standard $\mathcal{N}=4$ supergravity, while it can be introduced if we take
a different, $\mathcal{N}=3$, starting point.
\paragraph{Solvable Lie algebra decompositions}
To discuss this point, it is convenient to use the solvable Lie algebras
language. A non--compact coset manifold $G_1/H_1$ is  consistently embedded as
a sub-manifold of another non--compact coset  $G_2/H_2$ iff the solvable Lie
algebra $\mathrm{Solv}(G_1/H_1)$ generating the  first coset is a subalgebra of the
solvable Lie algebra  $\mathrm{Solv}(G_2/H_2)$ generating the second coset.
\par
It is then not difficult to see that the ``matter'' part
$\mathrm{USp}(3,3)/(\mathrm{SU}(3)\times{U}(1))$ of the scalar manifold of the
gravitino multiplet (\ref{masgravmanif}) can not be embedded into the scalar
manifold of $\mathcal{N}=4$ vector multiplets $\mathrm{SO}(3,6)/(\mathrm{SO}(3)
\times \mathrm{SO}(6))$, since
\begin{equation}
  \mathrm{Solv}\left(\frac{\mathrm{USp}(3,3)}{\mathrm{SU}(3)\times{U}(1)}\right)
  \not\subset \mathrm{Solv}\left(\frac{\mathrm{SO}(3,6)}{\mathrm{SO}(3) \times
  \mathrm{SO}(6)} \right)~.
\label{noway}
\end{equation}
Indeed, the 12 generators that we should select among those of the solvable
Lie algebra for  $\mathrm{SO}(3,6)/(\mathrm{SO}(3) \times \mathrm{SO}(6))$,
given in eq.s (\ref{cartano}--\ref{tuttaalg2}) are those of isospin $J=2$ or
isospin $J=0$ with respect to the diagonal $\mathrm{SO}(3)_R$ $R$--symmetry
group\footnote{This follows from the assignment to the $\mathbf{6}$ and
$\bar\mathbf{6}$ representations of $\mathrm{SU}(3)$ of the coset generators of
$\mathrm{USp}(3,3)/(\mathrm{SU}(3)\times{U}(1))$ discussed above; indeed,
$\mathrm{SO}(3)_R$ is maximally embedded into $\mathrm{SU}(3)$ and eq.
(\ref{picchio}) applies.} (\ref{O3diag2}).  Such generators correspond to the
matrices of eq. (\ref{cartano}) (3 Cartan generators) or to the matrices of eq.
(\ref{tuttaalg}) with the constraint $E=E^T$ (6 nilpotent generators) or,
finally, to those of eq. (\ref{tuttaalg2}) with $D=U^T$ (3 nilpotent
generators). The problem is that these constraints are \emph{not} preserved by
the commutation relations, namely they do \emph{not} define a proper
subalgebra.
\par
Let us instead suppose that the manifold of the scalars of the 3 vector
multiplets be the one that arises in the coupling to $\mathcal{N}=3$ supergravity,
namely $\mathcal{M}^{(\mathcal{N}=3)}_{(3)}$ of eq. (\ref{su3n/su3u1sun}). This space
\emph{can} contain $\mathrm{USp}(3,3)/(\mathrm{SU}(3)\times{U}(1))$  as a
sub-manifold, as
\begin{equation}
  \mathrm{Solv}\left(\frac{\mathrm{USp}(3,3)}{\mathrm{SU}(3)\times{U}(1)}\right)
  \subset
  \mathrm{Solv}\left(\frac{\mathrm{SU}(3,3)}{\mathrm{SU}(3) \times \mathrm{SU}(3)\times
  \mathrm{U}(1)}
  \right)~.
\label{yesway}
\end{equation}
The solvable Lie algebra generating the manifold
$\mathrm{SU}(3,3)/(\mathrm{SU}(3) \times \mathrm{SU}(3) \times \mathrm{U}(1))$
was studied in detail in  \cite{Andrianopoli:1998wi}, and its structure is
recalled in appendix \ref{appesolv} for the reader's convenience. It is
immediate to select among the  18 generators listed in eq. (\ref{6x6matrices})
of that appendix those that belong to the subalgebra  $\mathrm{USp}(3,3)
\subset \mathrm{SU}(3,3)$. They are twelve:
\begin{equation}
 g_i\,,~ h_i~~(i=1,2,3)\,,~ \mathbf{X}^+_1\,,~
 \mathbf{Y}^+_1\,,~\mathbf{Z}^+_1\,,~\mathbf{X}^-_2\,,~
 \mathbf{Y}^-_2\,,~\mathbf{Z}^-_2~,
\label{purgone}
\end{equation}
and it is easy to check from eq. (\ref{Alekal}) that they close an algebra
among themselves.  It is the solvable Lie algebra generating
$\mathrm{USp}(3,3)/\left( \mathrm{SU}(3) \times\mathrm{U}(1)\right)$.
\par
The consequence of the statements in eq.s (\ref{noway}) and (\ref{yesway})
is that there should exist a \emph{shadow extension} of $\mathcal{N}=4$ supergravity
constructed according to the recipe outlined below.
%
\paragraph{A conjecture about shadow $\mathcal{N}=4$ supergravity}
We start from $\mathcal{N}=3$ supergravity coupled to $3+n$ vector multiplets.
In this case the scalar manifold $\mathcal{M}^{(\mathcal{N}=3)}_{(n)}$
was given in eq. (\ref{su3n/su3u1sun})
and has an obvious $\mathrm{SU}(3,3)/(\mathrm{SU}(3)\times
\mathrm{SU}(3) \times \mathrm{U}(1))$ submanifold containing the essential
degrees of freedom necessary for the super-Higgs phenomenon.
Next we consider a
\emph{massless} $\mathcal{N}=3$ gravitino multiplet, with the following field
content:
\begin{center}
\begin{tabular}{ccc}
\hline\hline
  Spin & Number of fields & $\mathrm{SU}(3)$ representation\\
  \hline
  $\ft 32$ & 1 & $\mathbf{1}$ \\
  1 & 3 & $\mathbf{3}$ \\
  $\ft 12$ & 3 & $\overline{\mathbf{3}}$\\
  0 & 2 & $\mathbf{1}$\\
  \hline\hline
\end{tabular}
\end{center}
and we try to couple it to the theory we already have. The two scalar fields of
the massless gravitino multiplet span the coset manifold
$\mathrm{SU}(1,1)/\mathrm{U}(1)$ and in this way we have, altogether, the coset
manifold:
\begin{equation}
  \frac{\mathrm{SU}(1,1)}{\mathrm{U}(1)} \times
  \frac{\mathrm{SU}(3,3+n)}{\mathrm{SU}(3) \times \mathrm{SU}(n) \times
  \mathrm{U}(1)}~.
\label{totalcosso}
\end{equation}
Our conjecture is that, using the composite connections and the vielbein of the
above coset (\ref{totalcosso}), we should be able to construct a gauged
supergravity action supersymmetric under $3+1$ local supersymmetries. This is
the \emph{shadow extension} of $\mathcal{N}=4$ supergravity we have many times invoked.
\par
The catch of this construction, which we postpone to a future publication, is
given by the $\mathrm{SU}(3) \times \mathrm{U}(1)$  charge assignments of the
gravitinos: indeed these charges determine  the coupling to the scalar sector,
the gravitino mass--matrices and the  fermion shifts.
\par
In standard $\mathcal{N}=4$ supergravity, the three unbroken gravitinos and the fourth
broken one sit together in the $4$--dimensional fundamental representation of
$\mathrm{SU}(4)$. When we split such a representation with respect to the
$\mathrm{SU}(3) \times \mathrm{U}(1)$ subgroup, the  $\mathrm{U}(1)$--charge of
the singlet is $3$ times the $\mathrm{U}(1)$  charge of the triplet. We think
that this fact is at the origin of the  bound (\ref{boundo}) on the conformal
dimension of the broken gravitino.
\par
If we start instead from the coset manifold (\ref{totalcosso}), there are two
$\mathrm{U}(1)$ charges to be assigned, corresponding to the two
$\mathrm{U}(1)$ factors appearing in the $H$--subgroup, and nothing a priori
prevents us from  independent assignments for the triplet and singlet
gravitinos. Obviously, if our choices are incompatible with an $\mathrm{SU}(4)$
symmetry, we cannot expect any $\mathcal{N}=4$ supersymmetric vacuum from such
a theory. But this is precisely what we are looking for: in all vacua the
singlet gravitino will become massive by eating the goldstino and $6$ vector
fields will become its massive partners by eating up the $6$ scalars of the
$6$-dimensional nilpotent subalgebra of
$\mathrm{Solv}[\mathrm{SU}(3,3)/( \mathrm{SU}(3) \times \mathrm{SU}(3)\times
\mathrm{U}(1))]$.   If our conjecture is true, it is quite obvious that the
bound (\ref{boundo}) on the conformal dimension of the broken gravitino will be
replaced by a new one depending on the choice of the $U(1)$--charges, and that
the critical value $E_0=3$ can now be reached for suitable choices.
\paragraph{Possible extensions to higher supergravities}
A comment that we should make in relation with our
conjecture is that it might be applied also to the case of other
extended supergravities, for instance $\mathcal{N}=8$.
\par
Here the scalar manifold is
$\mathcal{M}^{(\mathcal{N}=8)}=\mathrm{E}_{7(7)}/\mathrm{SU}(8)$, and the
theory can be also regarded as the coupling of $\mathcal{N}=2$ supergravity to
15 $\mathcal{N}=2$ vector multiplets and $6$ $\mathcal{N}=2$ massless gravitino
multiplets. This gives an $\mathrm{SU}(2) \times \mathrm{SU}(6) \times
\mathrm{U}(1)$ symmetry and the question is  whether in the broken phase  it is
absolutely necessary to enforce the larger  $\mathrm{SU}(8)$ local symmetry.
\par
We are tempted to assume that the correct answer is no. In this case the $6$
massive gravitinos can assume $\mathrm{U}(1)$--charges unrelated to those of
the unbroken ones, leading to less restrictive values for their conformal
dimensions $E_0$. Eventually this destroys the $\mathrm{E}_{7(7)}$ symmetry.
Alternatively, we can break $\sn=8$ into $\sn=4$,$\sn=5$ or $\sn=6$ plus a
complementary number of gravitino multiplets and formulate similar questions.
This may be the solution of a puzzle recently discovered by Ferrara and
Sokatchev \cite{Ferrara:2000ki}. It appears that ordinary $\mathcal{N}=8$
supergravity is able to produce only a subset of the possible BPS states
allowed by $\mathrm{Osp}(8\vert 4)$ representation theory, namely those
obtained by tensoring only one kind of singleton representations while two are
algebraically available. BPS states preserving a fraction of supersymmetry
different from 1/2, 1/4 or 1/8 seem to be forbidden by the $\mathrm{E}_{7(7)}$
duality symmetry of the theory. Their construction would involve the use of
both kind of singletons. It is tempting to conjecture that these missing BPS
states might be solutions of some \emph{shadow extension} of $\mathcal{N}=8$
supergravity, constructed along the same lines of thought we have outlined
above.
%
\section{CFT interpretation of the spin $\ft 32$ shadow multiplet}
\label{cft}
In \cite{noialtro} we have already constructed the field theory realization,
in the CFT dual of an ${\cal N}=3$ compactification, of the universal long
gravitino multiplet which constitutes the main focus of the present
paper.
It is given by the following composite operator:
\begin{equation}
{\cal SH}={\rm Tr}\big[\underbrace{\Theta_\Sigma\otimes\Theta_\Sigma\otimes
\Theta_\Sigma}_{J=0}\big]={\rm Tr}\left[\Theta_\Sigma^+\Theta_\Sigma^{\,0\,}
\Theta_\Sigma^-\right]\, ,
\label{CFTinte}
\end{equation}
where $\Theta_\Sigma$ is the $J\!=\!1$ short superfield that, by definition,
contains the field strength of the world--volume gauge theory:
\begin{equation}
  \Theta_\Sigma = \left(\begin{array}{c}
  Y\\
  \Sigma\\
  -Y^\dagger
\end{array}\right)+{\cal O}(\theta^0)\,.
\label{pincopallo}
\end{equation}
{}From the $\theta$ expansion of the superfield (\ref{CFTinte}),
we retrieve the field theory interpretation of the various Kaluza
Klein modes appearing in the multiplet.
As stressed in \cite{noialtro}, the {\it breathing mode} of the
internal manifold $X^7$, namely the scalar component of zero
isospin and conformal dimension $6$ corresponds to the following
gauge theory operator:
\begin{equation}
\int d^2\theta^+ d^2\theta^- d^2\theta^{\,0\,}
{\cal SH}=Tr\left[3\ii H\overline HP+\ft{1}{4}\epsilon^{\lambda\mu\nu}
\epsilon^{\rho\sigma\tau}F_{\lambda\mu}F_{\nu\rho}F_{\sigma\tau}\right]+\mbox{derivative terms}\,,
\label{FFF}
\end{equation}
where $H$ and $P$ are auxiliary fields of the world--volume gauge multiplet.
In other words, the operator (\ref{FFF}) is the $\mathcal{N}=3$
supersymmetrisation of the third power of the gauge field strength \footnote{It should be noticed that,
since vector multiplets are not conformal in three
dimensions, the vector multiplet fields in the previous expression should be considered as re-expressed
in terms of the fundamental degrees of freedom at the conformal point (via equations of motion or
Hodge dualization, as requested by the specific example).}:
$$
  \epsilon^{\lambda\mu\nu}\epsilon^{\rho\sigma\tau}
  F_{\lambda\mu}F_{\nu\rho}F_{\sigma\tau}\,,
$$
whose dimension, as a consequence of our analysis, is shown to be
protected from quantum corrections.
On the field theory side, this suggests the existence of some new no
renormalization theorem yet to be discovered.
A closely similar situation appears in type  IIB  $\mathrm{AdS}_5$
compactifications, where the volume mode of the internal manifold
corresponds to the CFT operator $F^4$, of dimension 8, which is known
to satisfy some non-renormalization theorem.
This consideration suggests that the operator (\ref{FFF}) could originate
by the low energy expansion of an analogue of the Dirac Born Infeld for
the $M2$-brane, as well as the operator $F^4$ comes from the $\alpha'$
expansion of the DBI Lagrangian of the $D3$-brane
(see\cite{noialtro}).
\section{Conclusions and discussion}
There are many directions where the analysis we started in this paper may lead.
We already stressed that one basic suggestion of our work is the existence of
shadow supergravities. Having exhaustively discussed this point in the previous
sections, we can devote these conclusions to the quantum field theory aspects
that originally motivated our work.
\par
One immediate question is the fate of shadow multiplets in compactifications
that are not of Freund-Rubin form. We known that for AdS$_5$, for example,
interesting (both for supergravity and CFT) backgrounds \cite{warner} are not
of Freund-Rubin form, having a non-trivial warp factor and internal
antisymmetric tensor fields. If we could show that a pairing exists also for
these compactifications, it would be interesting to study these cases as  well
and understand what changes in the form and the quantum numbers of the
\emph{universal} volume multiplet. Notice, for example, that the known  $\mathcal{N}=4$
gauged supergravities admit a plethora of $\mathcal{N}=3$ vacua with $E_0<3$. It would
be quite interesting to find the 11 dimensional solution corresponding to these
critical points.
\par
The $\mathcal{N}=4\rightarrow \mathcal{N}=3$ spontaneous symmetry breaking in four-dimensional
gauged supergravity is also interesting because it provides examples of RG
flows between 3d CFT's \cite{RG}. This is indeed the reason that originally
motivated this work. The case $\n010$ is quite special. The critical point
is at infinite distance from the origin in moduli space. Since the volume mode
is singular (either vanishes or diverges),
the $\mathcal{N}=4$ and $\mathcal{N}=3$ compactifications are related only through a change of
topology. It is not completely clear what is the right interpretation from the
quantum field theory point of view. We have made no attempt to identify the
parent $\mathcal{N}=4$ theory in terms of the elementary degrees of freedom of the CFT
associated with $\n010$, but this is certainly an aspect that deserves
further interest.
\par
We explicitly exhibited the form of the \emph{universal} volume multiplet in
terms of CFT operators in the case of $\n010$. It is obtained by tensoring
three massless multiplets. It is the three-dimensional counterpart of the
AdS$_5$ volume multiplet that contains an operator roughly of the form $F^4$,
with dimension 8. The shadowing mechanism, which works also for AdS$_5$
compactifications, then guarantees that this multiplet has {\it canonical}
dimension. This may be expected in view of non-renormalization theorems which
are conjectured to hold for operators like  $F^4$. We may speculate that
something similar happens in three dimensions with the $E_0=6$ scalar operator.
Since, in three dimensions, the CFT are not continuously connected to free
theories but are just IR limit of some non-conformal gauge theory, it is
difficult to give a general form for this \emph{universal} operator. The
explicit result reported in Section \ref{cft} might help.
\par
When this paper was nearly finished we learnt of the very recent
paper \cite{popus} where the question of consistent truncations of
$\mathrm{AdS}_4 \times X^7$ compactifications is addressed. It would
certainly be interesting  to
inquire about the relation between consistency of the truncation and
shadowing.
\par
We leave all these interesting questions for future work.
\paragraph{Acknowledgements}
\vskip 0.3cm
We are very grateful to L. Castellani,
R. D'Auria, Sergio Ferrara, C. Reina, A. Tomasiello,
M. Trigiante and A. Zampa for important discussions.
\appendix
\section{Conventions and notations}
\label{appendo}
We follow the conventions of \cite{ricpie11} for $11$--dimensional Supergravity and
of \cite{noi321,univer,spectfer,castn010} for its Kaluza Klein
compactifications. In $D=11$ the flat metric is mostly minus
$$\eta_{AB}=(+,\underbrace{-,\dots,-}_{\mbox{ 10 times}} )$$ so that the flat
metric for the internal $7$--dimensional geometry is purely negative
$\eta_{\alpha\beta}=-\delta_{\alpha\beta}$. Since we use differential
form language the only indices we introduce are the flat ones
($A,B,C,\dots,=0,8,9,10,1,2,\dots,7$). We also split them
into:
$$A=\underbrace{a}_{0,8,9,10},\underbrace{\alpha}_{1,\dots,7}.$$
With the negative metric the gamma matrices in $7$--dimensions are
purely real and antisymmetric and are named $\tau_\alpha$:
$$ \left \{ \tau_\alpha \, , \, \tau_\beta \, \right \} = - \delta_{\alpha \beta
} $$
\section{Explicit form of the harmonics}
\label{appenc}
In this Appendix, we give some discussions and expressions
regarding the harmonics of the ``super-Higgs'' shadow gravitino multiplet which
were not included in section \ref{miracle}. We refer to table \ref{tabellas}
for notations.
\par
$\cdot$ \emph{The harmonic of the massive $\pi$ scalars with $E=5$}
When we discussed the mechanism of shadowing we announced that it
originates from two general features:
\begin{enumerate}
  \item The fact that the same harmonic is associated with two different
  Kaluza Klein fields
  \item The fact that, via Killing spinor multiplication, each Bose/Fermi harmonic generates
  a new  Fermi/Bose harmonic with predetermined eigenvalue.
\end{enumerate}
We have already taken advantage of the second type of {\sl shadowing}
when we have constructed the harmonic (\ref{Tharmo}) of the
$Z$--vectors. Let us use it once again to show that the shadow of the
massless gravitinos contains also the pseudoscalars $\pi$ with scale
dimension $E=5$. It suffices to use eq.s (4.47b),(4.49) and (4.50) of
\cite{univer} that instruct us how to construct an eigenstate of
the operator (\ref{1cube})  with eigenvalue:
\begin{equation}
  M_{(1)^3}=-\ft 14 \, \left(  M_{(\ft 12)^3}+8\right)
\label{m1c-m1/2}
\end{equation}
from each eigenstate of the internal Dirac operator (\ref{unmezcube})
of eigenvalue $M_{(\ft 12)^3}$. From the massless gravitino harmonics
($M_{(\ft 12)^3}=-16$), via such a relation, we obtain the  harmonics
(\ref{upsilon}), namely:
\begin{equation}
  \Upsilon^{(AB)}_{\alpha\beta\gamma}=
  \overline{\eta}^A \,\tau_{\alpha\beta\gamma} \,
  \eta^B
\label{upsilon2}
\end{equation}
which are bilinear in Killing spinors and belong to the eigenvalue
$M_{\, (1)^3}=-2$ of (\ref{unmezcube}). These are the
harmonics of the pseudoscalar particles of
energy $E=5$ and isospin $J=2 \oplus J=0$ displayed in  table
(\ref{brokengravitino}). Indeed the three index $\tau$--matrix is
symmetric so that we have symmetry in the $\mathrm{SO}(3)$ R--symmetry indices
$A,B$. Decomposing into the traceless and trace parts we obtain the
${\bf J}=2$ and ${\bf J}=0$ states.
\par
$\cdot$ \emph{The harmonic of the massive $E=9/2$  spinors }
We can now return to the $E=\ft 92$ spinors we have left aside. Their
harmonic is of the transverse type $\Xi_\alpha$ and must be an
eigenvalue of the operator (\ref{raritsch}) with eigenvalue $M_{\ft
32(\ft 12)^2}=4$. Recalling eq.s (5.7-5.12) of \cite{univer} we see
that, with suitable coefficients $a,b,c$ the  fermionic combination:
\begin{equation}
  \Omega_\alpha = a \, \tau_{\alpha \mu \nu \rho } \eta \, Y_{\mu \nu \rho
  } + b \, \tau _{\mu \nu } \eta \,Y_{\alpha \mu \nu } + c\,
  \tau _{\mu \nu \rho } \eta \, D_\alpha  Y_{\mu \nu \rho }
\label{combinatio}
\end{equation}
has   the desired eigenvalue
\begin{equation}
 4 \,=\,  M_{\ft 32(\ft 12)^2} = -4\left( M_{(1)^3} +4\right)
\label{desireig}
\end{equation}
if   $Y_{\alpha \beta \gamma }$ is identified with the bosonic harmonic
(\ref{upsilon}) of eigenvalue $M_{(1)^3}=-2$. Inserting the values of
$a,b,c$ given in eq.s (5.9a-5.9b) of \cite{univer} we consider
therefore the following  $\mathrm{SO}(3)_R$ 3-index
tensor  which is also an $\mathrm{SO}(7)$  $\tau$-traceless spinor-vector
\begin{eqnarray}
\Omega^{A(BC)}_\alpha & = & 14 \,
\tau_{\alpha\mu\nu\rho}\,\eta^A \,
\overline{\eta}^B \, \tau^{\mu\nu\rho} \,
\eta^C \, + \, 48 \, \tau^{\mu\nu}
\, \eta^A \, \overline{\eta}^B \,
\tau_{\alpha\mu\nu} \eta^C \nonumber\\
\null & \null & 2 \, \tau^{\mu\nu\rho} \, \eta^A
\, \overline{\eta}^B \, \tau_{\alpha\mu\nu\rho} \,
\eta^C
\label{omeg32}
\end{eqnarray}
A priori we are multiplying an isospin
 $J=1$ with $J=2$ or $J=0$, so that we can
obtain $J=3$, $J=2$, $J=1$ and  $J=1$ a second time, according to the
numerology and $7+5+3+3=18= 3 \times (5+1)$. However
Fierz identities on the $\eta$--spinors should imply that we get only an isospin J=2 state and an
isospin J=1 state since this  is  what is required by the structure
of the $\mathrm{AdS}_4$ spin $3/2$ massive multiplet displayed in table \ref{tabellas}
where no $J=3$ does appear. This is completely verified by explicit
calculations that were  numerically  performed
introducing an explicit basis for $\tau$ matrices and choosing
Killing spinors $\eta^A$ in three arbitrary directions.
The projection on the $J=3$ state corresponds to
full symmetrization in the indices $(ABC)$ and then removal
of the trace: the result of this projection is identically zero.
Hence it turns out that, in the $\mathrm{SO}(3)_R$ indices the tensor $\Omega^{A(BC)}$
has automatically the following symmetry:
\begin{equation}
\begin{array}{l}
\begin{array}{|c|c|}
\hline
             B & C \\
\hline
\end{array} \\
\begin{array}{|c|}
             A \\
\hline
\end{array}
\end{array}\\
\label{pistola}
\end{equation}
We obtain the isospin $J=2$ projection by defining
\begin{equation}
  \widehat{\Omega}_{\alpha}^{(XY)}=\ft{1}{2}\,
  \left( \epsilon^{XTZ} \, \Omega^{T(ZY)}_\alpha \,
   + \,
\epsilon^{YTZ} \, \Omega^{T(ZX)}_\alpha \right)
\label{omej2}
\end{equation}
while we obtain the isospin $J=1$ projection by defining
\begin{equation}
  \widehat{\Omega}_{\alpha}^{(A)}=
\sum_{B=1}^3\Omega^{A(BB)}_\alpha
\label{omej1}
\end{equation}
Indeed, because of the symmetry (\ref{pistola}) the
symmetric tensor $\widehat{\Omega}_{\alpha}^{(XY)}$
is automatically traceless in the indices
$XY$ and hence a pure $J=2$ representation. This can also be
numerically verified.
\par
$\cdot$ \emph{The harmonic of the massive $E=7/2$  spinors }
Also these states have a harmonic of the transverse type $\Xi_\alpha$. The
eigenvalue, this time is $M_{\ft 32(\ft 12)^2}=-16$. Looking at
fig.\ref{molecola}, we see that the transverse spinor harmonics communicate
either with the $1$--forms $Y_\alpha $ or with the $2$--forms $Y_{\alpha \beta
}$ or with the $3$--forms $Y_{\alpha \beta \gamma }$ or, finally with the
symmetric tensors $Y_{(\alpha \beta )}$. We are descending the multiplet from
higher to lower spins since we want to express everything in terms of Killing
spinors using the harmonics we have so far already constructed. Hence the mass
relation which is relevant to us at this point is that given in eq. (4.70) of
\cite{univer}, namely:
\begin{equation}
  M_{(1)^20}=( M_{\ft 32(\ft 12)^2} +8) ( M_{\ft 32(\ft 12)^2}+4)
\label{dimentican}
\end{equation}
Indeed it is satisfied by $M_{(1)^20}=96$ and $M_{\ft 32(\ft
12)^2}=-16$, which shows that in terms of a harmonic $\Theta^A_\alpha
$ which is eigenstate of the operator (\ref{raritsch}) with
eigenvalue $M_{\ft 32(\ft 12)^2}=-16$ we could construct the
harmonic (\ref{Tharmo}) of the $E=4,J=1$ vector fields of type $Z$.
Such a relation was given in eq.s (4.67b-4.69) of \cite{univer}. We
are interested in the inverse relation which expresses
$\Theta^A_\alpha$ in terms of the two form (\ref{Tharmo}) and of
Killing spinors. Such an inverse relation was not given in
\cite{univer} but it can be easily derived. In our explicit case we
find eq.(\ref{truccolo}), namely:
\begin{equation}
  \Theta^A_\alpha \equiv \epsilon ^{ABC} \, \left(
  \,\frac{3}{16}\,\tau _{\alpha \mu \nu \rho
  } \, \eta^B \, D_\mu \, T^C_{\nu\rho}+\frac{9}{2} \,\tau
  _\mu \, \eta^B \, T^C_{\alpha \mu } \right)
\label{truccolo2}
\end{equation}
So also this harmonic is universal and depends only on Killing
spinors
\section{The $\mathrm{SU}(3,3)/\mathrm{SU}(3)\times \mathrm{SU}(3)
\times U(1)$  solvable Lie algebra}
\label{appesolv}
For the reader's convenience we report in this appendix the structure
of the solvable Lie algebra associated with the manifold
$\mathrm{SU}(3,3)/\mathrm{SU}(3) \times \mathrm{SU}(3) \times U(1)$ that was already presented
in \cite{Andrianopoli:1998wi}.
Since the coset manifold $\mathrm{SU}(3,3)/\mathrm{SU}(3)\times U(3)$ is a {\it Special
K\"ahler} manifold  the elements of $Solv$ can
be  described in the Alekseevski's formalism for K\"ahlerian
algebras \cite{aleks} whose general structure is given by the following general
algebraic relations:
\begin{eqnarray}
\mathrm{Solv}\, &=&\, F_1\, \oplus\, F_2\, \oplus\, F_3\, \oplus\,
{\bf X}\, \oplus\, {\bf Y}\, \oplus\, {\bf Z}\nonumber\\ F_i\, &=&\,
\{{\rm h}_i\, ,\,{\rm g}_i\}\quad i=1,2,3\nonumber\\ {\bf X}\, &=&\,
{\bf X}^+\, \oplus\, {\bf X}^-\nonumber\\ {\bf Y}\, &=&\, {\bf Y}^+\,
\oplus\, {\bf Y}^-\nonumber\\ {\bf Z}\, &=&\, {\bf Z}^+\, \oplus\,
{\bf Z}^-\nonumber\\
\left[{\rm h}_i\, ,\,{\rm g}_i\right]\, &=&\,
2{\rm g}_i\quad i=1,2,3\nonumber\\
\left[F_i\, ,\,F_j\right]\, &=&\, 0 \quad i\neq j\nonumber\\
\left[{\rm h}_3\, ,\,{\bf Y}^{\pm}\right]\, &=&\,\pm{\bf Y}^{\pm}
\nonumber\\
\left[{\rm h}_3\, ,\,{\bf X}^{\pm}\right]\, &=&\,\pm{\bf X}^{\pm}
\nonumber\\
\left[{\rm h}_2\, ,\,{\bf Z}^{\pm}\right]\, &=&\,\pm{\bf Z}^{\pm}
\nonumber\\
\left[{\rm h}_1\, ,\,{\bf Z}^{\pm}\right]\, &=&\,{\bf Z}^{\pm}
\nonumber\\
\left[{\rm h}_1\, ,\,{\bf Y}^{\pm}\right]\, &=&\,{\bf Y}^{\pm}
\nonumber\\
\left[{\rm g}_1\, ,\,{\bf Y}\right]\, &=&\,\left[{\rm g}_1\, ,\,{\bf Z}
\right]\, =\, 0\nonumber\\
\left[{\rm g}_3\, ,\,{\bf Y}^{+}\right]\, &=&\,\left[{\rm g}_2\, ,\,{\bf Z}^{+}
\right]\, =\,\left[{\rm g}_3\, ,\,{\bf X}^{+}\right]\, =\, 0\nonumber\\
 \left[{\rm g}_3\, ,\,{\bf Y}^{-}\right]\, &=&\,{\bf Y}^+\, ;\,\,
\left[{\rm g}_2\, ,\,{\bf Z}^{-}\right]\, =\,{\bf Z}^+\, ;\,\,
\left[{\rm g}_3\, ,\,{\bf X}^{-}\right]\, =\,{\bf X}^+\nonumber\\
\left[F_1\, ,\,{\bf X}\right]\, &=&\,\left[F_2\, ,\,{\bf Y}\right]\, =\,
\left[F_3\, ,\,{\bf Z}\right]\, =\, 0\nonumber\\
\left[{\bf X}^-\, ,\,{\bf Z}^{-}\right]\, &=&\, {\bf Y}^-
\label{Alekal}
\end{eqnarray}
Explicitly the corresponding $\mathrm{SU}(3,3)$ matrices are listed below \footnote{We want to express
our gratitude to Mario Trigiante for providing us with this explicit representation
of the generators and for his invaluable advice on this algebraic point} :
 \begin{eqnarray}
 {\rm h}_1\, =\,
\left( \matrix{ 0 & 0 & 0 & 1 & 0 & 0 \cr 0 & 0 & 0 & 0 & 0 & 0 \cr 0
& 0 & 0 & 0 & 0 & 0 \cr 1 & 0 & 0 & 0 & 0 & 0 \cr 0 & 0 & 0 & 0 & 0 &
0 \cr 0 & 0 & 0 & 0 & 0 & 0 \cr }\right)\,\,\,\,{\rm g}_1\,&=&\,
\left( \matrix{ {i\over 2} & 0 & 0 & {{-i}\over 2} & 0 & 0 \cr 0 & 0 & 0 & 0 & 0 &
   0 \cr 0 & 0 & 0 & 0 & 0 & 0 \cr {i\over 2} & 0 & 0 & {{-i}\over 2}
    & 0 & 0 \cr 0 & 0 & 0 & 0 & 0 & 0 \cr 0 & 0 & 0 & 0 & 0 & 0 \cr
    }\right)
\nonumber\\
{\rm h}_2\, =\, \left( \matrix{ 0 & 0 & 0 & 0 & 0 & 0 \cr 0 & 0 & 0 &
    0 & 1 & 0 \cr 0 & 0 & 0 & 0 & 0 & 0 \cr 0 & 0 & 0 & 0 & 0 & 0 \cr
    0 & 1 & 0 & 0 & 0 & 0 \cr 0 & 0 & 0 & 0 & 0 & 0 \cr
    }\right)\,\,\,\,{\rm g}_2\,&=&\,
\left( \matrix{ 0 & 0 & 0 & 0 & 0 & 0 \cr 0 & {i\over 2} & 0 & 0 & {{-i}\over 2} &
   0 \cr 0 & 0 & 0 & 0 & 0 & 0 \cr 0 & 0 & 0 & 0 & 0 & 0 \cr 0 &
   {i\over 2} & 0 & 0 & {{-i}\over 2} & 0 \cr 0 & 0 & 0 & 0 & 0 & 0
   \cr }\right)\nonumber\\ {\rm h}_3\, =\, \left( \matrix{ 0 & 0 & 0 &
   0 & 0 & 0 \cr 0 & 0 & 0 & 0 & 0 & 0 \cr 0 & 0 & 0 & 0 & 0 & 1 \cr 0
   & 0 & 0 & 0 & 0 & 0 \cr 0 & 0 & 0 & 0 & 0 & 0 \cr 0 & 0 & 1 & 0 & 0
   & 0 \cr }\right)\,\,\,\,{\rm g}_3\,&=&\,
\left(   \matrix{ 0 & 0 & 0 & 0 & 0 & 0 \cr 0 & 0 & 0 & 0 & 0 & 0 \cr 0 & 0 &
   {i\over 2} & 0 & 0 & {{-i}\over 2} \cr 0 & 0 & 0 & 0 & 0 & 0 \cr 0
    & 0 & 0 & 0 & 0 & 0 \cr 0 & 0 & {i\over 2} & 0 & 0 & {{-i}\over 2}
    \cr } \right)\nonumber\\ {\bf X}^+_1\, =\, \left( \matrix{ 0 & 0 &
    0 & 0 & 0 & 0 \cr 0 & 0 & {i\over 2} & 0 & 0 & {{-i}\over 2} \cr 0
    & {i\over 2} & 0 & 0 & {{-i}\over 2} & 0 \cr 0 & 0 & 0 & 0 & 0 & 0
    \cr 0 & 0 & {i\over 2} & 0 & 0 & {{-i}\over 2} \cr 0 & {i\over 2}
    & 0 & 0 & {{-i}\over 2} & 0 \cr }\right)\,\,\,\,{\bf X}^+_2\,&=&\,
\left( \matrix{ 0 & 0 & 0 & 0 & 0 & 0 \cr 0 & 0 & {1\over 2} & 0 & 0 & -{1\over 2}
    \cr 0 & -{1\over 2} & 0 & 0 & {1\over 2} & 0 \cr 0 & 0 & 0 & 0 & 0
    & 0 \cr 0 & 0 & {1\over 2} & 0 & 0 & -{1\over 2} \cr 0 & -{1\over
    2} & 0 & 0 & {1\over 2} & 0 \cr } \right)\nonumber\\ {\bf X}^-_1\,
    =\, \left( \matrix{ 0 & 0 & 0 & 0 & 0 & 0 \cr 0 & 0 & {i\over 2} &
    0 & 0 & {i\over 2} \cr 0 & {i\over 2} & 0 & 0 & {{-i}\over 2} & 0
    \cr 0 & 0 & 0 & 0 & 0 & 0 \cr 0 & 0 & {i\over 2} & 0 & 0 & {i\over
    2} \cr 0 & {{-i}\over 2} & 0 & 0 & {i\over 2} & 0 \cr }
    \right)\,\,\,\,{\bf X}^-_2\,&=&\,
\left(  \matrix{ 0 & 0 & 0 & 0 & 0 & 0 \cr 0 & 0 & {1\over 2} & 0 & 0 & {1\over 2}
    \cr 0 & -{1\over 2} & 0 & 0 & {1\over 2} & 0 \cr 0 & 0 & 0 & 0 & 0
    & 0 \cr 0 & 0 & {1\over 2} & 0 & 0 & {1\over 2} \cr 0 & {1\over 2}
    & 0 & 0 & -{1\over 2} & 0 \cr } \right)\nonumber\\ {\bf Y}^+_1\,
    =\, \left(\matrix{ 0 & 0 & {i\over 2} & 0 & 0 & {{-i}\over 2} \cr
    0 & 0 & 0 & 0 & 0 & 0 \cr {i\over 2} & 0 & 0 & {{-i}\over 2} & 0 &
    0 \cr 0 & 0 & {i\over 2} & 0 & 0 & {{-i}\over 2} \cr 0 & 0 & 0 & 0
    & 0 & 0 \cr {i\over 2} & 0 & 0 & {{-i}\over 2} & 0 & 0 \cr
    }\right)\,\,\,\,{\bf Y}^+_2\,&=&\,
\left(   \matrix{ 0 & 0 & {1\over 2} & 0 & 0 & -{1\over 2} \cr 0 & 0 & 0 & 0 & 0 & 0
    \cr -{1\over 2} & 0 & 0 & {1\over 2} & 0 & 0 \cr 0 & 0 & {1\over
   2} & 0 & 0 & -{1\over 2} \cr 0 & 0 & 0 & 0 & 0 & 0 \cr -{1\over 2}
   & 0 & 0 & {1\over 2} & 0 & 0 \cr }\right)\nonumber\\ {\bf Y}^-_1\,
   =\, \left(\matrix{ 0 & 0 & {i\over 2} & 0 & 0 & {i\over 2} \cr 0 &
   0 & 0 & 0 & 0 & 0 \cr {i\over 2} & 0 & 0 & {{-i}\over 2} & 0 & 0
   \cr 0 & 0 & {i\over 2} & 0 & 0 & {i\over 2} \cr 0 & 0 & 0 & 0 & 0 &
   0 \cr {{-i}\over 2} & 0 & 0 & {i\over 2} & 0 & 0 \cr }
   \right)\,\,\,\,{\bf Y}^-_2\,&=&\,
\left(  \matrix{ 0 & 0 & {1\over 2} & 0 & 0 & {1\over 2} \cr 0 & 0 & 0 & 0 & 0 & 0
    \cr -{1\over 2} & 0 & 0 & {1\over 2} & 0 & 0 \cr 0 & 0 & {1\over
   2} & 0 & 0 & {1\over 2} \cr 0 & 0 & 0 & 0 & 0 & 0 \cr {1\over 2} &
   0 & 0 & -{1\over 2} & 0 & 0 \cr }\right)\nonumber\\ {\bf Z}^+_1\,
   =\, \left( \matrix{ 0 & {i\over 2} & 0 & 0 & {{-i}\over 2} & 0 \cr
   {i\over 2} & 0 & 0 & {{-i}\over 2} & 0 & 0 \cr 0 & 0 & 0 & 0 & 0 &
   0 \cr 0 & {i\over 2} & 0 & 0 & {{-i}\over 2} & 0 \cr {i\over 2} & 0
   & 0 & {{-i}\over 2} & 0 & 0 \cr 0 & 0 & 0 & 0 & 0 & 0 \cr }
   \right)\,\,\,\,{\bf Z}^+_2\,&=&\,
\left( \matrix{ 0 & {1\over 2} & 0 & 0 & -{1\over 2} & 0 \cr -{1\over 2} & 0 & 0 &
   {1\over 2} & 0 & 0 \cr 0 & 0 & 0 & 0 & 0 & 0 \cr 0 & {1\over 2} & 0
   & 0 & -{1\over 2} & 0 \cr -{1\over 2} & 0 & 0 & {1\over 2} & 0 & 0
   \cr 0 & 0 & 0 & 0 & 0 & 0 \cr } \right)\nonumber\\ {\bf Z}^-_1\,
   =\, \left( \matrix{ 0 & {i\over 2} & 0 & 0 & {i\over 2} & 0 \cr
   {i\over 2} & 0 & 0 & {{-i}\over 2} & 0 & 0 \cr 0 & 0 & 0 & 0 & 0 &
   0 \cr 0 & {i\over 2} & 0 & 0 & {i\over 2} & 0 \cr {{-i}\over 2} & 0
   & 0 & {i\over 2} & 0 & 0 \cr 0 & 0 & 0 & 0 & 0 & 0 \cr
   }\right)\,\,\,\,{\bf Z}^-_2\,&=&\,
\left( \matrix{ 0 & {1\over 2} & 0 & 0 & {1\over 2} & 0 \cr -{1\over 2} & 0 & 0 &
   {1\over 2} & 0 & 0 \cr 0 & 0 & 0 & 0 & 0 & 0 \cr 0 & {1\over 2} & 0
   & 0 & {1\over 2} & 0 \cr {1\over 2} & 0 & 0 & -{1\over 2} & 0 & 0
   \cr 0 & 0 & 0 & 0 & 0 & 0 \cr }\right)
\label{6x6matrices}
\end{eqnarray}


\end{document}